\documentclass[reprint,superscriptaddress,floatfix,preprintnumbers, nofootinbib,hyperref]{revtex4-2} 
\pdfoutput=1
\usepackage[title]{appendix}
\usepackage[colorlinks=true,breaklinks=true]{hyperref}
\usepackage[normalem]{ulem}
\usepackage[utf8]{inputenc}
\hypersetup{allcolors=[rgb]{0.0 0.0 0.6},linkcolor=[rgb]{0.75 0.05 0.05}}
\usepackage{amsmath,amssymb}
\usepackage{epsfig}  
\usepackage{graphicx}   
\usepackage{slashed}       
\usepackage{url}
\usepackage{color}
\usepackage{multirow}
\usepackage{comment}
\usepackage{amssymb}
\usepackage[version=3]{mhchem}
\usepackage{orcidlink}

\DeclareMathOperator{\cm}{cm}

\DeclareMathOperator{\s}{s}

\DeclareMathOperator{\kpc}{kpc}

\DeclareMathOperator{\Hz}{Hz}
\DeclareUnicodeCharacter{2212}{-}

\allowdisplaybreaks

\bibpunct{[}{]}{,}{n}{}{,}
\definecolor{ForestGreen}{RGB}{34,139,34}

\begin{document}

\title{Gravitational-wave signals for supernova explosions of three-dimensional progenitors}

\author{Alessandro\,Lella~\orcidlink{0000-0002-3266-3154}}
\email{alessandro.lella@unipd.it}
\affiliation{Dipartimento Interateneo di Fisica  ``Michelangelo Merlin'', Via Amendola 173, 70126 Bari, Italy}
\affiliation{Istituto Nazionale di Fisica Nucleare - Sezione di Bari, Via Orabona 4, 70126 Bari, Italy}
\affiliation{Dipartimento di Fisica e Astronomia, Universit{\`a} degli Studi di Padova, Via Marzolo 8, 35131 Padova, Italy}
\affiliation{Istituto Nazionale di Fisica Nucleare (INFN), Sezione di Padova, Via Marzolo 8, 35131 Padova, Italy}

\author{Giuseppe\,Lucente~\orcidlink{0000-0002-3266-3154}}
\email{lucenteg@slac.stanford.edu}
\affiliation{SLAC National Accelerator Laboratory,
2575 Sand Hill Rd, Menlo Park, CA 94025, USA}

\author{Daniel\,Kresse~\orcidlink{0000-0003-1120-2559}}
\email{danielkr@mpa-garching.mpg.de}
\affiliation{Max-Planck-Institut f{\"u}r Astrophysik, Karl-Schwarzschild-Str.~1, 85748 Garching, Germany}

\author{Robert\,Glas~\orcidlink{0000-0002-7040-9472}}
\email{rglas@mpa-garching.mpg.de}
\affiliation{Max-Planck-Institut f{\"u}r Astrophysik, Karl-Schwarzschild-Str.~1, 85748 Garching, Germany}

\author{Hans-Thomas\,Janka~\orcidlink{0000-0002-0831-3330}}
\email{thj@mpa-garching.mpg.de}
\affiliation{Max-Planck-Institut f{\"u}r Astrophysik, Karl-Schwarzschild-Str.~1, 85748 Garching, Germany}

\author{Alessandro\,Mirizzi~\orcidlink{0000-0002-5382-3786}}
\email{alessandro.mirizzi@ba.infn.it}
\affiliation{Dipartimento Interateneo di Fisica  ``Michelangelo Merlin'', Via Amendola 173, 70126 Bari, Italy}
\affiliation{Istituto Nazionale di Fisica Nucleare - Sezione di Bari, Via Orabona 4, 70126 Bari, Italy}%

\begin{abstract}
Core-collapse supernovae (SNe) are sources of gravitational waves (GWs) produced by hydrodynamical instabilities and highly time-dependent anisotropies of the neutrino radiation. In this work we analyze both contributions to the GW signal for two state-of-the-art three-dimensional (3D) SN models computed with the \textsc{Prometheus-Vertex} neutrino-hydrodynamics code. In contrast to the far majority of models analyzed for GWs so far, our core-collapse simulations were started with 12.28\,$M_\odot$ (18.88\,$M_\odot$) progenitors, whose final hour (7\,min) of convective oxygen-shell burning was computed in 3D and featured a vigorous oxygen-neon shell merger. The corresponding large-scale asymmetries in the oxygen layer are conducive to buoyancy-aided neutrino-driven explosions. The models were continuously evolved in 3D from the pre-collapse evolution until 5.11\,s (1.68\,s) after the core bounce. The GW signals result from the well-known dynamical phenomena in the SN core such as prompt postshock convection, neutrino-driven convection, the standing accretion shock instability, proto-neutron star oscillations, and anisotropic ejecta expansion. They do not exhibit any new or specific features that can be unambiguously connected to the powerful pre-collapse activity in the progenitors, but we identify interesting differences compared to results in the literature. We also discuss measurement prospects by interferometers, confirming that GW signals from future Galactic SNe will be detectable with existing and next-generation experiments working in the frequency range $f\sim 1-2000\,$Hz. 
\end{abstract}
\date{\today}
\maketitle

\section{Introduction}

Gravitational waves (GWs) were predicted in 1916 by Einstein's theory of general relativity. Beyond previous indirect signatures, the first detection of this signal was achieved in 2015 from a binary black-hole merger~\cite{LIGOScientific:2016aoc}. After that, GWs from other astrophysical binary systems were observed (see, e.g., Ref.~\cite{LIGOScientific:2017vwq}), opening a new window for multi-messenger astrophysics~\cite{Ronchini:2022gwk}. 
In this context, the detection of GW signals from core-collapse supernova (SN) explosions constitutes one of the next frontiers for GW astronomy \cite{Kalogera+2021}.
The production of GWs in stellar collapse events is triggered by a time-variable global deformation, for example connected to rotation or jets, as well as highly time-dependent hydrodynamical instabilities~\cite{Wheeler:1966tg,Finn:1990qf,Ott:2008wt,Murphy:2009dx,Mueller:2012sv} and anisotropic neutrino emission~\cite{Epstein:1978dv,Turner:1978jj}. 
Notably, general relativity predicts that the asymmetric ejection of matter or emission of radiation could cause a non-oscillatory permanent change in the spacetime metric. Since the shape of the strain depends on the properties of the source phenomenon, the gravitational perturbation of the spacetime should carry memory of the powerful astrophysical events that produce the GW emission. For this reason this effect is commonly known as \emph{gravitational memory effect}~\cite{Zeldovich:1974gvh,cite-key}. 
 
Significant production of GWs in collapsing stars, even without rotation, is caused by anisotropic mass motions during different phases of the evolution: prompt convective overturn due to negative entropy and lepton-number gradients takes place in the postshock layer during a few 10\,ms after core bounce; convection and the standing accretion-shock instability (SASI; \cite{Blondin:2002sm,Foglizzo:2002hi,Foglizzo:2005xr,Foglizzo:2006fu,Blondin:2006fx,Blondin+2007,Foglizzo:2011aa}) produce violent non-radial flows in the neutrino-heating region between the protoneutron star (PNS) and the stalled SN shock emitting GWs of typically 100--200\,Hz; anisotropic mass accretion onto the PNS before and after the onset of the explosion instigates g-, p-, and f-mode oscillations of the PNS at higher frequencies; PNS convection and the impact of intermittently accreted (fallback) matter create a long-lasting f-mode vibration of the PNS with a characteristically increasing frequency well beyond 1000\,Hz; and the asymmetrically expanding ejecta lead to a very-low frequency signal contribution (below a few 10\,Hz) and a persistent displacement (``memory'') of the GW strain (for recent reviews, see~\cite{Ott:2008wt,Kotake2013,KotakeKuroda2017_Handbook,Abdikamalov:2020jzn,Kalogera+2021,Mezzacappa:2024zph}). Although the corresponding GW signals are highly case-dependent and stochastic, many of the features that might be distinguishable in the observed wave strains can be clearly linked to the peculiar processes that determine the evolution of the collapse, the onset of the explosion, and the propagation of the SN shock. Rapid rotation in collapsing stellar cores can imprint additional characteristic features on the GW emission such as prominent, quickly damped large-amplitude oscillations due to the asymmetric core bounce \cite{Moenchmeyer+1991,Dimmelmeier+2002}, (quasi-)periodic post-bounce modulations caused by spiral waves and triaxial deformation~\cite{Shibagaki+2020,Takiwaki:2021dve}, and frequency shifts or frequency-band smearing of the PNS oscillation modes~\cite{Powell+2023,Powell:2024nvv}. Moreover, anisotropic neutrino emission is expected to yield a unique low-frequency contribution to the GW signal~\cite{Epstein:1978dv,Turner:1978jj,Burrows:1995bb,Mueller+1997}.
Thus, the detection of the GW signal from a future Galactic SN, in coincidence with the associated neutrino burst, would provide a once-in-a-lifetime opportunity to obtain deeper insight into the complex physical processes that cause and accompany a core-collapse SN explosion. 

Predicting GW signals from SNe requires self-consistent multidimensional simulations, in which the mutual feedback between matter evolution and neutrino transport is taken into account. A huge body of work has been conducted since the first analyses of 2D and 3D models with still highly approximate inclusion of neutrino effects in the early 1990's \cite{Moenchmeyer+1991,Burrows:1995bb,Mueller+1997}. Many steps making gradual upgrades of the hydrodynamics codes, transport treatments and interaction descriptions for neutrinos, and of the nuclear equation-of-state (EoS) in the PNS have led to the current state-of-the-art multidimensional SN simulations. These have meanwhile reached the frontier of full 4$\pi$ modeling in three dimensions, include either approximate or detailed general relativistic effects, and span evolution times up to several seconds after core bounce. In the course of this work, GW amplitudes, spectrograms, and spectral energy distributions have been investigated for their detectability and for their dependence on a large variety of aspects: the emission phases, spatial regions, and oscillation modes that contribute to the GW signals; the influence of spatial dimensions and viewing directions, of stellar progenitor properties, success or failure of the explosion, rotation, and magnetic fields; the impact of non-standard particle physics and of different EoSs for the nuclear matter in PNSs, including features connected to hadron-quark phase transitions (e.g., \cite{Dimmelmeier+2002,Mueller:2003fs,Sumiyoshi:2005ri,Kotake:2005zn,Ott+2007,Dimmelmeier+2007,Dimmelmeier+2008,Marek:2008qi,Murphy:2009dx,Kotake:2009em,Kotake:2009rr,Scheidegger+2010a,Scheidegger+2010b,Reisswig+2011,Ott+2012,Mueller:2012sv,Andresen:2016pdt,Gossan+2016,Pan:2017tpk,Richers:2017joj,Andresen:2018aom,Morozova:2018glm,Radice:2018usf,Powell:2018isq,Torres-Forne:2019zwz,Pajkos:2019nef,Mezzacappa+2020,Mezzacappa+2023,Shibagaki+2020,Andresen:2020jci,Vartanyan:2020nmt,Shibagaki+2021,Andersen:2021vzo,Takiwaki:2021dve,Pan+2021,Kuroda+2022,Afle:2023mab,Bugli+2023,Vartanyan:2023sxm,Choi:2024irp,Powell:2024nvv,Ehring+2024,Jakobus+2025,Powell+2025,Schnauck+2025,Murphy+2025,Sykes+2025}). 

Besides the GW emission generated by nonspherical mass motions, the low-frequency GW ``memory'' components generated by asymmetric emission of neutrinos \cite{Burrows:1995ww,Mueller+1997,Muller:2011yi} have attracted particular recent interest \cite{Mukhopadhyay:2021gox,Mukhopadhyay:2021zbt,Richardson:2021lib,Richardson+2024,Richardson+2025}. Their detection for a future Galactic SN by proposed space-based deci-Hertz GW interferometers \cite{Kawamura:2008zz,Kawamura+2021} may offer a unique opportunity to probe the gravitational memory effect predicted by general relativity.

Expanding this timely line of research, we analyze here the GW signal from two state-of-the-art 3D SN models for progenitor stars with zero-age-main-sequence (ZAMS) masses of 12.28\,$M_\odot$ and 18.88\,$M_\odot$, respectively ($M_\odot = 1.989\times 10^{33}$\,g being the solar mass). The novel aspect of our work is that both of these models were started from 3D progenitors that have been evolved continuously in 3D from their final phase of convective oxygen-shell burning through core collapse and the onset of neutrino-driven explosions until several seconds after the core bounce. Both of these models experience violent oxygen-neon shell mergers that lead to large-scale and large-amplitude velocity and density asymmetries as well as chemical composition inhomogeneities in the convectively burning oxygen shell at the onset of iron-core collapse. The substantial perturbations in the stellar matter falling into the stalled SN shock are conducive to the onset of neutrino-driven explosions, because they accelerate and amplify the growth of convective and SASI instabilities in the postshock layer \cite{Couch+2013,Mueller+2015,Mueller+2017,Bollig:2020phc}. One might suspect that these perturbations in the pre-collapse progenitor might also have an impact on the amplitude and time structure of the GW signal. Our analysis shows, however, that the GW emission exhibits the well-known features witnessed in neutrino-driven explosion simulations. Although certain signal properties might be connected to large-scale vorticity in the convective oxygen shell, we cannot identify any obvious, easily discernible diagnostic features that permit to extract information on vigorous mass motions in the oxygen-dominated composition shells of a SN progenitor from a future GW measurement.

Our paper is structured as follows. 
In Sec.~\ref{sec:overview} we provide a summary of the main properties of our 3D pre-collapse and 3D SN explosion models analyzed for GW emission.
In Sec.~\ref{sec:matter} we describe the GW signals produced by non-radial mass motions, and in Sec.~\ref{sec:neutrino} we present a similar evaluation of the GW signals produced by anisotropic neutrino emission both in the time and frequency domains. Then, in Sec.~\ref{sec:comparison}, we compare our GW results to those from similar SN models in the literature.
Finally, in Sec.~\ref{sec:conclu} we assess the detection perspectives of these signals by current and future GW interferometers and draw our conclusions. 
In Appendix~\ref{app:code} we report results to verify our post-processing code for the GW extraction by reproducing previously published matter GW signals~\cite{Andresen:2020jci} for two additional 3D core-collapse simulations, for which we also present the corresponding newly evaluated neutrino-induced GW signals.

\begin{figure*}
    \centering
    \includegraphics[width=0.9\textwidth]{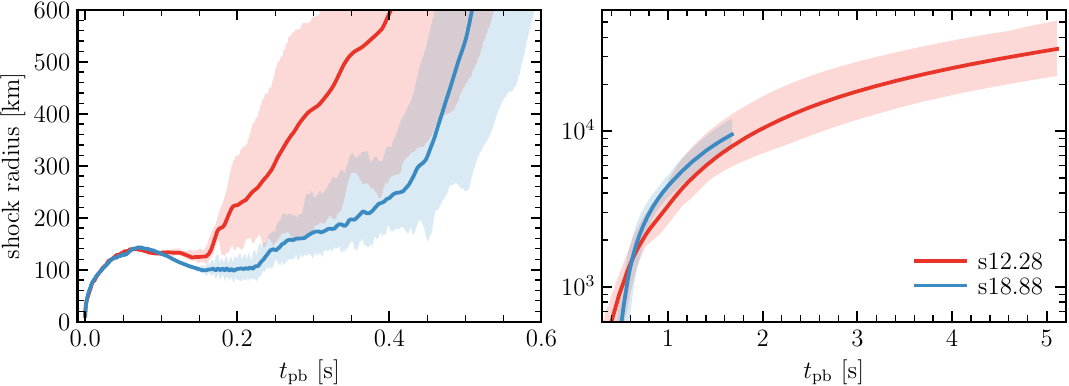}
    \caption{Evolution of the shock radius versus post-bounce time $t_\mathrm{pb}$ during the first 600\,ms after bounce (\textit{left}) and until the end of our simulations (\textit{right}) for models s12.28 (red) and s18.88 (blue). The solid lines show the spherically averaged shock radius, the shaded bands indicate the range between the minimum and maximum shock radii. Note the different scales on the axes of both plots.}
    \label{fig:ShockRadii}
\end{figure*}

\begin{table*}[t!]
\centering
	\caption{
    Characteristic properties of the two simulations considered in this work, namely the employed nuclear EoS, the explosion time $t_{\rm exp}$ (defined as the time after bounce when the spherically averaged shock radius exceeds 400\,km), the final post-bounce time $t_{\rm fin}$, and the progenitor compactness $\xi_M$ computed by Eq.~\eqref{Eq:compactness} for different values of the enclosed mass of $M=1.5$, 1.75, 2.0, and 2.5\,$M_\odot$. The last two columns contain the total energy radiated via GWs produced by asymmetric mass motions ($E_{\rm GW}^{\rm M}$) and anisotropic neutrino emission ($E_{\rm GW}^\nu$) until time $t_{\rm fin}$.
    }
    \begin{tabular}{l c c c c c c c c c}
    \hline
    Model\,\,\,\,\,\,\,\,& \,\,\,\,\,\,\,\,EoS \,\,\,\,\,\,\,\,& \,\,\,\,\,\,\,\,$t_{\rm exp}$\,\,\,\,\,\,\,\, & \,\,\,\,\,\,\,\,$t_{\rm fin}$\,\,\,\,\,\,\,\, & \,\,\,\,\,\,\,\,$\xi_{1.5}$\,\,\,\,\,\,\,\,  &  \,\,\,\,\,\,\,\,$\xi_{1.75}$\,\,\,\,\,\,\,\,   & \,\,\,\,\,\,\,\,$\xi_{2.0}$\,\,\,\,\,\,\,\,  &  \,\,\,\,\,\,\,\,$\xi_{2.5}$\,\,\,\,\,\,\,\,  &  \,\,\,\,\,\,\,\,$E_{\rm GW}^{\rm M}~(t_{\rm fin})$\,\,\,\,\,\,\,\, &  \,\,\,\,\,\,\,\,$E_{\rm GW}^\nu~(t_{\rm fin})$\,\,\,\,\,\,\,\,\\
    &  & [s] & [s] &  &    &  &   & [$M_\odot\,c^2$] &  [$M_\odot\,c^2$]\\
    \hline
    s12.28 & SFHo & 0.292 & 5.110 & 0.515 & 0.201 & 0.118 & 0.032 & $6.4\times 10^{-10}$ & $4.2\times 10^{-11}$ \\
    s18.88 & LS220 & 0.469 & 1.675 & 0.992 & 0.780 & 0.473 & 0.283 & $1.0\times 10^{-9}$ & $7.0\times10^{-11}$  \\
    \hline
	\end{tabular}
	\label{tab:SNmodels}
\end{table*}

\section{Overview of three-dimensional models}
\label{sec:overview}

In this work we analyze the GW signals for two successfully exploding 3D core-collapse SN models, namely s12.28 and s18.88, computed with the \textsc{Prometheus-Vertex} neutrino-hydrodynamics code~\cite{Rampp:2002bq,Buras:2005rp}. Here, the Newtonian hydrodynamics equations are solved by the \textsc{Prometheus} hydrodynamics module~\cite{Fryxell+1991,Mueller+1991,Kifonidis:2003fv} and the energy and velocity $[\mathcal{O}(v/c)]$ dependent three-flavor neutrino transport is solved by the \textsc{Vertex} transport module~\cite{Rampp:2002bq}. The transport in three dimensions is handled by a ray-by-ray-plus (RbR+) approach, which computes only radial neutrino flux components based on the assumption that the neutrino phase space distribution is (approximately) axially symmetric around the radial direction.\footnote{The good overall agreement of neutrino properties and hydrodynamic evolution in 3D simulations with the RbR+ approximation and fully multidimensional transport was shown in~\cite{Glas:2018oyz} (see also Appendix~\ref{app:neutrinosignals} for more discussion of this aspect).} General relativistic corrections are used in the gravitational potential (\cite{Rampp:2002bq}, with Case~A of \cite{Marek+2006}) and neutrino transport \cite{Rampp:2002bq,Buras:2005rp}. A comprehensive summary of the physics inputs in the latest version of the \textsc{Prometheus-Vertex} code is provided in Ref.~\cite{Fiorillo:2023frv}. 

\begin{figure*}
    \centering
    \includegraphics[width=\textwidth]{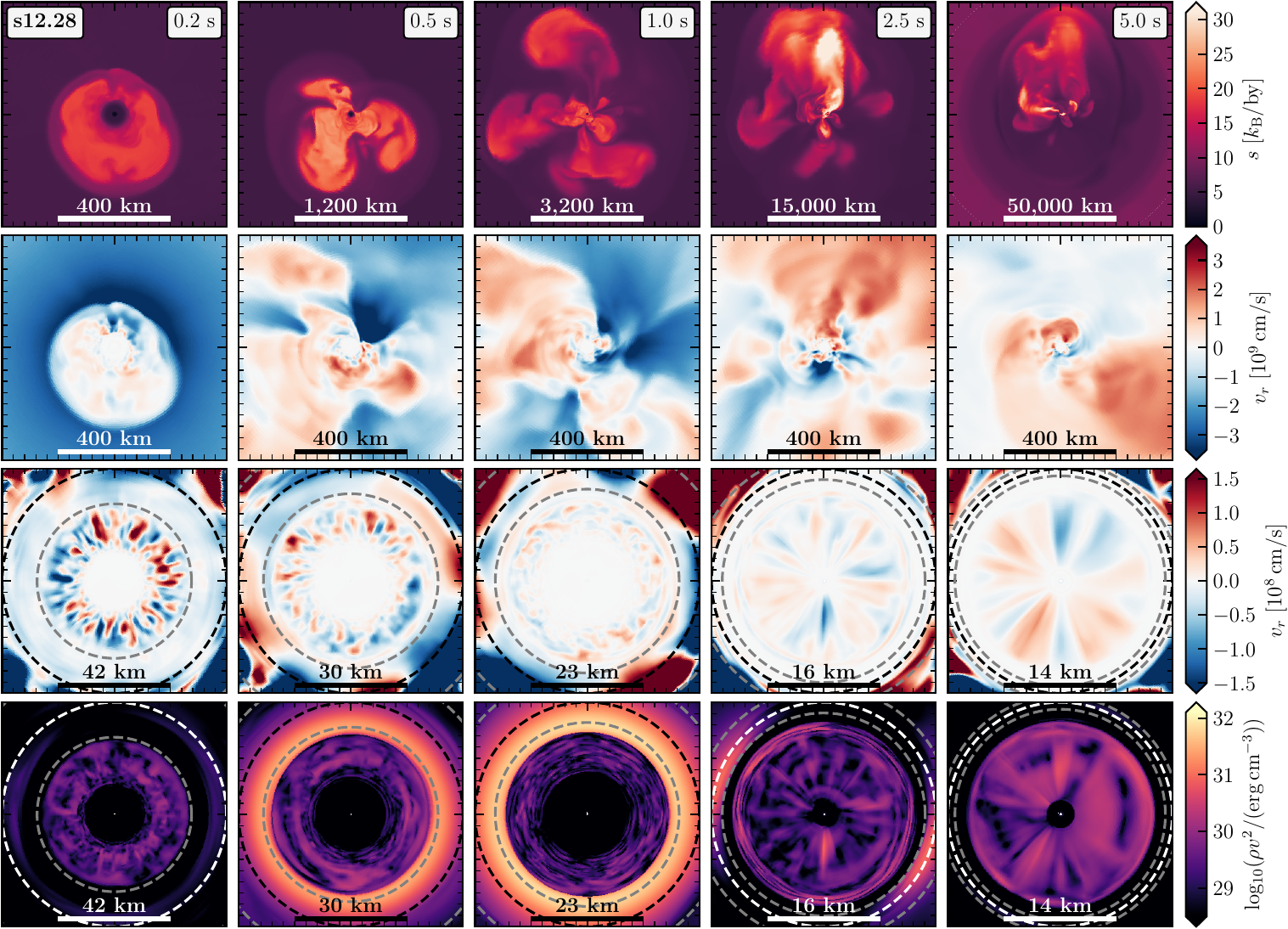}
    \caption{Snapshots of model s12.28, showing cross-sectional cuts ($x$-$y$ plane) of the specific entropy in the postshock region (\textit{top row}), the radial gas velocity in the surroundings of the PNS (\textit{second row}), and the radial flow velocity (\textit{third row}) and kinetic energy density in and near the PNS (\textit{bottom row}), all color-coded according to the color bars on the right side of each row. The dashed lines, with growing radius, correspond to contours for spherically averaged matter densities of $3\times10^{12}$, $1\times10^{11}$, and $5\times10^{9}$\,g\,cm$^{-3}$. Note the different and time-dependent length scales indicated by yard sticks. The post-bounce times of the snapshots in each column are given in the \textit{top panels}. In the \textit{upper two rows} one can recognize outward-rising high-entropy plumes of neutrino-heated matter separated by lower-entropy accretion downflows. In the \textit{lower two rows} one can witness a shell with growing radial depth where convection takes place in the PNS interior, surrounded by a gravity-wave perturbed, convectively stable accretion layer. Intense yellow and orange in this layer in the \textit{bottom panels} indicate rapid rotation due to angular momentum received from accretion downflows.}
    \label{fig:Dynamics_s12.28}
\end{figure*}
\begin{figure*}
    \centering
    \includegraphics[width=0.9\textwidth]{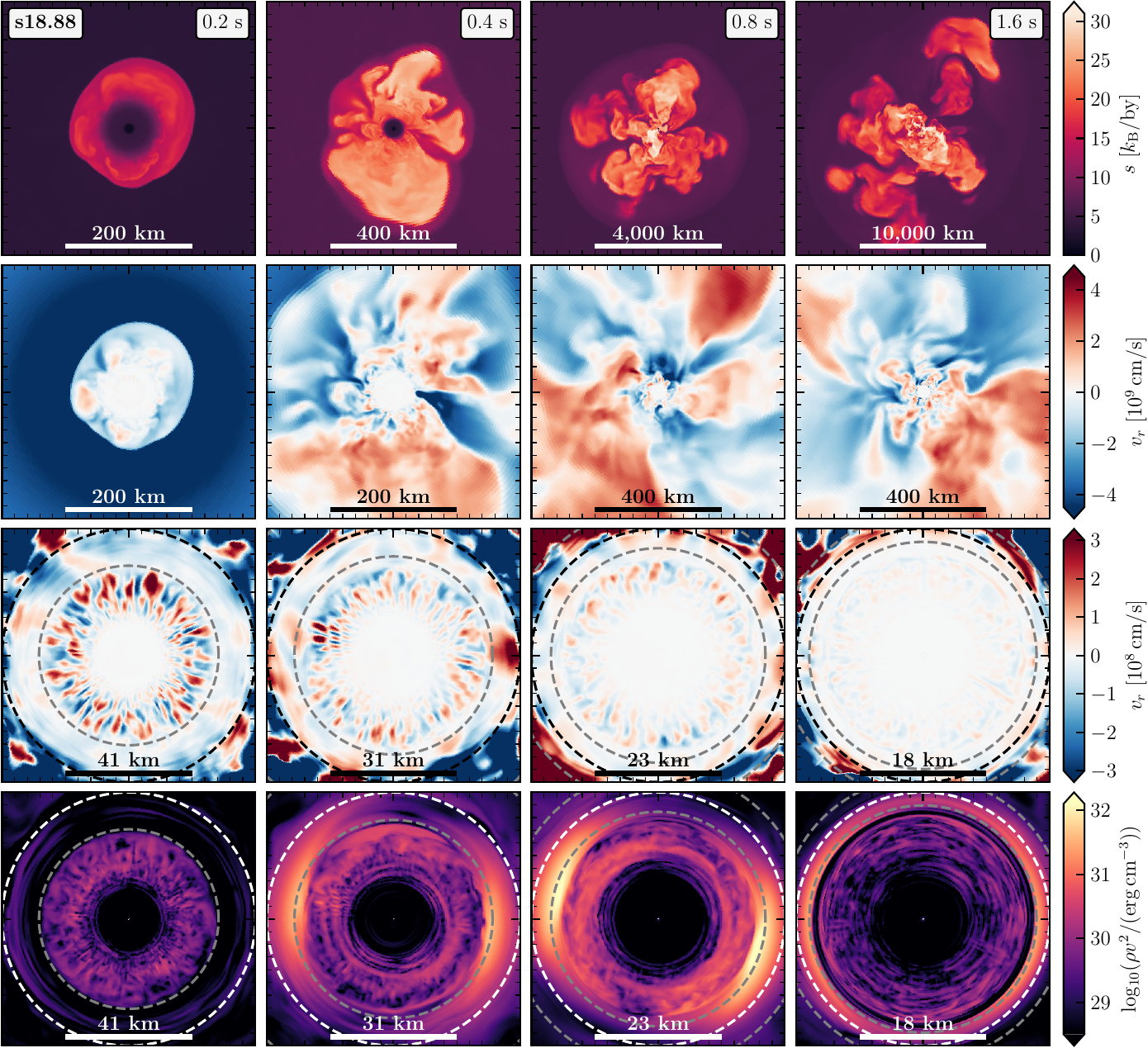}
    \caption{Snapshots of model s18.88, showing cross-sectional cuts ($x$-$y$ plane) of the specific entropy in the postshock region (\textit{top row}), the radial gas velocity in the surroundings of the PNS (\textit{second row}), and the radial flow velocity (\textit{third row}) and kinetic energy density in and near the PNS (\textit{bottom row}), all color-coded according to the color bars on the right side of each row. The dashed lines, with growing radius, correspond to contours for spherically averaged matter densities of $3\times10^{12}$, $1\times10^{11}$, and $5\times10^{9}$\,g\,cm$^{-3}$. Note the different and time-dependent length scales indicated by yard sticks. The post-bounce times of the snapshots in each column are given in the \textit{top panels}. In the \textit{upper two rows} one can recognize outward-rising high-entropy plumes of neutrino-heated matter separated by lower-entropy accretion downflows. In the \textit{lower two rows} one can witness a shell with growing radial depth where convection takes place in the PNS interior, surrounded by a gravity-wave perturbed, convectively stable accretion layer. Intense yellow and orange in this layer in the \textit{bottom panels} indicate rapid rotation due to angular momentum received from accretion downflows.}
    \label{fig:Dynamics_s18.88}
\end{figure*}

The s12.28 simulation, which was already discussed in Refs.~\cite{Janka:2024xbp,Janka2025}, was launched from a non-rotating progenitor with a ZAMS mass of $M_{\rm ZAMS}=12.28\,M_\odot$~\cite{Sukhbold:2017cnt,Ramalatswa+2026} and was evolved with the SFHo nuclear EoS of Steiner, Fischer, and Hempel~\cite{Hempel:2011mk,Steiner:2012rk}. For the s18.88 simulation with a non-rotating progenitor of $M_{\rm ZAMS}=18.88\,M_\odot$~\cite{Sukhbold:2017cnt,Yadav:2019rmo} the EoS of Lattimer and Swesty~\cite{Lattimer:1991nc} (LS220, using an incompressibility modulus of 220\,MeV) was applied and some of its results were reported in Refs.~\cite{Bollig:2020phc,Sieverding+2023,Janka:2024xbp,Janka2025}. Both simulations were performed with an axis-free polar Yin-Yang grid \cite{Kageyama+2004,Wongwathanarat+2010} and have an angular resolution of 3.5$^\circ$ (s12.28) and 2$^\circ$ (s18.88), respectively. The Yin-Yang grid ensures uniform angular resolution on the entire $4\pi$ sphere and thus avoids resolution differences in the different axis directions of polar grids.

Multidimensional initial conditions at the onset of iron-core collapse were generated for both SN runs by self-consistently simulating the final phase of convective oxygen-shell burning of the progenitor stars in 3D with the \textsc{Prometheus} code. These pre-collapse simulations were performed for seven minutes for the s18.88 model~\cite{Yadav:2019rmo} and for a full hour for model s12.28~\cite{Ramalatswa+2026}. The simulations were sufficiently long to follow the development of a vigorous oxygen-neon shell merger in both cases. The shell mergers happened shortly before the gravitational instability of the iron core when the increasingly stronger convective oxygen burning entrained an initially unburned, neon-rich layer embedded by the oxygen shell of the 1D progenitor model.\footnote{Such a shell-merger phenomenon could be quite common in far-evolved massive stars \cite{Rizzuti+2024,Whitehead+2026}. For example, high-resolution X-ray spectroscopy of the galactic SN remnant of Cassiopeia~A (Cas~A) reveals peculiar abundances of chemical elements that are typically produced during oxygen and neon burning. This has recently been interpreted as evidence that an oxygen-neon shell merger might also have happened during the final hours before the core collapse in the progenitor of the Cas~A SN~\cite{Sato+2025,XRISM:2025ujc}.} The enhanced energy release by nuclear burning led to the creation of large-scale asymmetries with big amplitudes in the density distribution, velocity field, and chemical composition. 

When these asymmetries in the infalling oxygen shell reached the stalled SN shock, they amplified the hydrodynamic instabilities in the neutrino-heated gain layer behind the shock. The stronger postshock convection pushed the shock to larger radii, thus increased the volume and mass in the gain layer and consequently also the neutrino energy transfer. This positive feedback engendered further shock expansion and triggered the onset of the SN explosions in both the s12.28 and s18.88 models. The corresponding shock evolution as a function of post-bounce time is displayed in Fig.~\ref{fig:ShockRadii}. The counterparts of these simulations that were started from the 1D pre-collapse models as adopted from the stellar evolution calculations in Ref.~\cite{Sukhbold:2017cnt} did not explode.

\begin{figure*}
    \centering
    \includegraphics[width=0.97\textwidth]{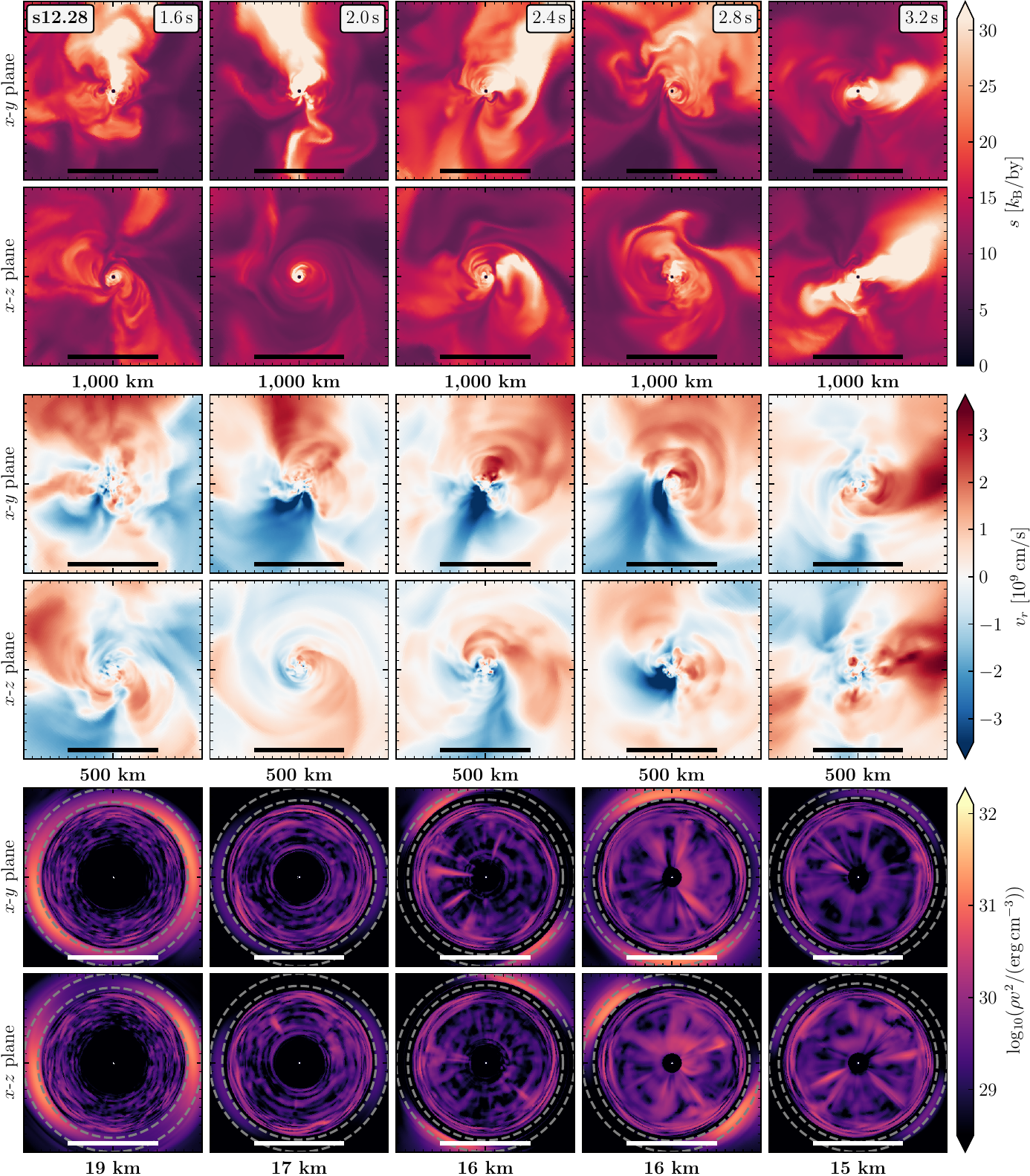}
    \vspace{-5pt}
    \caption{Snapshots of accretion and PNS dynamics in model s12.28 between 1.6\,s and 3.2\,s after bounce. The panels show cross-sectional cuts in the $x$-$y$ and $x$-$z$ planes (labeled on the left of each row) for the specific entropy (\textit{top two rows}), the radial gas velocity in the vicinity of the PNS (\textit{middle two rows}), and the kinetic energy density in and near the PNS (\textit{bottom two rows}). The color coding is defined by the color bars on the right. The outer (inner) dashed circle marks the contour for a spherically averaged matter density of $10^{11}$\,g\,cm$^{-3}$ ($10^{13}$\,g\,cm$^{-3}$). Note the different and time-dependent length scales indicated by the yard sticks. The post-bounce times of the snapshots in each column are given in the \textit{top panels}. In the \textit{upper four rows} one can recognize outward-rising high-entropy plumes of neutrino-heated matter separated by lower-entropy accretion downflows. In the \textit{bottom two rows} one can witness a shell with growing radial depth where convection takes place in the PNS interior, surrounded by a gravity-wave perturbed, convectively stable accretion layer. Intense yellow and orange in this layer in the \textit{bottom panels} indicate rapid rotation due to angular momentum received from accretion downflows.}
    \label{fig:LateDynamics_s12.28}
\end{figure*}

Table~\ref{tab:SNmodels} provides an overview of the main properties of the two SN models including the employed EoS, the time when the explosion sets in, $t_{\rm exp}$, the time when the simulation with detailed neutrino transport was stopped, $t_{\rm fin}$, and the compactness parameter~\cite{OConnor:2010moj}, 
\begin{equation}
    \xi_M \equiv \frac{M/M_\odot}{r(M)/1000~{\rm km}}\,,
    \label{Eq:compactness}
\end{equation}
for different values of the enclosed mass $M$ [$r(M)$ is the corresponding radius inside the star]. Moreover, the table also lists the total GW energy connected to nonspherical mass motions, $E_{\rm GW}^{\rm M}$ (see Sec.~\ref{sec:matter}), and the total energy radiated via GWs caused by anisotropic neutrino emission $E_{\rm GW}^\nu$ (see Sec.~\ref{sec:neutrino}). Additional details of the results of both SN simulations considered in this work can be found in Refs.~\cite{Janka:2024xbp,Janka2025}.

Figures~\ref{fig:Dynamics_s12.28} and~\ref{fig:Dynamics_s18.88} display the time evolution of the two SN models with a focus on the GW producing regions ---asymmetric ejecta, accretion downflows onto the PNS, and convection in the PNS interior--- by cross-sectional cuts in the $x$-$y$ plane of the 3D computational grid on different spatial scales. The top panels show the large-scale structure of the highly aspherical explosion with outward rising, high-entropy plumes of neutrino-heated matter and low-entropy accretion downdrafts enveloped by the deformed SN shock that can be recognized as an entropy discontinuity (though barely visible in some of the panels). The panels in the second row are enlargements of the vigorously turbulent closer surroundings of the PNS, visualizing the impact of the supersonic downflows on the PNS surface and the re-ejection of most of the infalling matter after absorption of energy from neutrinos.  

The shock revival that marks the onset of the explosion and the initial shock expansion are strongly influenced by the pre-collapse asymmetries in the convective oxygen layer. These asymmetries are carried inward by the collapsing stellar matter exterior to the stalled shock. As discussed in Refs.~\cite{Mueller+2017,Bollig:2020phc}, the shock preferentially expands into angular directions with lower densities ahead of the shock. Such regions correspond to reduced mass-accretion rates and thus weaker ram pressure, and they correlate with convective updrafts in the oxygen-burning layer. In this way large-scale convective structures in the oxygen shell leave an imprint in a global asphericity of the expanding SN shock that can persist from the early phase or even from the beginning of the explosion to the late-time evolution.

The extremely turbulent flows around the PNS do not possess the same long-term stability, but instead they exhibit variability of their geometry on time scales of seconds. Figures~\ref{fig:Dynamics_s12.28} and~\ref{fig:LateDynamics_s12.28} show this fact for model s12.28, whose post-bounce evolution was followed for more than 5\,s. For example, between $\sim$\,0.2\,s and $\gtrsim$\,1.0\,s after bounce, massive accretion downdrafts penetrate inward to the PNS from the $(+x)$-$(+y)$ quadrant, whereas there is a prominent, wide-angle outflow in the $+y$ direction from $\sim$\,1.4\,s until $\sim$\,3\,s, shifting slowly to the $+x$ direction, where it is strongest at later times until it moves more to the $-y$ direction near the end of the simulation. One should note that the flow pattern near the PNS does not closely resemble the flow geometry on the large scales, where the nonspherical distribution of the ejecta and the deformed SN shock adopt an effectively stable shape after about 1\,s (see Figs.~\ref{fig:Dynamics_s12.28} and~\ref{fig:Dynamics_s18.88}).    

\begin{figure*}[t!]
    \centering
    \includegraphics[width=0.9\textwidth]{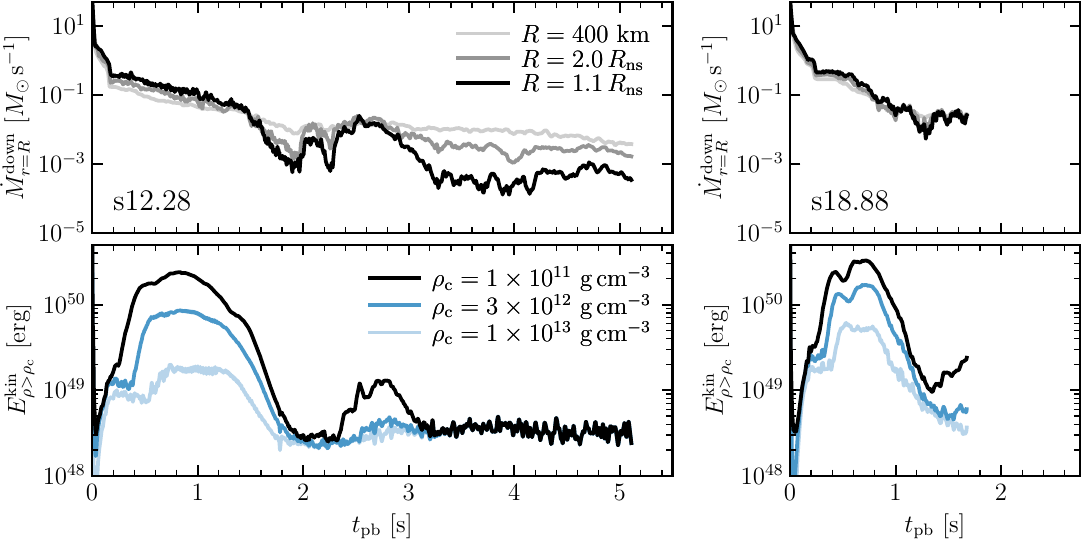}
    \caption{Mass-inflow rate (\textit{upper panels}) and PNS kinetic energy (\textit{lower panels}) as functions of post-bounce time $t_\mathrm{pb}$ in models s12.28 (\textit{left}) and s18.88 (\textit{right}). The mass-inflow rate is evaluated in accretion downflows (i.e., flows with negative radial velocity $v_r<0$) at a radius of $R = 400$\,km (light gray), at twice the time-dependent PNS radius $R_\mathrm{ns}$ (dark gray), and at 1.1 times the PNS radius (black), where the PNS radius is defined at a spherically averaged density of $10^{11}$\,g\,cm$^{-3}$. The kinetic energy inside the PNS is measured in volumes bounded by the PNS radius (black) and outer radii corresponding to spherically averaged densities of $3 \times 10^{12}$\,g\,cm$^{-3}$ (dark blue) and $10^{13}$\,g\,cm$^{-3}$ (light blue), respectively. The lowest density roughly corresponds to the PNS surface, the last two densities are near the outer edge of the PNS convection layer (see Figs.~\ref{fig:Dynamics_s12.28}--\ref{fig:LateDynamics_s12.28}).}
    \label{fig:Mdot_Ekin}
\end{figure*}

All phases when the inflow-outflow geometry around the PNS changes are correlated with dramatic variations of the mass-accretion rate onto the PNS, especially in the time intervals of $1.5\,\mathrm{s} \lesssim t_\mathrm{pb} \lesssim 1.9\,\mathrm{s}$ and between $t_\mathrm{pb} \sim 2.4$\,s and $\sim$\,3.2\,s (Fig.~\ref{fig:Mdot_Ekin}). In model s18.88 similar effects are visible between 0.8\,s and 1.4\,s with a migration of the main accretion downdrafts from the $(-x)$-$(+y)$ and $(+x)$-$(-y)$ quadrants to the $(-x)$-$(-y)$ and $(+x)$-$(+y)$ quadrants (Fig.~\ref{fig:Dynamics_s18.88}). In Sec.~\ref{sec:Time-domain analysis} we will see that such time intervals coincide with episodes when the GW matter-memory signals experience changes in their long-term trends.

The variations of the accretion geometry around the PNS are likely to be connected to convective cell patterns in the 3D oxygen-burning layer, which is swept up by the outward propagating SN shock. However, these variations are not a unique feature of SN simulations started from 3D progenitor models. Generally, the interaction of the expanding plumes of neutrino-heated matter with the narrow accretion downdrafts is so violent that the flow pattern in the close vicinity of the PNS is strongly affected by this hydrodynamic interaction. The accretion flows reaching the PNS surface are highly time-variable as suggested by the mass-inflow rate arriving at a radius of 1.1\,$R_\mathrm{ns}$ in Fig~\ref{fig:Mdot_Ekin} ($R_\mathrm{ns}$ is the PNS radius defined at the location where the spherically averaged density is $10^{11}$\,g\,cm$^{-3}$). Moreover, the downdrafts carry large amounts of angular momentum, which leads to their deflection on the way inward (see panels in the upper four rows of Fig.~\ref{fig:LateDynamics_s12.28}) and an enormous spin-up of the accretion layer of the PNS (bottom panels of Figs.~\ref{fig:Dynamics_s12.28}--\ref{fig:LateDynamics_s12.28}). The corresponding rotational kinetic energy at densities between $10^{11}$\,g\,cm$^{-3}$ and a few $10^{12}$\,g\,cm$^{-3}$ dominates the kinetic energy of convection in the PNS interior (at $\rho \gtrsim 10^{13}$\,g\,cm$^{-3}$) by a factor of 5 and more (Fig.~\ref{fig:Mdot_Ekin}). Although all of these phenomena should be influenced to some extent by the convective mass motions in the oxygen-burning layer, the flow dynamics around the PNS is too complex to identify characteristic effects or properties that are unambiguously and exclusively linked to the pre-collapse 3D structures in the progenitor star. Therefore we will see that the GW matter signal does not exhibit any obvious features that provide information about violent hydrodynamic processes in the deep interior of massive stars prior to their core collapse.

\begin{figure*}[t!]
    \centering
    \includegraphics[width=1\textwidth]{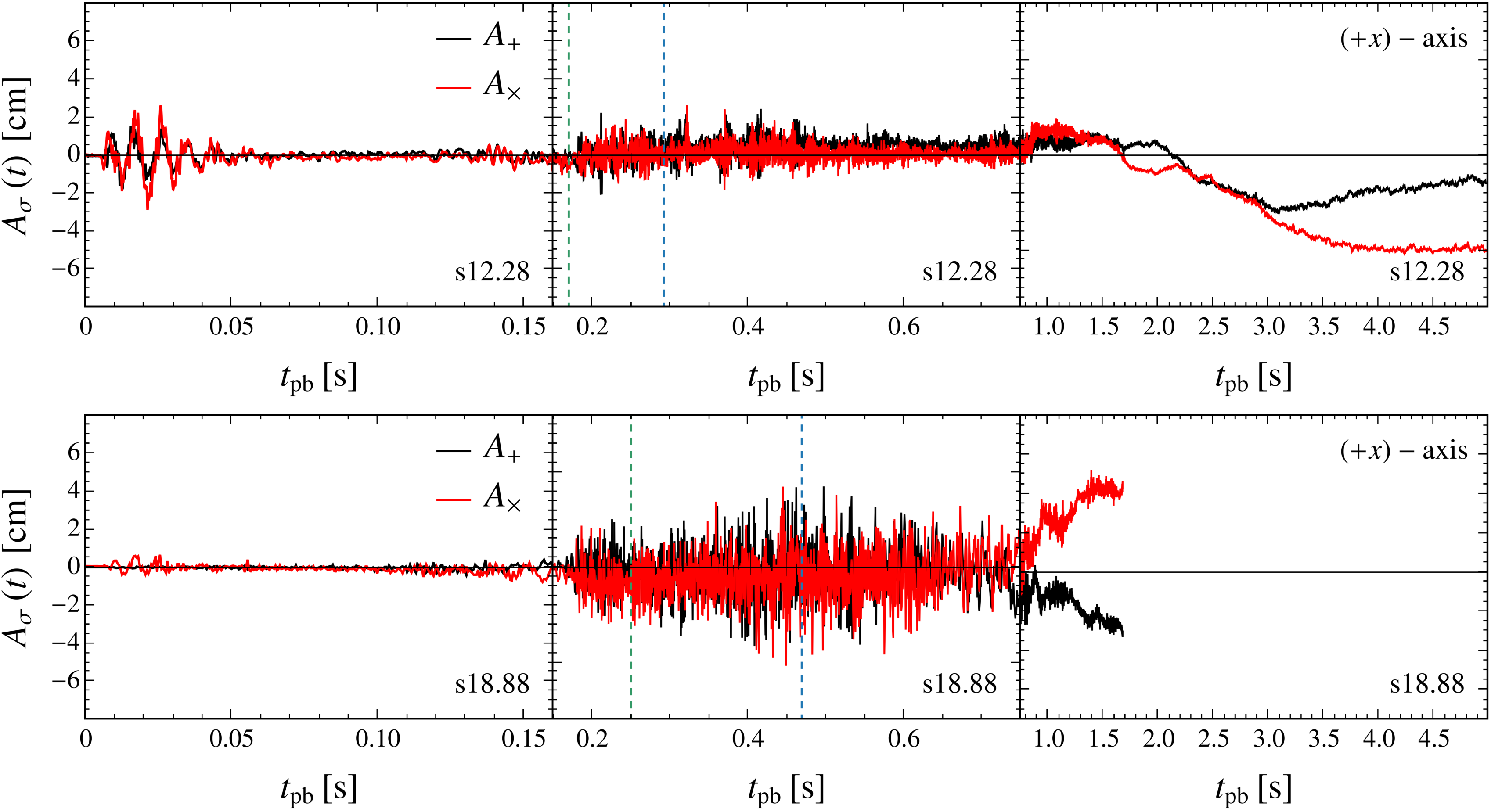}\\
    \includegraphics[width=1\textwidth]{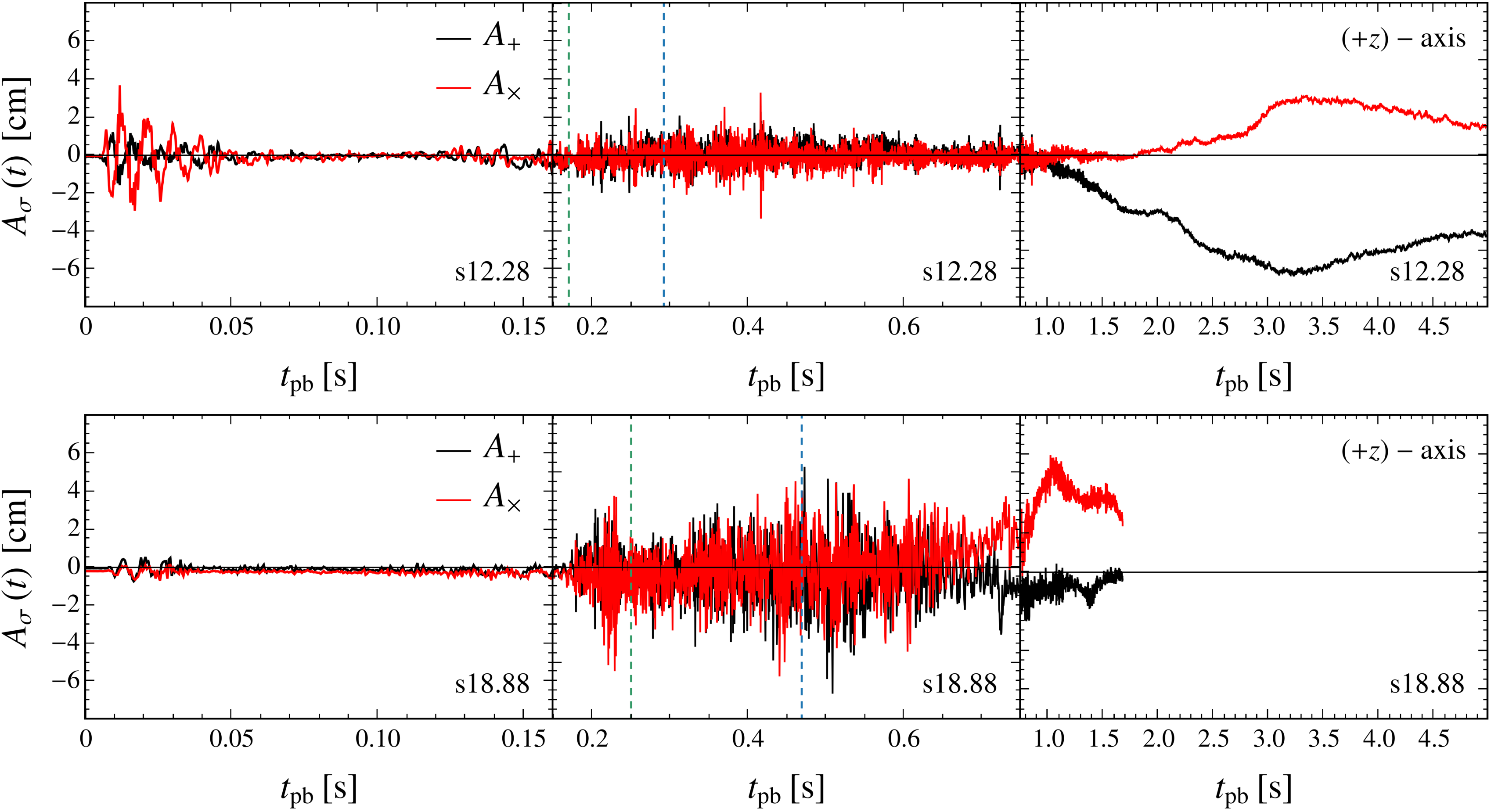}    
    \caption{GW amplitudes from mass motions in models s12.28 (\emph{upper panels}) and s18.88 (\emph{lower panels}) as functions of post-bounce time $t_\mathrm{pb}$. Black and red curves correspond to the two possible GW polarization states. As examples, we show here the results for an observer located along the $(+x)$-axis in the equatorial plane of the source [$(\phi,\theta)=(0,\pi/2)$] ({\em upper two plots}) and for an observer above the north pole, i.e., along the $(+z)$-axis [$(\phi,\theta)=(0,0)$] ({\em lower two plots}). The \emph{left panels} display the early post-bounce phase where the GW emission is dominated by prompt postshock convection. The \emph{middle panels} focus on the long-lasting and loud stochastic signal caused by SASI activity, strong convection in the neutrino-heated postshock layer, and PNS accretion before and shortly after the onset of the explosions. The dashed vertical blue lines mark these explosion times $t_\mathrm{exp}$ according to the values in Table~\ref{tab:SNmodels} ($t_\mathrm{exp} \approx 0.29$\,s for s12.28 and $\approx$\,0.47\,s for s18.88), while the dashed vertical green lines indicate the earlier times when the stalled SN shocks start to expand outward (see Fig.~\ref{fig:ShockRadii}). Finally, the \emph{right panels} display the GW amplitudes during the later time when the matter memory signals develop due to the anisotropic expansion of the shock front and explosion ejecta.}
    \label{Fig:GW_hydro_signal}
\end{figure*}

\section{Gravitational waves from hydrodynamical mass motions}
\label{sec:matter}

\subsection{Signal in the time domain}
\label{sec:Time-domain analysis}

The extraction of the GW signals connected to non-radial mass flows has been performed by strictly following the treatment described in Refs.~\cite{Misner:1973prb, Mueller+1997, Muller:2011yi, Andresen:2016pdt}. We verified our newly implemented post-processing analysis for the matter-induced GW signals by reproducing published GW results in the literature (Ref.~\cite{Andresen:2020jci}) for two other 3D core-collapse simulations of an exploding 9.0\,$M_\odot$ progenitor and a non-exploding 20\,$M_\odot$ progenitor; see Appendix~\ref{app:matter}.

In the slow-motion limit, the GW amplitudes of both independent (plus and cross) polarizations can be expressed as linear combinations of the second time derivatives of the transverse-traceless mass quadrupole-moment tensor components in the spherical coordinate system of the simulation (see Ref.~\cite{Misner:1973prb} for details). In the following, we will consider as an illustrative example an observer located along the $(+x)$-axis of the simulation frame [$(\phi,\theta)=(0,\pi/2)$], for which case the relation between the GW amplitudes and the quadrupole moment is particularly simple. Namely, one has
\begin{equation}
    \begin{split}
         &A_+=\frac{G}{c^4}\,(\Ddot{Q}_{zz}-\Ddot{Q}_{yy}) \, ,\\
        &A_\times=-2\,\frac{G}{c^4}\,\Ddot{Q}_{yz}\,. 
    \end{split}
    \label{Eq:amplitudes}
\end{equation}
The corresponding expressions for an observer along the north-polar, i.e., $(+z)$-axis, direction [$(\phi,\theta)=(0,0)$] are
\begin{equation}
    \begin{split}
         &A_+=\frac{G}{c^4}\,(\Ddot{Q}_{xx}-\Ddot{Q}_{yy}) \, ,\\
        &A_\times=2\,\frac{G}{c^4}\,\Ddot{Q}_{xy}\,. 
    \end{split}
    \label{Eq:amplitudes2}
\end{equation}

For an observer located in an arbitrary $(\phi,\theta)$ direction, the general expressions of the quadrupole-moment components in polar coordinates can be found in Ref.~\cite{Oohara:1997qd}. We make use of the fact that in a Cartesian orthonormal basis (spatial indices $i,j=1,2,3$) the second-order time derivatives of the quadrupole-moment tensor can be written as~\cite{Muller:2011dvt,Andresen:2020jci}
\begin{equation}
    \Ddot{Q}_{ij}={\rm STF} \left[2 \int d^3 x\,\rho\,\left(v_i v_j- x_i\partial_j\varphi \right)\right] ,
    \label{Eq:Quadrupole}
\end{equation}
where $\varphi$ is the effective general relativistic gravitational potential, $\rho$ the mass-density, and $v_i$ and $\partial_i$ are the velocity components and partial derivatives in the Cartesian basis considered. STF denotes the operator projecting the quadrupole-moment tensor onto its symmetric, trace-free part.\footnote{Using Eq.~(\ref{Eq:Quadrupole}) to determine the GW amplitudes leads to viewing-direction dependent offsets of $A_+$ and $A_\times$ from zero at $t_{\rm pb}=0$. These numerical artifacts are caused by discretization errors connected to the mapping of quantities from the polar grid used in our simulations to the Cartesian coordinates of the GW analysis. They can be removed by multiplying the diagonal elements of $\ddot{Q}_{ij}$ with correction factors that are close to unity and depend on the spatial resolution of the computational grid. These ``geometrical'' correction factors are constant in time and independent of the viewing angle, but lead to corresponding corrections of the amplitudes that depend on the observer direction and can vary with time. Remaining, very small amplitude displacements (of order 0.1\,cm) from zero at $t_{\rm pb}=0$ are corrected by adding or subtracting additional, direction-dependent but time-independent shifts.}

Moreover, we point out here that the evaluation of $\ddot{Q}_{ij}$ according to Eq.~(\ref{Eq:Quadrupole}) differs from the one employed in Ref.~\cite{Vartanyan:2023sxm}, where $\ddot{Q}_{ij}$ was computed by taking the numerical time derivative of $\dot{Q}_{ij}$ (see Appendix there). This numerical differentiation will depend on how the simulation outputs are sampled in time, and it may imply the risk of an increased level of numerical noise at high frequencies.

\begin{figure*}[t!]
    \centering
    \includegraphics[width=1\textwidth]{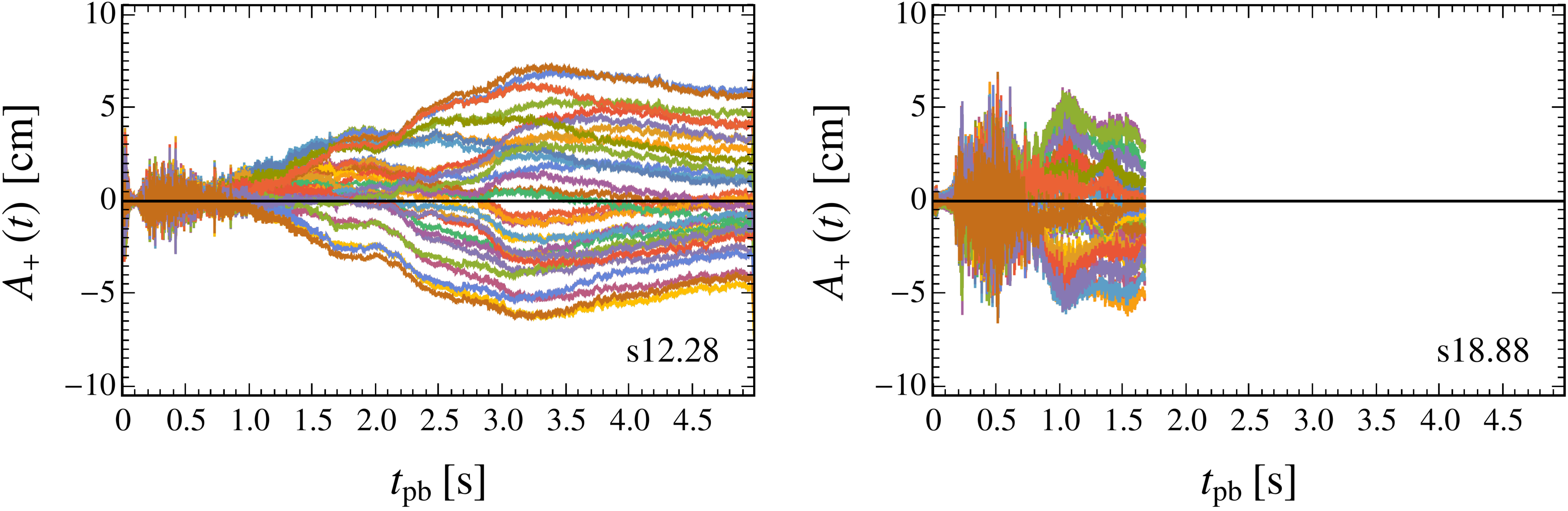}
    \caption{GW amplitudes associated with the $A_+$ polarization mode for the matter signal of model s12.28  (\emph{left panel}) and model s18.88 (\emph{right panel}), respectively. The different colors depict different randomly chosen viewing angles of the large-distance observer.}
    \label{Fig:GW_hydro_signal_directions}
\end{figure*}

Figure~\ref{Fig:GW_hydro_signal} displays the temporal evolution of $A_+$ (black lines) and $A_\times$ (red lines) as functions of post-bounce time $t_{\rm pb}$ for the s12.28 (upper panels) and the s18.88 (lower panels) models. In the early phase of the GW emission (left panels), the s12.28 model exhibits quasi-periodic oscillations of the amplitudes of both polarizations around zero with maxima up to 2--3\,cm at $t_{\rm pb} \lesssim 50$~ms. As described in Refs.~\cite{Burrows:1995bb, Marek:2008qi, Muller:2011dvt}, this feature is associated with prompt postshock convection. In contrast, model s18.88 shows only a weak indication of such an effect in $A_\times$ (reaching maxima of nearly 1\,cm) and essentially none in $A_+$ when observed from the $(+x)$-direction, although in both simulations the collapse through core bounce and shock formation until shock stagnation was computed in full 3D, starting from the 3D initial structures prior to core collapse. From other viewing directions $A_+$ in the first 50\,ms after bounce is similarly large as $A_\times$, also with maxima up to $\sim$\,1\,cm (see Fig.~\ref{Fig:GW_hydro_signal_directions}). But overall, the prompt-postshock GW signal in s18.88 is much weaker than in s12.28. The main reason for this difference is the use of the LS220 EoS in s18.88. Despite essentially identical shock propagation until $t_\mathrm{pb} \sim 50$\,ms in both of our models (see Fig.~\ref{fig:ShockRadii}), prompt postshock convection with LS220 occurs only in a narrow, Ledoux-unstable layer, where negative entropy and lepton-number gradients were formed when the decelerating SN shock turned into an accretion shock. This was discussed in detail in Ref.~\cite{Marek:2008qi} on the basis of 2D simulations with different nuclear EoSs including the LS180 EoS of~\cite{Lattimer:1991nc} with an incompressibility modulus of 180\,MeV (3D simulations with the LS220 EoS show very similar properties of the prompt postshock convection;~\cite{Kresse+2026}). Since the narrow convective layer involves relatively little mass, the GW signal becomes correspondingly weak. We cannot exclude that the considerably higher mass-accretion rate in model s18.88 (upper panels in Fig.~\ref{fig:Mdot_Ekin}) has an additional damping influence on the prompt convective overturn.

After this GW emission phase connected to prompt postshock convection, both models possess a more quiescent period between $t_{\rm pb}\sim50$\,ms and roughly 160\,ms, during which hydrodynamic instabilities (SASI and convection) grow in the neutrino-heated hot-bubble volume. These instabilities are further boosted when the convectively perturbed oxygen-burning shell of the 3D progenitor model begins to fall through the SN shock at $t_\mathrm{pb}\sim 170$\,ms in s12.28 and $t_\mathrm{pb}\sim 230$\,ms in s18.88, nearly coincident with a steep drop of the mass-accretion rate correlated with a density decline at the Si/O shell interface (Fig.~\ref{fig:Mdot_Ekin}). At about these times the SN shocks begin to expand outward to ultimately initiate the successful SN blasts. Convection in the postshock volume gives rise to a strong stochastic GW signal that exhibits a phase of slightly higher maximum amplitudes in the time interval between $t_{\rm pb}\sim0.35$\,s and $\sim$\,0.6\,s, which brackets the onset of the explosion in model s18.88 (see middle panels of Fig.~\ref{Fig:GW_hydro_signal}). However, in both exploding models a period of strong high-frequency GW activity lasts until at least 1.5\,s after bounce.

Besides a low-frequency GW component with frequencies of up to a few 100\,Hz connected to the convective (and SASI) mass motions between PNS and SN shock \cite{Andresen:2016pdt}, a high-frequency signal component (with frequencies from $\sim$\,500\,Hz increasing to several 1000\,Hz) is caused by supersonic accretion downflows that strike the PNS surface~\cite{Mueller:2012sv, Vartanyan:2023sxm}. This phenomenon instigates PNS oscillation eigenmodes \cite{Murphy:2009dx,Marek:2008qi,Mueller:2012sv,Morozova:2018glm,Torres-Forne:2019zwz}, which are commonly classified on the basis of the dominant restoring force, namely pressure for p-modes and fundamental (f-) modes and buoyancy for g-modes~\cite{Cowling:1941nqk,Rodriguez:2023nay}. 

Only short SASI episodes make visible, weak imprints on the GW signal in both models between $\sim$\,120\,ms and $\sim$\,180\,ms, where low-amplitude, quasi-periodic modulations with frequencies around 50--200\,Hz are visible in Fig.~\ref{Fig:GW_hydro_signal}. Overall, the low-mass model s12.28 shows smaller GW amplitudes than s18.88 (up to about 2--3\,cm compared to 5--7\,cm peak values in the latter case; see Figs.~\ref{Fig:GW_hydro_signal} and~\ref{Fig:GW_hydro_signal_directions}) in the time interval between 0.15\,s and 1.5\,s. This difference is a consequence of less massive downflows of matter as typical for progenitor models with lower compactness values~(see Table~\ref{tab:SNmodels}). 

The right panels of Fig.~\ref{Fig:GW_hydro_signal} reveal that in the time interval between $t_{\rm pb} \sim 0.7$\,s and $\sim$\,1\,s the effects caused by accretion downdrafts abate and the high-frequency modulation of the GW emission becomes significantly weaker. Instead, at $t_{\rm pb}\gtrsim 1$\,s a gradual and slow rise of the GW amplitude towards a long-lasting non-zero value sets in. This signal feature is connected to the asymmetric expansion of the shock and SN ejecta and coincides with the time when the ejecta on large scales attain a stable nonspherical geometry as mentioned above. The large displacement of the GW amplitude away from the zero level is typical for the phase when asymmetric explosions develop \cite{Burrows:1995bb, Murphy:2009dx, Muller:2011yi, Mueller:2012sv} and can be considered as a GW ``memory'' that manifests itself in a long-term deformation of spacetime.

Superimposed on the overall tendency to a growing amplitude excursion one can also witness low-frequency (Hz or sub-Hz) modulations that lead to non-monotonic phases and trend inversions. These are probably caused by the changes of the flow geometry around the PNS described in connection with Figs.~\ref{fig:Dynamics_s12.28}--\ref{fig:LateDynamics_s12.28}. Moreover, a long-term trend of a monotonically decreasing excursion of the GW amplitudes in model s12.28 after about 3.5\,s may be explained by the fact that the outward propagating shock sweeps up more and more mass from the spherically distributed matter of the outer stellar layers. During the late GW emission at $t\gtrsim 1.5$\,s in model s12.28, one still notices the presence of small-scale oscillations on the GW amplitudes connected to long-persistent high-frequency activity in the PNS with a nearly constant strength. This signal feature is partly driven by accretion downflows to the PNS, which keep hitting the PNS surface. However, this explanation does not appear very likely at $t_\mathrm{pb} \gtrsim 3$\,s. At these very late times PNS convection may be a more plausible GW source, because the rate of mass incidence onto the PNS has declined to very low values ($\lesssim 10^{-3}\,M_\odot\,\mathrm{s}^{-1}$) and the kinetic energy inside the PNS (at densities above $10^{11}$\,g\,cm$^{-3}$) is solely determined by PNS convection on a fluctuating level around a basically constant average value of $\sim$\,$3.5\times 10^{48}$\,erg (Fig.~\ref{fig:Mdot_Ekin}). 

In Fig.~\ref{Fig:GW_hydro_signal_directions} we show the matter GW amplitudes $A_+$  for both of our models as functions of $t_{\rm pb}$ for a large number of arbitrarily chosen viewing angles. The left panel refers to model s12.28 and the right panel is for s18.88. As discussed before, at early times, $t_{\rm pb}\lesssim 0.7$\,s, the signals are effectively stochastic and there is no clear preference of any observer direction dependence, in particular when one also takes into account the cross-polarization amplitude $A_\times$. Only during SASI episodes the GW signal can exhibit characteristic differences depending on the observer position perpendicular or in the plane of the SASI mass motions (see, e.g., \cite{Andresen:2018aom}). SASI activity in our models, however, is constrained to narrow time intervals (of duration $\sim$\,60\,ms) before the inward falling Si/O interface passes the SN shock and the GW emission gains substantially in strength. During the first second after bounce peaks in the amplitudes of both polarization states do not exceed $\sim$\,3.5\,cm for s12.28 and $\sim$\,7\,cm for s18.88 in any direction. At later times the stable large-scale morphology of the highly nonspherical explosion implies that the excursions of $A_+$ and $A_\times$ from the zero-level and their time variations discussed above are strongly viewing-angle dependent. In model s12.28 maximum amplitudes during this late-time evolution reach up to 4 times the typical values during the initial second before the already mentioned secular trend of slow, gradual decline sets in.

\begin{figure}[!]
    \centering
    \includegraphics[width=1\columnwidth]{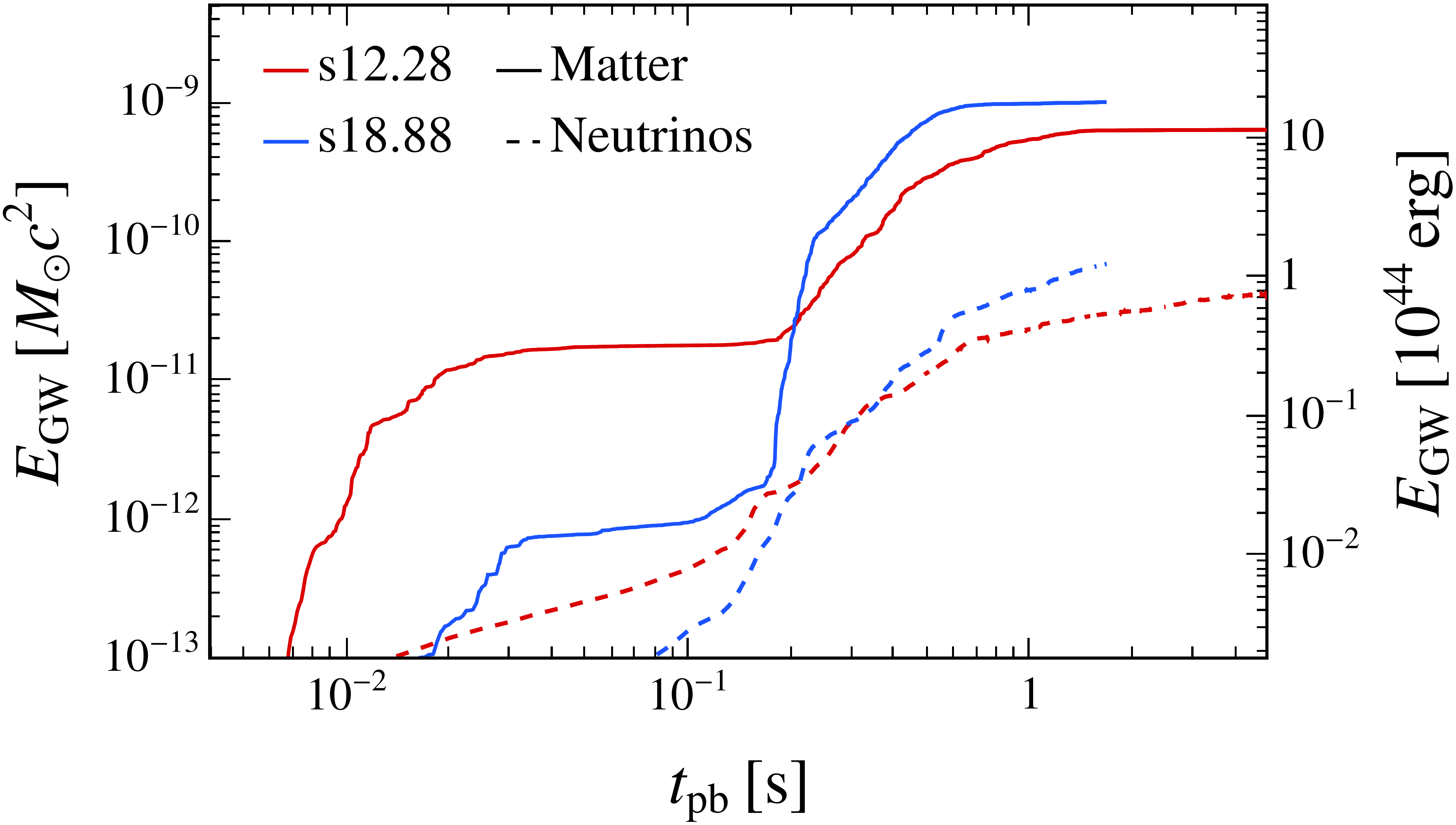}
    \caption{Cumulative energy released in GWs induced by non-radial mass flows (solid lines) and anisotropic neutrino emission (dashed lines) as a function of post-bounce time $t_{\rm pb}$ for the s12.28 (red) and s18.88 (blue) models.
    }
    \label{Fig:Energy_both}
\end{figure}

The total energy radiated via GWs produced by hydrodynamic mass motions in SNe until a time $t$ can be computed as~\cite{Misner:1973prb,Muller:2011yi,Vartanyan:2023sxm}
\begin{equation}
    E_{\rm GW}^{\rm M}(t)=\frac{G}{5\,c^5}\int_0^t dt' \sum_{ij}{\dddot{Q}^2_{ij}\,}(t') \,,
\label{Eq: GWEnergyMatter}
\end{equation}
where $\dddot{Q}_{ij}$ denotes the third time derivative of the symmetric trace-free part of the quadrupole-moment tensor, computed by numerical time differentiation of $\ddot{Q}_{ij}$ from Eq.~(\ref{Eq:Quadrupole}). Once more we repeat that in quantitative terms the detailed results from Eqs.~(\ref{Eq:Quadrupole}) and~(\ref{Eq: GWEnergyMatter}) may depend on the way how the simulation data are sampled in the time domain and how the time derivatives of $\ddot{Q}_{ij}$ and $\dddot{Q}_{ij}$ are numerically evaluated. Figure~\ref{Fig:Energy_both} displays the time evolution of the cumulative energy released in matter-induced GWs (solid lines) for the two SN models considered in this work. The growth behavior is characterized by periods of rapid increases associated with the different emission phases described above, until the growth asymptotes to nearly final values at $t_{\rm pb} \gtrsim 1$\,s. In both SN models most of the GW power is radiated around the onset of the explosion and shortly afterwards, when the stochastic signals are produced by  vigorous SASI and convective activity in the postshock volume and massive accretion downflows onto the PNS. 

However, the evolution in both models differs in important aspects. The initial rise of the GW energy occurs later in the $18.88\,M_\odot$ model than in the case of the $12.28\,M_\odot$ simulation. This difference can be understood by the presence of well developed prompt postshock convection in the first 50\,ms after bounce in s12.28, which produces roughly $2\times 10^{-11}\,M_\odot c^2$ of GW energy until $t_{\rm pb}\sim0.1\,\s$, whereas in model s18.88 the postshock convection shortly after the core bounce is very weak. Figure~\ref{Fig:Energy_both} also reveals that 95\% of the GW energy is radiated until $t_{\rm pb} \approx 1.3\,\s$ and $t_{\rm pb} \approx 0.6\,\s$ in s12.28 and s18.88, respectively. When the simulations are stopped, the GW energy released by mass motions is ${E^{\rm M}_{\rm GW}\simeq 6.4\times10^{-10}\,M_{\odot}c^2\approx 1.1\times10^{45}}$\,erg in model s12.28, and ${E^{\rm M}_{\rm GW}\approx 1.0\times 10^{-9}\,M_{\odot}c^2\approx 1.8\times10^{45}}$\,erg in model s18.88 (Table~\ref{tab:SNmodels}).

\begin{figure*}[t!]
    \centering
    \includegraphics[width=1\textwidth]{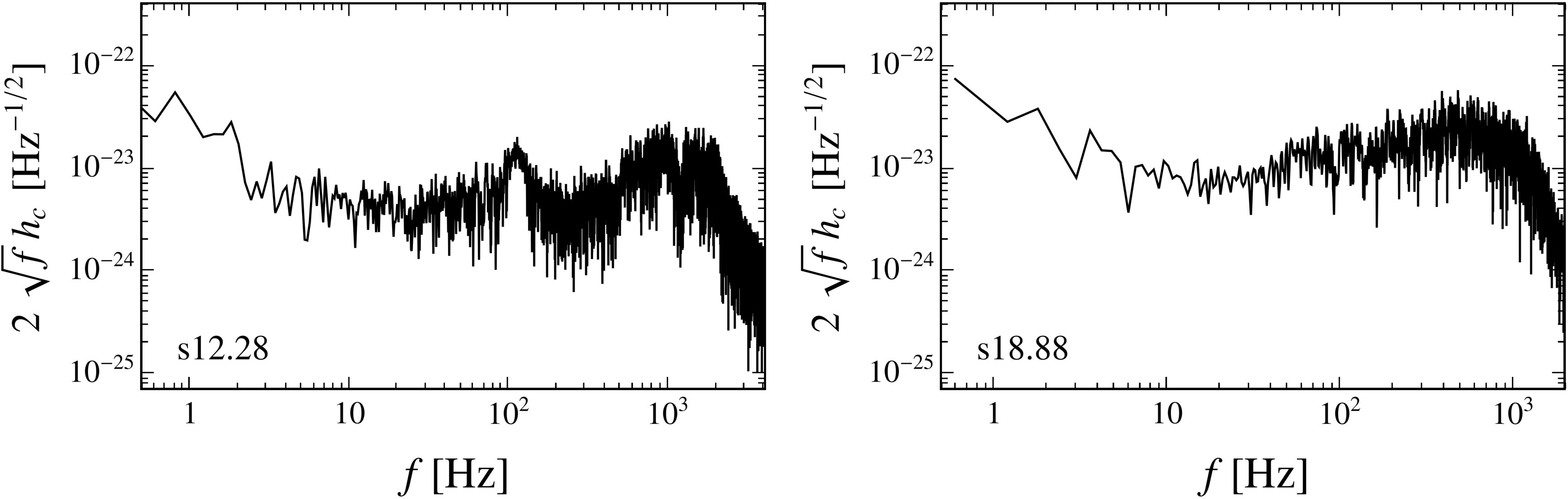}
    \caption{Amplitude spectral densities [ASDs; Eqs.~(\ref{Eq:ASD})  and~(\ref{Eq:CharStrain})] for the matter-induced GW signal vs.\ frequency $f$ for an observer located along the $(+x)$-axis in the equatorial plane of the source [$(\phi,\theta)=(0,\pi/2)$] at a distance $D=10\,\kpc$. The \emph{left} and \emph{right panels} correspond to SN models s12.28 and s18.88, respectively. Note that the power above 1000\,Hz in the case of s18.88 is underestimated due to aliasing effects (see also Fig.~\ref{Fig:Spectrograms}).}
    \label{Fig:GW_hydro_freq}
\end{figure*}

\subsection{Signal in the frequency domain}
\label{Sec:Hydro_freq}

It is useful to introduce  the amplitude spectral density~(ASD),  employed to compute the signal-to-noise ratio for GW detectors~(see e.g., Ref.~\cite{Maggiore:2007ulw}). The ASD is defined as
\begin{equation}
    {\rm ASD}=2\,\sqrt{f}\,h_c
    \label{Eq:ASD}
\end{equation}
in terms of the frequency $f$ and the characteristic strain~\cite{Takami:2014tva}
\begin{equation}
h_c(f) =\sqrt{0.5\,\left(|\Tilde{h}_+(f)|^2+|\Tilde{h}_\times(f)|^2\right)}\, ,
\label{Eq:CharStrain}
\end{equation}
where $\Tilde{h}_{+,\times}$ are the Fourier transforms of the dimensionless GW strains
\begin{equation}
h_{+,\times} = \frac{A_{+,\times}}{D}\,,
\end{equation}
with $D$ being the distance to the source. In our analysis we will adopt $D=10$\,kpc for a typical distance to a Galactic SN. The Fourier transforms $\Tilde{h}_{\sigma}(f)$ of the GW strains $h_{\sigma}(t)$ (where $\sigma = +,\times$) are given by
\begin{equation}
    \Tilde{h}_{\sigma}(f)=\int_{-\infty}^{+\infty} h_{\sigma}(t)\,e^{2\pi i f t}\,dt\,.
\label{Eq:ContinuousFT}
\end{equation}
Since the strains are sampled at discrete time steps, a reliable evaluation of these quantities requires the introduction of the discrete Fourier transform formalism as described in Ref.~\cite{Andresen:2018aom}:
\begin{equation}
    \widetilde{X}_k(f_k)=\frac{T}{N}\sum_{n=1}^N x_n\,e^{i\,2\pi k n/N}\,,
    \label{Eq:DFT}
\end{equation}
where $x_n$ is the time series obtained by sampling the signal at $N$ equidistant discrete time instants labeled by the index $n$, and $f_k=k/T$ is the frequency of bin $k$, with $T$ being the total duration of the analyzed signal. Because the simulation outputs are not exactly equally spaced in time, the discrete Fourier transform formalism is applied after interpolating the GW strains to equidistant time intervals with sampling steps of $\delta t=0.2\,$ms and $\delta t=0.5\,$ms for models s12.28 and s18.88, respectively. Moreover, to avoid the introduction of non-physical high-frequency noise due to a sharp cut-off of the signals at the boundaries of the evaluated time interval, we apply a Turkey window function~\cite{doi:10.1137/1019120} that drives the GW signal smoothly to zero over a time scale of $\sim200\,$ms at the boundaries of the given time domain.

Additionally, in order to analyze the time evolution of the GW signals for their dominant modes in frequency space during the sequence of emission phases discussed before, we apply the so-called short-time Fourier transform (STFT)~\cite{1184347} to the waveforms in Fig.~\ref{Fig:GW_hydro_signal}. This is obtained by convolving the GW amplitudes with a sliding Kaiser window (shape parameter $\beta=2.5$) of $50\,{\rm ms}$ width and applying the discrete Fourier transform defined in Eq.~\eqref{Eq:DFT} to each time window. Thus, for a given signal $x_n$ sampled at discrete time steps $\Delta t$, the STFT reads~\cite{1184347}
\begin{equation}
    \mathrm{STFT}[x]\,(t_n, f_k)=\frac{T}{N}\sum_{j=0}^{N-1} x_j\,\,w_{j-n}\,\,e^{i\,2\pi k j/N}\,,
    \label{eq:STFT}
\end{equation}
where $w_{j-n}$ is the window function centered at the $n$-th time instant $t_n=n\Delta t$, while the other quantities involved are defined as in Eq.~\eqref{Eq:DFT}.

\begin{figure*}[t!]
    \centering
    \includegraphics[width=1\textwidth]{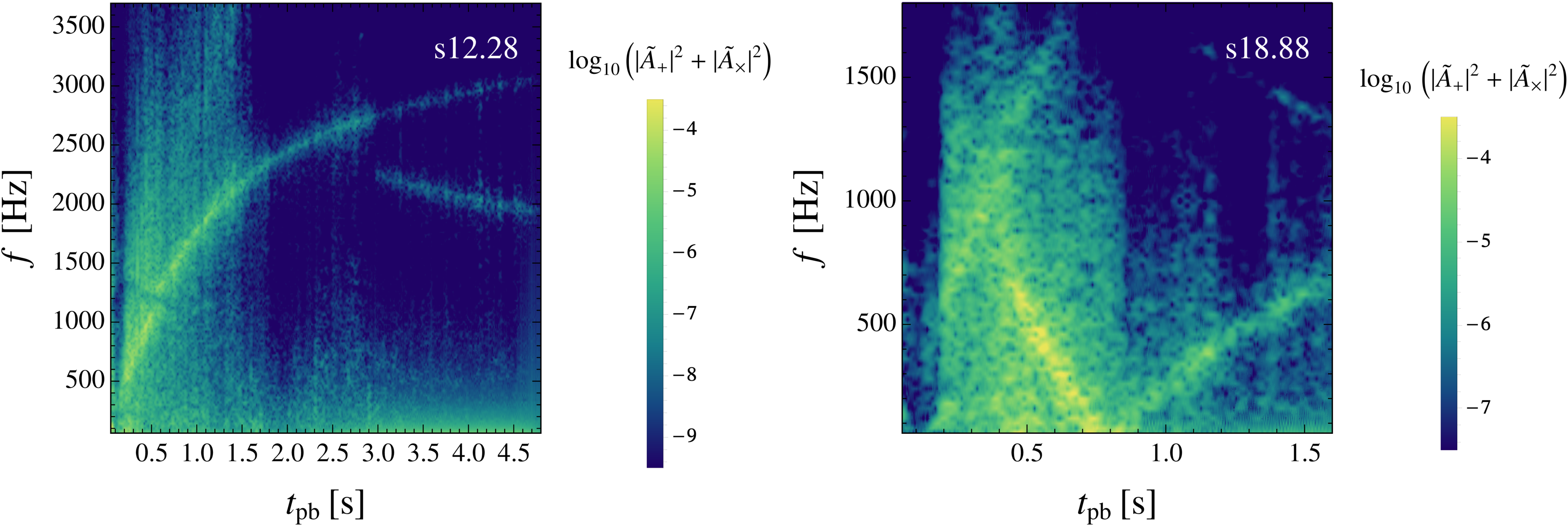}
    \caption{Amplitude spectrograms (in units of cm$^2$s$^2$) for the GW matter signals of models s12.28 ({\em left panel}) and s18.88 ({\em right panel}), for an observer along the $(+x)$-axis in the equatorial plane of the source [$(\phi,\theta)=(0,\pi/2)$]. Aliasing artifacts due to the sampling rate of output data with time intervals of 0.2\,ms for s12.28 and 0.5\,ms for s18.88 are visible when the GW frequency reaches 2500\,Hz and 1000\,Hz, respectively.
    }
    \label{Fig:Spectrograms}
\end{figure*}

Figure~\ref{Fig:GW_hydro_freq} displays the ASDs for our models s12.28 (left panel) and s18.88 (right panel) as functions of the frequency $f$, assuming an observer at a distance of $D=10\,\kpc$, who is located in the equatorial plane along the $(+x)$-axis of the coordinate frame of the source. Figure~\ref{Fig:Spectrograms} provides the corresponding amplitude spectrograms in terms of the squared modulus of the STFT of the wave amplitudes, $|\widetilde{A}_{\sigma}|^2=|{\rm STFT}[A_{\sigma}]|^2$, summed over the $+$ and $\times$ polarization modes.

The maximum frequencies chosen for all of these plots are geared to the time sampling of the simulation outputs, $\delta t$, which implies a Nyquist frequency $f_\mathrm{Ny}=1/(2\delta t)$. Since $\delta t = 0.2$\,ms for s12.28 and 0.5\,ms for s18.88, the corresponding Nyquist frequencies are ${f_\mathrm{Ny} \sim 2500}$\,Hz and $f_\mathrm{Ny} \sim 1000$\,Hz, respectively. When the GW signals exceed these frequencies, the ASDs and spectrograms exhibit artifacts by higher-frequency emission that is mirrored into the lower-frequency domain (see the discussion in \cite{Andresen:2016pdt}). We will adequately address these issues, which are particularly strong for s18.88, when we discuss our GW ASDs and spectrograms in the following. 

The ASDs in Fig.~\ref{Fig:GW_hydro_freq} show that most of the GW power associated with the strain is primarily concentrated in a broad maximum around frequencies of $f\sim\mathcal{O}(10^3)$\,Hz. In the case of model s12.28, a prominent peak extends from $f\sim 600$\,Hz to $f\sim 2000\,\Hz$. In contrast, the power maximum in model s18.88 is flatter and located at slightly lower frequencies between about 400\,Hz and 1000\,Hz. This shift towards lower frequencies is a consequence of the appearance of aliasing artifacts around the Nyquist frequency of $f_\mathrm{Ny}\sim 1000\,\Hz$ in s18.88. Indeed, it is expected from previous works that SN simulations for progenitors with higher core compactness yield ASD maxima at slightly higher frequencies than explosion models of less compact progenitors~\cite{Vartanyan:2023sxm,Choi:2024irp}. 

The ASD of s12.28 also exhibits the existence of a second, lower maximum at frequencies around 100--120\,Hz. This bump is associated with prompt postshock convection in the first 50\,ms after bounce, which has an oscillation period of slighty less than 10\,ms (see Fig.~\ref{Fig:GW_hydro_signal}). Since prompt postshock convection is much weaker in the case of s18.88, the local maximum around 100\,Hz is not clearly visible, also because it is additionally covered by the aliasing effects, which project higher-frequency power to lower frequencies (see more discussion below). We also point out the presence of a narrow gap in the spectrum of model s12.28 around $f\sim 1100$--1300\,Hz, which can be linked to an avoided crossing of f- and g-mode PNS oscillations~\cite{Torres-Forne:2019zwz,Morozova:2018glm,Vartanyan:2023sxm,Zha+2024}.

Generally the high-frequency spectral peaks in Fig.~\ref{Fig:GW_hydro_freq} are somewhat less pronounced for our models than those observed for most of the models analyzed in Refs.~\cite{Vartanyan:2023sxm,Choi:2024irp,Powell:2024nvv}. This is compatible with the slightly smaller GW amplitudes during the phase around the onset of the SN explosion and some 100\,ms afterwards, when the GW emission is fueled by hydrodynamic instabilities in the neutrino-heated postshock volume and post-explosion accretion onto the PNS (Sec.~\ref{sec:Time-domain analysis}). The smaller GW amplitudes lead to a lower amount of GW energy radiated in this period of the evolution.

Finally, the spectral power at low frequencies ($f\lesssim\mathcal{O}(10)\,\Hz$) in the ASDs and its ramp-up towards $f\sim 1$\,Hz for both SN models is directly connected to the matter ``memory'' effect observed in the GW signals at late times $t_{\rm pb}\gtrsim 1\,\s$ in Fig.~\ref{Fig:GW_hydro_signal}. 

The spectrograms in Fig.~\ref{Fig:Spectrograms} display in more detail how the GW power is radiated during the post-bounce evolution. As discussed before in the context of Fig.~\ref{Fig:Energy_both}, most of the energy released in matter-associated GWs is concentrated at $t_{\rm pb}<1.5\,\s$ for the s12.28 model (left panel) and at $t_{\rm pb}<0.8\,\s$ for s18.88 (right panel), i.e., at times when the GW emission is enhanced due to hydrodynamic instabilities in the postshock region and PNS accretion activity. Therefore these periods stick out in Fig.~\ref{Fig:Spectrograms} by their intense yellow and green colors.

Immediately after core bounce the GW spectrum is dominated by the emission from prompt postshock convection mainly at frequencies slightly above 100\,Hz plus some much weaker higher-frequency contributions up to a few 100\,Hz. From Figs.~\ref{Fig:GW_hydro_signal} and \ref{Fig:GW_hydro_freq} this is expected as well as the clearer presence of such emission in model s12.28. Then a relatively quiescent period of about 100\,ms with low-level broad-band activity follows.

In the subsequent phase of strong GW emission accompanying the vigorous fluid flows around the PNS, we observe the presence of low-frequency emission around 100--200\,Hz, first due to the short time intervals of SASI mass motions mentioned in Sec.~\ref{sec:Time-domain analysis}, and later due to convective overturn in the neutrino-heated layer behind the shock. This low-frequency band with strong emission extends to higher frequencies in a hazy region, which signals some faster flow activity, possibly connected to the fragmentation of convective downflows in the close vicinity of the PNS. 

The most evident structure in the spectrograms of both SN simulations, however, is a prominent, sharp emission component that rises in model s12.28 from frequencies of $f\sim 500\,\Hz$ up to $f\sim 2200\,\Hz$ until $t_{\rm pb}\sim 1.5\,\s$ and to more than $f\sim 3000\,\Hz$ until $t_{\rm pb}\sim 4.8\,\s$. This feature is directly associated with f/g-mode PNS oscillations excited by supersonic accretion downflows impinging on the PNS, and its increasing frequency is a consequence of the monotonic contraction of the PNS in response to the loss of energy and lepton number by neutrino emission (e.g., \cite{Murphy:2009dx, Marek:2008qi, Mueller:2012sv, Andresen:2016pdt, Torres-Forne:2019zwz, Andersen:2021vzo, Bruel:2023iye, Vartanyan:2023sxm}). Such a feature, though significantly less sharp between 0.3\,s and $\sim$\,0.45\,s after bounce, is also visible in the spectrogram of s18.88. However, when the f/g-mode frequency approaches the Nyquist frequency for the two models, $f_\mathrm{Ny}\sim2500\,$Hz at $t_{\rm pb}\sim 2.5\,\s$ in s12.28 and $f_\mathrm{Ny}\sim1000\,$Hz at $t_{\rm pb}\sim 0.4\,\s$ in s18.88, we observe a splitting into two branches, which are a direct result of aliasing effects, which project some of the emission at higher frequencies into the lower-frequency domain (see the discussion in Ref.~\cite{Andresen:2016pdt}). This mapping explains the downward reaching branches in both panels of Fig.~\ref{Fig:Spectrograms} and, after ``reflection'' on the time axis between $t_\mathrm{pb}\sim 0.7$\,s and $t_\mathrm{pb}\sim 0.9$\,s, also the upward pointing branch in the right panel for model s18.88.

Above the prominent f/g-mode strip, one can recognize weaker, broad-band emission extending to very high frequencies $f\gtrsim 3500\,\Hz$ in s12.28. This high-frequency ``haze'' is caused by a superposition of different PNS oscillation modes, whose precise nature is still unclear~\cite{Vartanyan:2023sxm}. Rotation in the PNS accretion layer due to the large angular momentum gained from accretion downflows (see Figs.~\ref{fig:Dynamics_s12.28}--\ref{fig:LateDynamics_s12.28} and discussion in Sec.~\ref{sec:overview}) may also play a role for the contributing oscillation modes and the exact structure of the GW emission spectrum at very high frequencies. The spectrogram for model s12.28 exhibits a fainter, narrow band at $f\sim1100-1300\,\Hz$ between roughly 0.3\,s and 0.7\,s. This gap-like feature is interpreted as a consequence of f- and p-mode PNS pulsations interfering with g-mode oscillations, as extensively discussed in Refs.~\cite{Morozova:2018glm, Torres-Forne:2019zwz, Torres-Forne+2019, Andersen:2021vzo, Vartanyan:2023sxm}.

After about 1.5\,s in model s12.28, the low-frequency contributions to the GW emission of up to several 100\,Hz (with weak extensions to more than 1000\,Hz) begin to become well separated from the sharp high-frequency f/g-mode strip. Interestingly, this low-frequency spectral band of GW activity continues to be present even at late times until the end of our simulation. It cannot be explained merely by the matter memory effect connected to the stable morphology of the asymmetric SN ejecta and shock at $t_\mathrm{pb}\gtrsim 1$\,s, which manifests itself in GW emission mainly at very low frequencies below some $\sim$10\,Hz with a spectral ramp-up towards $f\sim\mathcal{O}(1)$\,Hz (see Fig.~\ref{Fig:GW_hydro_freq} and also Refs.~\cite{Vartanyan:2023sxm,Choi:2024irp}). We think the broad-band emission of hundreds of Hz up to over 1000\,Hz is caused by persistent accretion downdrafts that create a highly turbulent PNS environment, where supersonic inward flows, which are deflected by their large angular momentum, collide with each other and with re-ejected matter. This chaotic hydrodynamic activity happens at distances between some 10\,km and a few 100\,km with velocities of several $10^9$\,cm\,s$^{-1}$, corresponding to time scales from ms to some ten ms or frequencies of tens of Hz to more than 1000\,Hz, creating a weak, broad-band background of GW ``noise''. Such a link is supported by a visible strengthening of the GW emission between $t_\mathrm{pb}\sim 2.1$\,s and $t_\mathrm{pb}\sim 2.9$\,s not only in the narrow, high-frequency f/g-mode strip, but also in the well separated lower-frequency broad-band region ranging up to $\gtrsim$\,1000\,Hz (Fig.~\ref{Fig:Spectrograms}, left panel). This is exactly the time interval when the mass inflow rate towards the PNS increases (see Fig.~\ref{fig:Mdot_Ekin}). It happens after a period of reduced accretion between about 1.6\,s and 2.1\,s, when also a phase of weaker GW emission can be witnessed in the low-frequency domain.

\begin{figure*}
    \centering
    \includegraphics[width=1\textwidth]{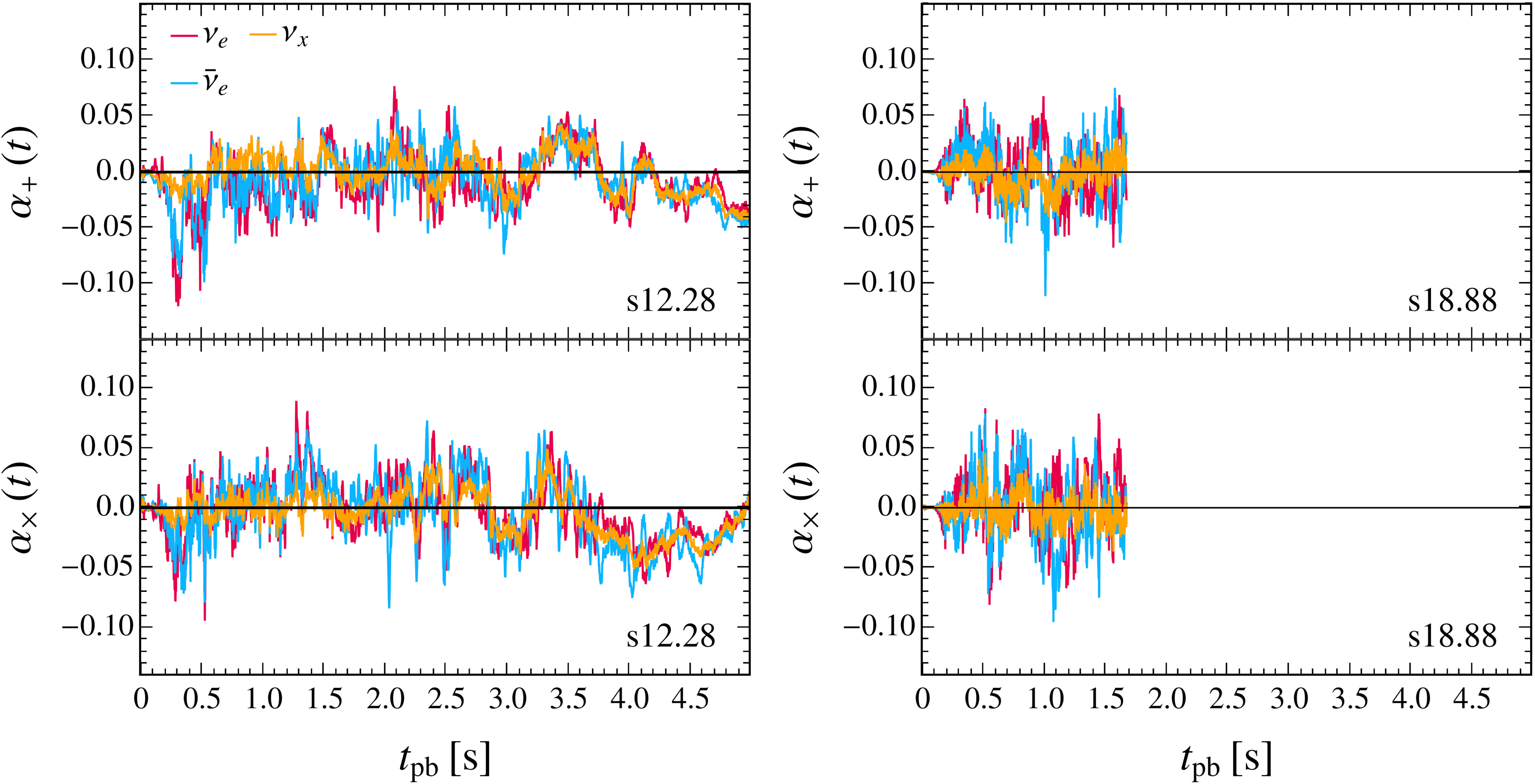}\\
    \vspace{0.7cm}
    \includegraphics[width=1\textwidth]{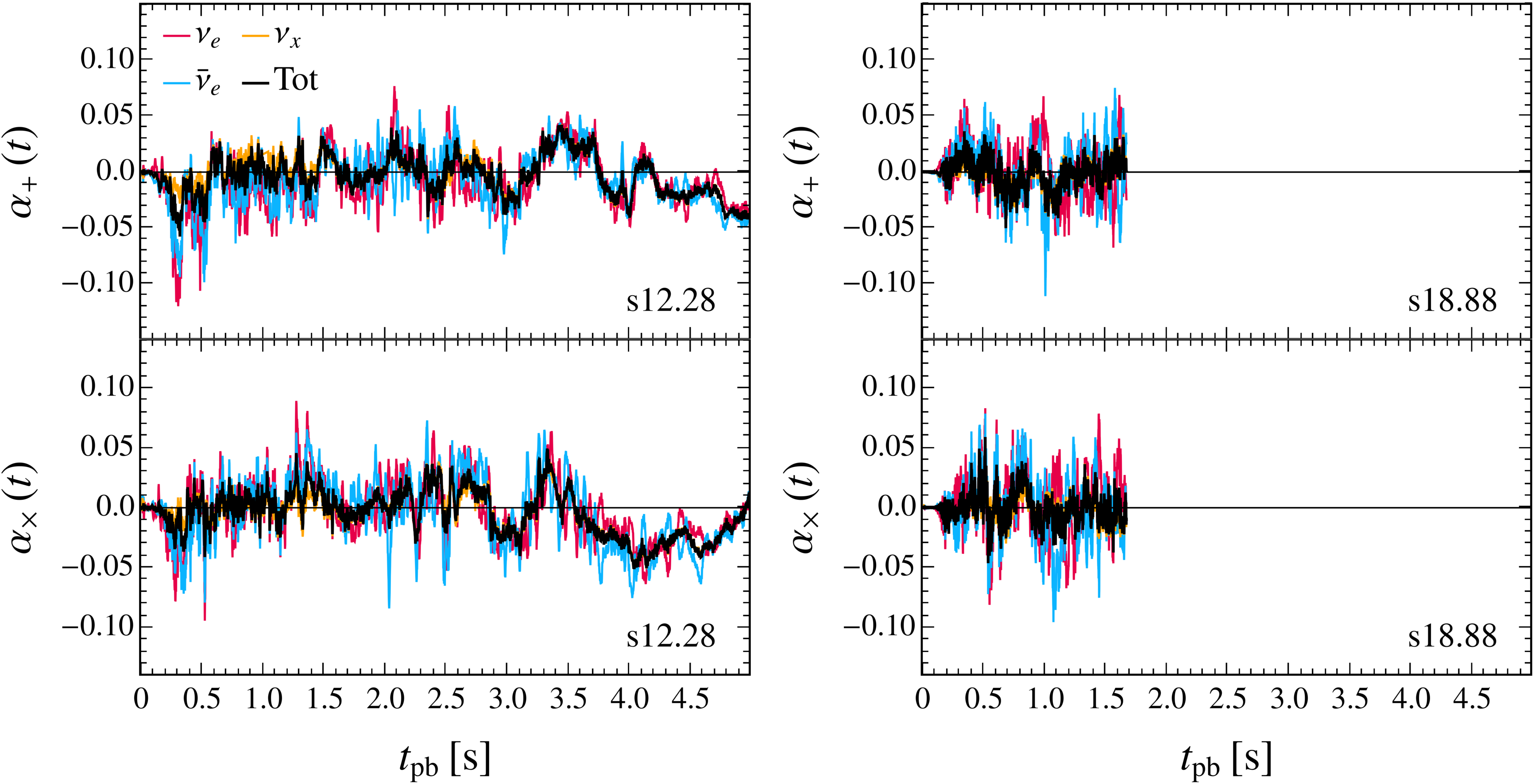}  
    \caption{Neutrino anisotropy parameters [Eq.~\eqref{Eq:Anisotropy}] for an observer located along the $(+x)$-axis in the equatorial plane of the source [$(\alpha,\beta)=(0,\pi/2)$]. The \emph{left} and \emph{right panels} refer to models s12.28 and s18.88, respectively, and the \emph{upper} and \emph{lower panels} of each plot display the $+$ and $\times$ GW polarization states, respectively. The different colors (as labeled in the upper left panels) correspond to the different neutrino species. According to our convention, $\nu_x$ represents a single species of the heavy-lepton neutrinos and antineutrinos. Since all of these are handled equally in the neutrino transport of the discussed SN models, the anisotropy parameters for all heavy-lepton neutrinos are the same. The \emph{top two plots} present only the results for the three neutrino species, whereas in the \emph{bottom two plots} the black solid line is added for the anisotropy of the total emission of SN neutrinos, i.e., considering in Eq.~\eqref{Eq:Anisotropy} the total neutrino luminosity as the sum of the contributions of neutrinos and antineutrinos of all three lepton flavors.}
    \label{Fig:GW_nu_alpha}
\end{figure*}

\begin{figure*}
    \centering
    \includegraphics[width=1\textwidth]{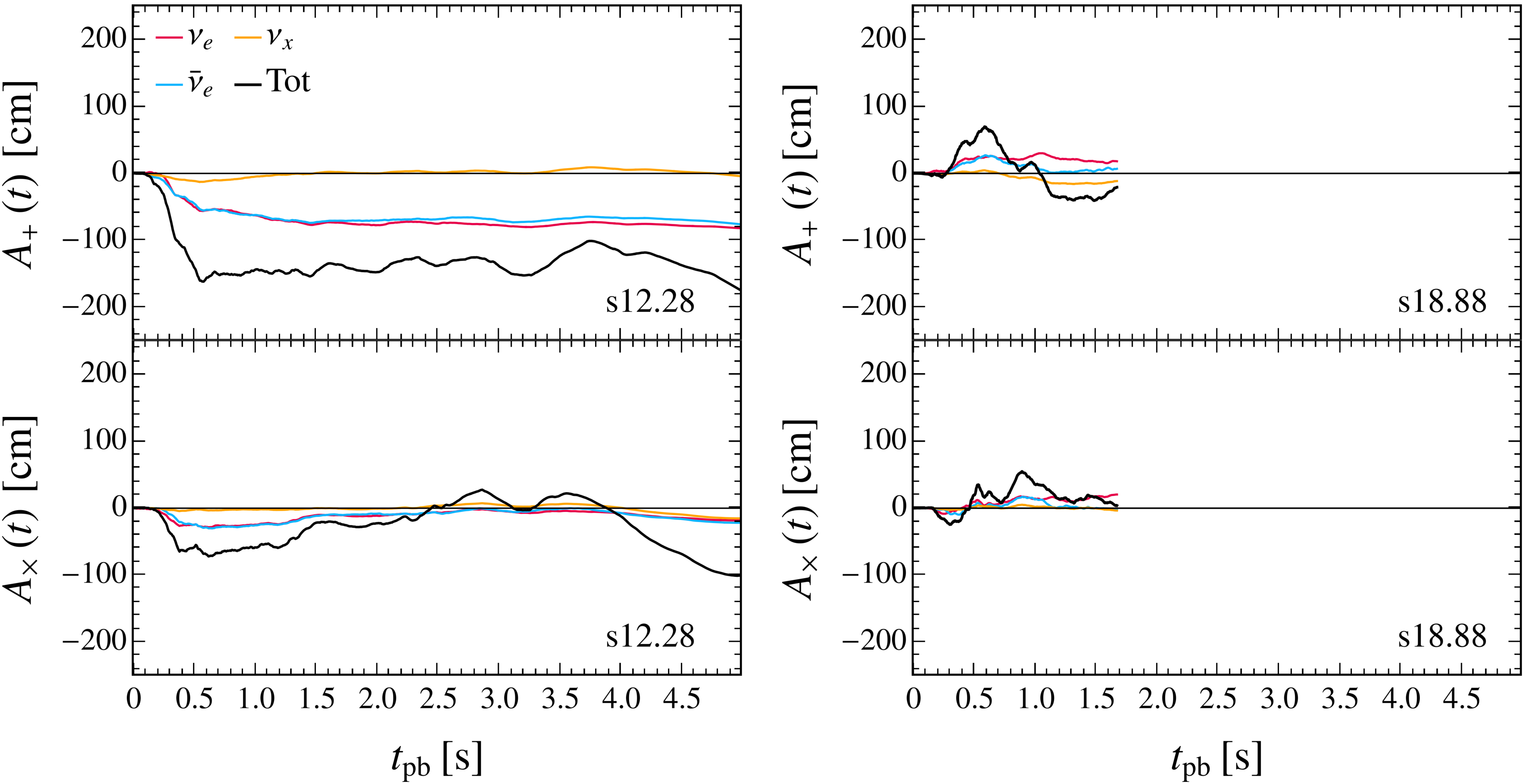} 
    \caption{GW amplitudes $A_{\sigma}=D\,h_{\sigma}$ of the asymmetric neutrino emission for an observer located along the $(+x)$-axis in the equatorial plane of the source [$(\alpha,\beta)=(0,\pi/2)$]. The \emph{left panels} refer to SN model s12.28 and the \emph{right panels} depict the case of s18.88. The \emph{upper} and \emph{lower panels} display the time evolution of the signal for the $+$ and $\times$ polarization states, respectively. The different colors (as labeled in the upper left panel) correspond to the results for the different neutrino species, where $\nu_x$ represents a single species of the heavy-lepton neutrinos and antineutrinos. The black solid lines represent the GW amplitudes for the total emission of all neutrinos and antineutrinos obtained by summing up the luminosities of all six neutrino species.}
    \label{Fig:GW_nu_signals}
\end{figure*}

\begin{figure*}
    \centering
    \includegraphics[width=1\textwidth]{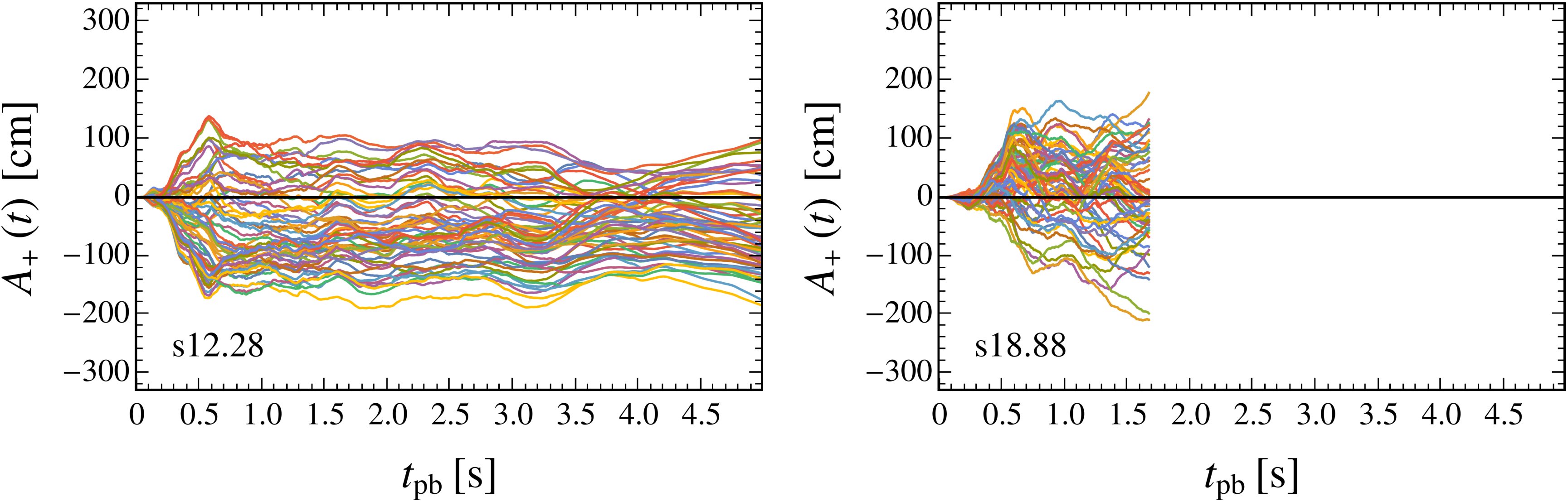} 
    \caption{GW amplitudes $A_+$ of the total neutrino emission for model s12.28 (\emph{left panel}) and model s18.88 (\emph{right panel}). The different colors depict different randomly chosen viewing angles of an observer at a large distance from the source.}
    \label{Fig:GW_nu_directions}
\end{figure*}

\section{Gravitational waves from anisotropic neutrino emission}
\label{sec:neutrino}

\subsection{Signal in the time domain}
\label{subsec:nu_timedomain}

The formalism to determine the GW signal connected to anisotropic neutrino emission is described in Refs.~\cite{Epstein:1978dv,Mueller+1997,Muller:2011yi, Mukhopadhyay:2021zbt}, to which we refer interested readers for all relevant details. Here, we simply subsume that the time profile of the GW strain, in the transverse-traceless gauge, can be written as
\begin{equation}
    h_{\sigma}(t)=\frac{2 G}{D c^4}\int_0^t dt'\,L_\nu(t')\,\alpha_{\sigma}(t')\,,
\label{Eq:Signal}
\end{equation}
where $\sigma=+,\times$ stands for the two possible polarization states of the metric perturbation, and $L_\nu(t)$ is the angle-averaged neutrino luminosity at infinity. Assuming a GW propagating in the $z$-direction of the source's coordinate frame, $h_+=h_{yy}=-h_{xx}$, and $h_\times=h_{xy}=h_{yx}$. In Eq.~(\ref{Eq:Signal}), $\alpha_{\sigma}(t)$ is the so-called anisotropy parameter, defined as
\begin{equation}
    \alpha_{\sigma}(t)=\frac{1}{L_\nu(t)}\int_{4\pi}d\Omega^\prime\,W_{\sigma}(\theta^\prime,\phi^\prime\,\alpha,\beta)\,\frac{dL_\nu(\Omega^\prime,t)}{d\Omega^\prime}\,,
\label{Eq:Anisotropy}
\end{equation}
where $d\Omega^\prime=d\cos{\theta^\prime}d\phi^\prime$ is the solid-angle element in the coordinate frame of the neutrino-radiating source and $\alpha$ and $\beta$ are the observer coordinates that define the viewing direction of the SN event.\footnote{The coordinates set-up we refer to is depicted in Fig.~10 of~\cite{Muller:2011yi}.} Moreover, $dL_\nu/d\Omega$ denotes the direction-dependent neutrino luminosity per solid angle. The angle-dependent functions $W_+$ and $W_\times$ are given in Refs.~\cite{Muller:2011yi,Vartanyan:2023sxm}. For illustration, we consider an observer located along the $(+x)$-axis in the equatorial plane of the source ($\alpha=0$, $\beta=\pi/2$) in the following. In this case the angular weight functions read:
\begin{equation}
    \begin{split}
        W_+&=\left(\cos^2{\theta}-\sin^2{\theta}\,\sin^2{\phi}\right)\,\frac{1+\sin{\theta}\cos{\phi}}{\cos^2{\theta}+\sin^2{\theta}\,\sin^2{\phi}}\,,\\
        W_\times&=-2\cos{\theta}\sin{\theta}\cos{\phi}\,\,\frac{1+\sin{\theta}\cos{\phi}}{\cos^2{\theta}+\sin^2{\theta}\,\sin^2{\phi}}\,.
    \end{split}
\end{equation}
The amplitudes $A_{\sigma}$ of the neutrino GW signals are again connected with the dimensionless and distance-dependent GW strains $h_{\sigma}$ [Eq.~\eqref{Eq:Signal}] via the expression,
\begin{equation}
A_{\sigma}= D\,h_{\sigma} \,,
\label{eq:nuampl}
\end{equation}
where $D$ is the distance of the source.

In Fig.~\ref{Fig:GW_nu_alpha} the time evolution of the anisotropy parameters $\alpha_{+}$ (upper panels) and $\alpha_{\times}$ (lower panels) is displayed as a function of the post-bounce time $t_{\rm pb}$ for both of our models, s12.28 (left) and s18.88 (right). The solid black lines represent the results for the total neutrino emission summed over all six species $\nu_e$, $\bar\nu_e$, and $\nu_x = \nu_\mu, \bar\nu_\mu, \nu_\tau, \bar\nu_\tau$.\footnote{Note that according to our convention, $\nu_x$ denotes a single type of heavy-lepton neutrinos and antineutrinos, all of which are treated identically in the neutrino transport of the discussed SN models. In contrast, ``$\nu_\mu$'' in Ref.~\cite{Choi:2024irp} (used there instead of ``$\nu_x$'') represents the bundling of all heavy-lepton neutrino species. Therefore our results for $\nu_x$ should be multiplied by a factor of 4 to compare non-normalized quantities with those for $\nu_\mu$ in Ref.~\cite{Choi:2024irp}.} In addition, we show the anisotropy parameters for the emission of $\nu_e$ in magenta, $\bar{\nu}_e$ in light blue, and $\nu_x$ in dark yellow. Just after the core bounce the values of $\alpha_{\sigma}$ are always compatible with zero, although the core collapse was computed in full 3D. Since no significant asymmetries are present in the collapsing iron cores, the neutrino emission remains essentially isotropic for the first 100--150\,ms after bounce. A small anisotropy with a tendency to grow develops during the $\sim$60\,ms time interval of SASI activity roughly around 100\,ms after bounce, but larger, time-dependent anisotropies in the neutrino emission up to typically several percent (and close to 10\% or even more in extreme, rare peaks) appear only when strong postshock convection sets in at $t_{\rm pb}\gtrsim 0.2\,$s.

One can witness variability of the anisotropy parameters in Fig.~\ref{Fig:GW_nu_alpha} on different time scales: high-frequency fluctuations with ${\cal O}(10\,\mathrm{ms})$ periods as well as longer-time-scale modulations with lengths of hundreds of ms to seconds.\footnote{Interestingly, we do not witness any clear correlations between features in the neutrino anisotropy parameters $\alpha_{+}$ and $\alpha_{\times}$ (Fig.~\ref{Fig:GW_nu_alpha}) and the phases of enhanced PNS accretion (Fig.~\ref{fig:Mdot_Ekin}), which increase the angular momentum in layers around the neutrinopheres. In contrast, in an analysis of PNS kicks in Ref.~\cite{Janka:2024xbp}, we found a strong correlation between phases of higher PNS accretion and excursions of the instantaneous neutrino dipole-asymmetry parameter $\alpha_\nu^\mathrm{tot}$ that is relevant for the neutrino-induced PNS kick (see Figs.~11 and~17 for model s12.28 in \cite{Janka:2024xbp}). We understand this discrepancy by the fact that at late times the accretion geometry is usually dominated by a dipolar configuration with a massive downflow in one hemisphere and an outflow in the other hemisphere (see Fig.~\ref{fig:LateDynamics_s12.28}). Since in the late evolution phase the asymmetry of the neutrino radiation seen at large distances is mostly caused by neutrino absorption in the dense downflows rather than asymmetric neutrino emission from the neutrinospheres (as discussed in~\cite{Janka:2024xbp}), the accretion asymmetry has a substantial impact on the neutrino dipole parameter $\alpha_\nu^\mathrm{tot}$ but not prominently on the neutrino anisotropy parameters $\alpha_\sigma$ for the neutrino GW production.} The neutrino GW strains and amplitudes of Eqs.~(\ref{Eq:Signal}) and~(\ref{eq:nuampl}), however, give rise to a low-frequency neutrino GW memory that is imprinted on spacetime by the anisotropically expanding flux of emitted neutrinos \cite{Epstein:1978dv,Turner:1978jj,Mukhopadhyay:2021zbt}. Because of the time integration performed in Eq.~(\ref{Eq:Signal}), excursions in $\alpha_\sigma$ with opposite signs cancel each other out and fluctuations with the same sign sum up to monotonic trends.

The amplitudes $A_\sigma$ of the neutrino GW signals are displayed as functions of time for our models in Fig.~\ref{Fig:GW_nu_signals}. The color scheme employed there is the same as in Fig.~\ref{Fig:GW_nu_alpha}. Since the amplitudes scale with the neutrino luminosity, we stress that in Fig.~\ref{Fig:GW_nu_signals} the results for $\nu_x$ are given for a single species of the heavy-lepton (anti)neutrinos. Governed by the behavior of the anisotropy parameters, the total GW amplitudes (for all neutrinos combined) show values close to zero till $t_{\rm pb}\sim 0.2\,\s$. Afterwards they exhibit a rise over periods of a few 100\,ms to reach absolute values between 50--60\,cm and roughly 150--160\,cm with subsequent small-amplitude fluctuations on time scales of hundreds of ms and longer-time variations with more extreme changes (and even sign inversions) over seconds. In model s12.28 the neutrino cooling was followed long enough such that both amplitudes seem to have reached a late phase of secular evolution, where the amplitudes display only slow, long-term trends that constitute the neutrino memory, which manifests itself in a permanent, nearly constant displacement in the spacetime metric.

Figure~\ref{Fig:GW_nu_signals} also reveals that the anisotropy parameters as well as the GW amplitudes for the emission of $\nu_e$ and ${\bar\nu}_e$ are significantly greater in magnitude than the amplitudes for the radiated $\nu_x$, in particular during the phase of massive accretion by the PNS ($t_\mathrm{pb}\lesssim 1.5$\,s in s12.28 and $t_\mathrm{pb}\lesssim 1.2$\,s in s18.88). This difference has two main reasons. First, before the onset of the explosion, the single $\nu_x$ luminosity is $\sim$\,50\% of the $\nu_e$ or $\bar{\nu}_e$ luminosity. Second, the emission of the heavy-lepton neutrino species is less affected by asymmetric accretion onto the PNS, and therefore it exhibits less extreme emission anisotropies, because they decouple from the PNS matter at deeper neutrinospheres. The production of neutrinos in the PNS's accretion layer mainly fuels to fluxes of electron neutrinos and antineutrinos, which causes peak values of their anisotropy parameters up to 0.05--0.1 (Fig.~\ref{Fig:GW_nu_alpha}). After the explosion has set in and the accretion abates ($t_\mathrm{pb}\gtrsim 1.5$\,s in s12.28 and $t_\mathrm{pb}\gtrsim 1.2$\,s in s18.88), the heavy-lepton neutrinos continue to escape from deeper regions in the PNS interior and therefore tend to be somewhat less perturbed by the remaining accretion flows to the PNS (see Fig.~\ref{Fig:GW_nu_alpha}, top two plots). Since the luminosities of the six neutrino species are more similar during this phase, however now with a slight dominance of those of $\nu_x$ (see Figs.~17 and~19 in \cite{Janka:2024xbp}), the GW amplitudes of all neutrino types contribute roughly equally to the changes of the total neutrino GW amplitude during this late-time evolution. An example is the growth of $A_\times$ for the total neutrino emission in model s12.28 at $t_\mathrm{pb} \gtrsim 4$\,s, in which all neutrino kinds are involved to roughly the same extent (Fig.~\ref{Fig:GW_nu_signals}, lower left panel). Large, persistent differences in the amplitudes of the $\nu_x$ compared to the $\nu_e$ and $\bar{\nu}_e$ ---as in the case of $A_+$ in s12.28 (Fig.~\ref{Fig:GW_nu_signals}, upper left panel)--- are therefore a consequence of the effects that occurred during the preceding phase of strong PNS accretion.
 
Figure~\ref{Fig:GW_nu_directions} presents the time evolution of $A_{+}$ for different, randomly chosen viewing directions of a distant observer for model s12.28 (left panel) and s18.88 (right panel). Although the overall behavior described above is similar in all directions, the plots show a strong dependence of the neutrino GW amplitudes on the observer's position. In particular, when the simulations were terminated, the late-time amplitudes in s12.28 cover the range of $-200\,\cm\lesssim A_+ \lesssim 100\,\cm$ and in model s18.88 they display an even larger spread within $-200\,\cm\lesssim A_+ \lesssim 200\,\cm$.

Finally, we consider the total energy carried away by neutrino-induced GWs. The total energy radiated in GWs caused by the anisotropic emission of neutrinos of all species and integrated over all viewing directions, can be computed as~\cite{Mueller+1997,Muller:2011yi} 
\begin{equation}
    E_{\rm GW}^{\rm \nu}(t)=\frac{G}{4\pi c^5}\int_0^t \!\!\! dt^\prime \int_{4\pi}\!\!\!d\Omega_{\alpha\beta}\left[l_+^2(t^\prime ,\alpha,\beta)+l_\times^2(t^\prime,\alpha,\beta)\right],
    \label{Eq: GWEnergyNu}
\end{equation}
where
\begin{equation}
    l_{+,\times}(t,\alpha,\beta)=L_\nu(t)\,\alpha_{+,\times}(t,\alpha,\beta)
\end{equation}
with $L_\nu$ and $\alpha_{+,\times}$ being the luminosity and anisotropy parameters for the summed contributions of neutrinos and antineutrinos of all three lepton flavors.

The dashed lines in Fig.~\ref{Fig:Energy_both} represent the cumulative energy radiated in neutrino GWs as a function of post-bounce time $t_{\rm pb}$ for model s12.28 (red) and model s18.88 (blue). In both cases a fast increase takes place in the time window from $\sim$\,200\,ms until $\sim$\,600\,ms, connected to the strong growth of the neutrino emission anisotropy during the period of extremely vigorous hydrodynamic instabilities in the postshock volume and asymmetric downdrafts onto the PNS with high mass accretion rates. This is exactly the time interval when the $\nu_e$ and $\bar\nu_e$ possess the highest accretion luminosities and their anisotropy parameters as well as GW amplitudes reach large values contemporaneously (Figs.~\ref{Fig:GW_nu_alpha}--\ref{Fig:GW_nu_directions}). After this fast growth phase until $t_\mathrm{pb}\sim600$\,ms, the neutrino GW energies have reached about $2\times 10^{-11}\,M_\odot c^2$ in s12.28 and approximately $3\times 10^{-11}\,M_\odot c^2$ in s18.88. Subsequently, they continue to rise in a slower evolution (because of gradually declining neutrino luminosities but fairly stable neutrino anisotropy values and GW amplitudes) to roughly twice these values until the end of our 3D SN simulations. The total neutrino GW energies then are $E_\mathrm{GW}^\nu\simeq 4.2\times10^{-11}\,M_\odot c^2\approx7.5\times10^{43}$\,erg for s12.28 and $E_\mathrm{GW}^\nu\simeq 7.0\times10^{-11}\,M_{\odot}c^2\approx1.3\times 10^{44}$\,erg for s18.88 (Table~\ref{tab:SNmodels}). In contrast to the GW energies associated with asymmetric mass motions, which have effectively saturated at that time, the neutrino GW energies still exhibit a trend of significant growth because of the ongoing neutrino emission. 

\begin{figure*}[t!]
    \centering
    \includegraphics[width=1\textwidth]{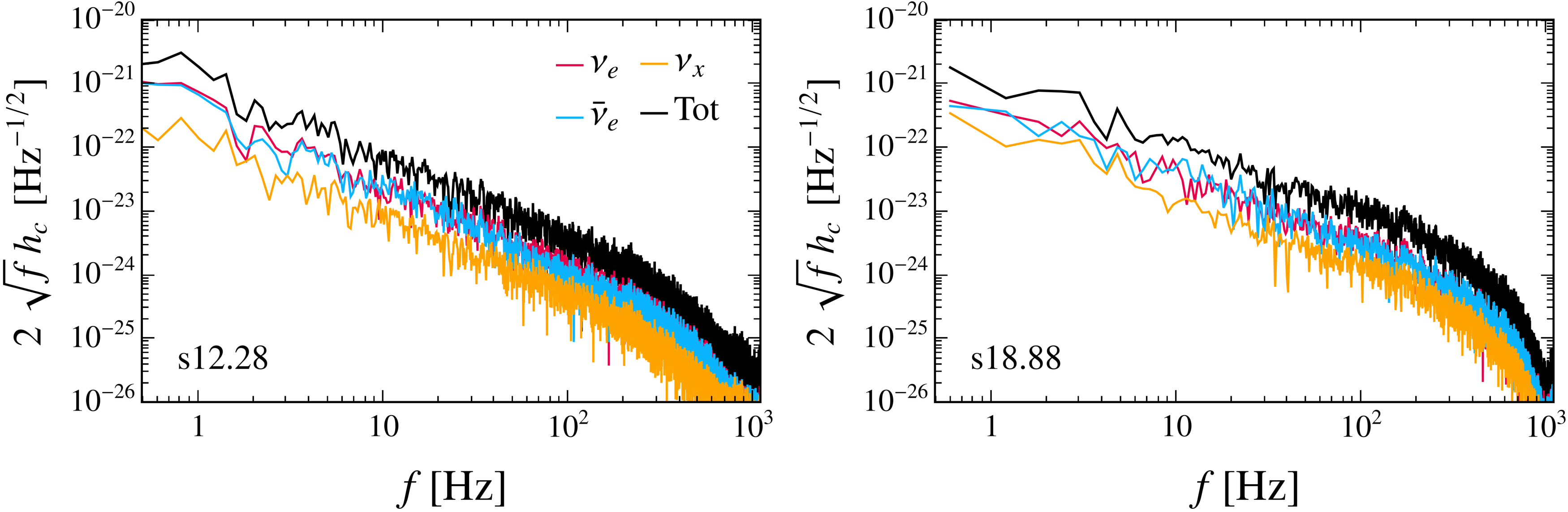}
    \caption{Amplitude spectral densities (ASDs) as functions of frequency $f$ for the GW signals due to anisotropic neutrino emission in SN models s12.28 (\emph{left panel}) and s18.88 (\emph{right panel}) for an observer located along the $(+x)$-axis in the equatorial plane of the source [$(\alpha,\beta)=(0,\pi/2)$] at a distance of $D=10$\,kpc. The different colors (as labeled in the left panel) display the contributions for $\nu_e$, $\bar{\nu}_e$, and $\nu_x$, i.e., a single species of the heavy-lepton (anti)neutrinos. The solid black line shows the result obtained by summing up the contributions of all the six different neutrino species.}
     \label{Fig:GW_nu_hc}

\end{figure*}

Interestingly, in both cases the total energies associated with the matter-induced GW emission are more than an order of magnitude higher than the GW energies due to anisotropic neutrino emission (Fig.~\ref{Fig:Energy_both}). This energy hierarchy exists despite similar masses being involved (some 0.1\,$M_\odot$ of moving stellar matter compared to an energy radiated in neutrinos that is equivalent to roughly 10\% of the PNS rest mass, i.e., $\sim$\,0.15\,$M_\odot$), neutrinos moving much faster with the speed of light, and the GW amplitudes associated with the neutrino emission being much larger (compare Figs.~\ref{Fig:GW_hydro_signal} and~\ref{Fig:GW_hydro_signal_directions} with Figs.~\ref{Fig:GW_nu_signals} and~\ref{Fig:GW_nu_directions}). The greater GW energies are explained by the higher frequencies of the GWs produced by mass flows. For the latter the GW spectrum peaks at frequencies well above those of the neutrino-GW spectrum, which drops steeply towards high frequencies (Figs.~\ref{Fig:GW_hydro_freq} and~\ref{Fig:GW_nu_hc}; see Sec.~\ref{sec:nu_freq_domain} for a discussion). For this reason the total GW energy released by core-collapse SNe is expected to be strongly dominated by the matter component, if the neutrino emission asymmetry is as small as in our simulations.

\subsection{Signal in the frequency domain}
\label{sec:nu_freq_domain}

The ASDs of our neutrino GW signals are computed by using again Eqs.~(\ref{Eq:ASD})--(\ref{Eq:DFT}) and neutrino outputs from the simulations that were stored with time steps of about 0.5\,ms for model s18.88 and for model s12.28 until $t_{\rm pb}\simeq 2.28$\,s; at later times the output rate in model s12.28 was reduced to time steps of approximately 1\,ms. For performing the discrete Fourier transforms, we mapped the GW strains to equidistant time intervals of $\delta t = 0.5\,$ms, corresponding to a Nyquist frequency of 1000\,Hz. Because of the coarser sampling of the neutrino data at $t_{\rm pb}\gtrsim 2.28$\,s in s12.28, $f_\mathrm{Ny} \approx 500$\,Hz is a more rigorous value for our neutrino GW analysis of both simulations.  

Figure~\ref{Fig:GW_nu_hc} displays the corresponding results as functions of the frequency $f$ for model s12.28 (left panel) and model s18.88 (right panel) for the different neutrino species as well as the total neutrino emission. Apart from fluctuations, the ASDs for neutrino-induced GWs decrease monotonically and those of the individual neutrino species look very similar to each other. Interestingly, the decline of the ASDs with increasing frequency can be well fitted by power-law functions that behave approximately like $f^{-1.2}$ for model s12.28 and $f^{-1}$ for s18.88 up to 200--300\,Hz. Then the spectra steepen up to 1000\,Hz, with a somewhat faster drop in the case of s18.88 (by a bit more than two orders of magnitude compared to 1.5 orders of magnitude in s12.28). This difference in the drop at high frequencies is connected to the more shallow slope of the ASD in s18.88, which implies somewhat higher values just before the spectrum finally plummets. The shapes of the neutrino ASDs are clearly different from those of the matter GW signals in Fig.~\ref{Fig:GW_hydro_freq}, where a lot of power is located at frequencies around and above 1000\,Hz. Because of the steep negative gradients of the neutrino ASDs with frequency $f$, the neutrino GW energies are dominated by the low-frequency contributions. This implies that the Nyquist frequency has less influence on the GW energy estimates than for the matter signals.

While $\nu_e$ and $\bar\nu_e$ possess ASDs that essentially overlap, the ASD of $\nu_x$ follows the other two in parallel but somewhat lower, because the GW amplitudes of the heavy-lepton neutrinos are smaller during most of the evolution (Fig.~\ref{Fig:GW_nu_signals}). Most power is concentrated at low frequencies of $f\sim 0.5$--5\,Hz. This is connected to the slow time variation of the amplitudes, which takes place on a time scale of 100--200\,ms around the onset of the explosions ($t_{\rm pb}\lesssim 0.5$\,s) and over time intervals of a few hundred ms to seconds at later times. 

Around 1\,Hz the ASDs for all neutrinos individually and in both models reach values between $\sim$\,$10^{-22}$\,Hz$^{-1/2}$ and $\sim$\,$10^{-21}$\,Hz$^{-1/2}$, with the ASD for one species of heavy-lepton neutrinos $\nu_x$ being the lowest. The ASDs of the total neutrino GW signals have maxima above $10^{-21}$\,Hz$^{-1/2}$, with a bit higher power in the case of model s12.28 because of its longer simulated evolution time. Moving the viewing angle away from the ($+x$)-axis direction [$(\alpha,\beta)=(0,\pi/2)$], where the neutrino GW amplitudes are relatively small compared to other observer positions (see Figs.~\ref{Fig:GW_nu_signals} and~\ref{Fig:GW_nu_directions}), has little impact on the ASDs. In the observer direction with [$(\alpha,\beta)=(-\pi/3,\pi/2)$], for example, the GW amplitudes reach values as high as $|A_+|\simeq120\,\cm$ and $|A_\times|\simeq200\,\cm$ for model s18.88, which are larger than those seen from the ($+x$)-axis direction. This viewing-angle dependence reflects in the ASDs mainly at frequencies around $\sim$\,1\,Hz, corresponding to the typical rise time of the GW amplitudes and to the time scale for slow, long-term changes. However, the difference in the low-frequency power of the ASDs is at most around a factor of 2. On the contrary, the ASDs for the two different viewing directions are effectively identical at frequencies $f\gtrsim10$ Hz, because the variations of the amplitudes on time scales of $\lesssim$\,0.1\,s are very similar.

\section{Comparison with 3D SN models in the literature}
\label{sec:comparison}

At this point we briefly discuss important differences of our GW signals compared to those in Refs.~\cite{Vartanyan:2023sxm,Choi:2024irp}, where also long-term 3D simulations of successful SN explosions, computed with the \textsc{Fornax} code, were evaluated. We constrain ourselves (mostly, with few exceptions) to these references, because (a) the set of models considered there consists of non-rotating progenitors from the same stellar evolution calculations of Refs.~\cite{Woosley+2015,Sukhbold+2016,Sukhbold:2017cnt} as our models, including some progenitors with the same or similar ZAMS masses and pre-collapse compactness values (the 9.0\,$M_\odot$ progenitor used in~\cite{Vartanyan:2023sxm,Choi:2024irp} is identical to the progenitor for our SN model s9 in Appendix~\ref{app:code}); (b) the models have been evolved in 3D with full neutrino transport but without magnetic fields for similarly long times after core bounce as our models; and (c) hydrodynamics, the treatment of gravity, a polar-grid-based numerical mesh, and neutrino transport and interactions in \textsc{Fornax} are presumably similarly elaborate as in our \textsc{Vertex} code and in our \textsc{Alcar} code applied for models s9 and s20 in Appendix~\ref{app:code}. 

In contrast to our models, however, the SN simulations of Refs.~\cite{Vartanyan:2023sxm,Choi:2024irp} were started from spherically symmetric pre-collapse conditions with no initial perturbations applied, except in one of the 9.0\,$M_\odot$ models there, where perturbations that should have featured convection in the initial model were imposed. This was justified by the argument that only for the lowest-mass progenitors that explode quickly such initial physical perturbations make a demonstrable difference in model aspects of the outcomes \cite{Choi:2024irp,Burrows+2024,Wang+2023}. This may apply to the \textsc{Fornax} simulations, which typically exhibit explosions quite early after core bounce. However, it cannot hold in general in view of the fact that only our simulations started from 3D initial conditions produce explosions, whereas those initiated with the 1D progenitors fail to do so. The only exception among the considered models is the s9 case discussed in Appendix~\ref{app:code},\footnote{In Appendix~\ref{app:code} our 3D simulation with the \textsc{Alcar} code~\cite{Glas:2018oyz,Glas:2018vcs} is considered. For a corresponding 3D SN simulation for the same 9\,$M_\odot$ progenitor with the \textsc{Vertex} code, see Refs.~\cite{Melson+2020,Stockinger+2020,Janka:2024xbp,Janka2025}.} and even in this case small random perturbations (on the level of 0.1 percent of the density) were imposed on the matter falling through the stalled shock wave to seed the growth of hydrodynamic instabilities in the postshock region.

\subsection{Comparison of matter GW signals}
\label{sec:comparison-matter}

The total energy radiated in GWs is by far dominated by the matter-associated GW signals and neutrino-induced GWs contribute typically only on the level of a few percent. Therefore we compare the GW energies in connection to the matter signals. 

Our SN models tend to release substantially less energy in GWs than the models considered in Refs.~\cite{Vartanyan:2023sxm,Choi:2024irp}, except for the s9 case in Appendix~\ref{app:code}, where our model radiates $5.56\times 10^{-11}\,M_\odot c^2$, whereas the authors of Ref.~\cite{Choi:2024irp} report $2.69\times 10^{-11}\,M_\odot c^2$ and $3.59\times 10^{-11}\,M_\odot c^2$, where interestingly the lower value is for their 9.0\,$M_\odot$ model {\em with} imposed initial perturbations. All the other exploding models in Refs.~\cite{Vartanyan:2023sxm,Choi:2024irp} produce between roughly 10 times and several 10 times more GW energy than our models even for explosions of progenitors with lower compactness values. In fact, the GW energies of our models are best compatible with those of the 9.25\,$M_\odot$ and 9.5\,$M_\odot$ cases in Ref.~\cite{Choi:2024irp}, which possess very small core-compactness values. These differences are unlikely to be explained solely by our too low choice of the data-output frequency, but they are consistent with the tendency of higher explosion energies in simulations with the \textsc{Fornax} code (see discussion in Refs.~\cite{Stockinger+2020,Bollig:2020phc,Janka2025}), which goes hand in hand with more vigorous mass motions in the postshock volume.

The lower GW energy for the 9.0\,$M_\odot$ simulations in Refs.~\cite{Vartanyan:2023sxm,Choi:2024irp} compared to our s9 model does not contradict this explanation. Because of a very steep density gradient at the edge of its iron core, the 9.0\,$M_\odot$ progenitor generally explodes fairly easily in multi-D simulations. Since it blows up extremely quickly with \textsc{Fornax}, i.e., very soon after bounce, the correspondingly rapid shock expansion shortens the post-explosion accretion onto the PNS. This weakens the GW emission already after $\sim$\,0.35\,s of post-bounce evolution; at $t_\mathrm{pb}\gtrsim 0.35$\,s the GW amplitudes in the \textsc{Fornax} models exhibit only tiny high-frequency variations (Fig.~3 in \cite{Vartanyan:2023sxm}), in contrast to our model s9 (Fig.~\ref{Fig:GW_test_s9}).

We can also link the lower GW energies in our simulations of s12.28 and s18.88 to some features of the matter GW signals that are distinctly different between our models and successfully exploding ones in Refs.~\cite{Vartanyan:2023sxm,Choi:2024irp}. The first prominent difference is that the latter models exhibit much higher $A_+$ GW amplitudes with peak values of 5--20\,cm during the first $\sim$\,50\,ms after bounce, whereas our simulations reach $\sim$\,3.5\,cm at most.\footnote{Considerably higher GW amplitudes for prompt postshock convection than in our models were also found in Ref.~\cite{Powell:2024nvv} with values up to $\sim$\,40\,cm in 3D SN simulations of a non-rotating 15\,$M_\odot$ progenitor with the \textsc{CoCoNuT-FMT} code, employing the SFHo and SFHx EoSs of~\cite{Hempel:2011mk,Steiner:2012rk}. In contrast, similarly small values as in our s18.88 model can be witnessed in the 3D SN models of Ref.~\cite{Mezzacappa+2023}, which were computed with the \textsc{Chimera} code and used the LS220 EoS, which was also applied in s18.88. Therefore these simulations are consistent with each other and confirm the weaker prompt convection and correspondingly lower GW amplitudes with the LS220 EoS.} As a consequence, in almost all of the \textsc{Fornax} models with more than 9\,$M_\odot$ the GW energy literally jumps to a level of more than several $10^{-11}\,M_\odot c^2$. In contrast, s12.28 marginally reaches $\sim$\,$2\times 10^{-11}\,M_\odot c^2$, and s18.88 with only $\sim$\,$10^{-12}\,M_\odot c^2$ remains significantly below because of its weaker prompt postshock convection connected to the use of the LS220 EoS (see Fig.~\ref{Fig:Energy_both} and Sec.~\ref{sec:Time-domain analysis}).
Also during the subsequent vigorous accretion phase, when most of the GW energy is emitted, the \textsc{Fornax} simulations exhibit somewhat higher GW amplitudes for longer periods of time. One example is the 15.01\,$M_\odot$ model, which has a similar core compactness to our s12.28 case and amplitudes of up to more than $\sim$\,5\,cm until $\sim$\,1.5\,s after bounce, whereas s12.28 has amplitudes of $\lesssim$\,2.5\,cm in the same time interval. Other examples are the \textsc{Fornax} models with 18.5, 19, 20, 23\,$M_\odot$, where the amplitudes exceed 5\,cm during extended time periods, whereas our s18.88 case with a similar or even higher compactness reaches this value (or marginally exceeds it) only for a short time (Figs.~\ref{Fig:GW_hydro_signal} and~\ref{Fig:GW_hydro_signal_directions}).    

We further note that we do not witness any systematic absence of oscillations in $A_\times$ during the prompt postshock convection phase during the first $\sim$\,50\,ms after bounce as reported for the \textsc{Fornax} models in Ref.~\cite{Vartanyan:2023sxm}. In our simulations we find that $A_+$ can be weak in some viewing directions, but usually the amplitudes for both polarizations are present with similar strengths (see Fig.~\ref{Fig:GW_hydro_signal}). Since the 3D nature of the progenitors in our simulations has an impact only when the convective oxygen-burning shell falls through the SN shock (i.e., typically at $t_\mathrm{pb}\gtrsim 0.2$\,s), we wonder whether this difference could be connected to the different numerical grids used, namely a dendritic grid with mesh coarsening in the \textsc{Fornax} models versus a Yin-Yang grid without any preferred axis directions in our computations. Also larger or smaller grid-induced numerical perturbations might play a role in explaining the spread of magnitudes found for the GW amplitudes shortly after core bounce with the \textsc{Fornax}, \textsc{CoCoNuT-FMT}~\cite{Powell:2024nvv}, \textsc{Chimera}~\cite{Mezzacappa+2023}, and our \textsc{Prometheus-Vertex} codes.   

The matter GW signals attributed to prompt postshock convection in Ref.~\cite{Vartanyan:2023sxm} also stick out by their timing in the \textsc{Fornax} models, where in all cases this emission starts at several 10\,ms after bounce, in some cases even at $t_\mathrm{pb}\gtrsim 50$\,ms. This is much later than in our models and in those of Refs.~\cite{Powell:2024nvv,Mezzacappa+2023}. In fact, these times are too late for prompt convection behind the decelerating core-bounce shock, because the shock stalls within only milliseconds after bounce, thus forming a negative entropy gradient in its downstream region that immediately triggers convective overturn. Typically this early convective activity and its GW production have ended after 30--50\,ms, as visible in Fig.~\ref{Fig:GW_hydro_signal}. Instead, the powerful GW activity at several 10\,ms post bounce in the \textsc{Fornax} models might originate from grid-induced numerical perturbations or it might indicate a strong transient associated with hydrodynamic relaxation that could be connected to a mapping from 1D collapse simulations to the 3D post-bounce modeling at too late times after core bounce. Accordingly, these earliest phases of vigorous GW emission are followed by only short periods of relative GW quiescence before enhanced GW emission develops again due to the hydrodynamic instabilities that occur in the neutrino-heated postshock layer.

Another difference, possibly linked to the phenomenon described above, concerns the start of the explosions in correlation with the vigor of the GW emission. In many of the \textsc{Fornax} models ---except the lowest-mass ($\sim$\,9\,$M_\odot$) cases--- the explosion sets in (defined by the time when the stalled shock begins to expand outward) during a phase of relatively weak matter-associated GW activity. This instant is immediately or soon after the earliest phase of strong GW production, but still before the GW emission regains more power by neutrino-driven postshock convection (see Figs.~3 and~4 in \cite{Vartanyan:2023sxm}). The corresponding moments of the onset of explosions in our simulations are $t_\mathrm{pb} \sim 170$\,ms for s12.28 and $t_\mathrm{pb} \sim 250$\,ms for s18.88, see Fig.~\ref{fig:ShockRadii}.\footnote{It is important to note that we usually define ---following the convention in previous publications--- the onset of the explosion, $t_\mathrm{exp}$, as the time after bounce when the angle-averaged shock radius exceeds 400\,km (which is roughly the time of no shock return). This is the reason why the times given here differ from those listed for $t_\mathrm{exp}$ in Table~\ref{tab:SNmodels}.} In contrast to the \textsc{Fornax} models, these points are within the time intervals of strong GW emission associated with non-radial hydrodynamic mass motions in the neutrino-heated postshock volume. (In this respect our models resemble the non-rotating 15\,$M_\odot$ models in \cite{Powell:2024nvv}.) While the \textsc{Fornax} models show a substantial growth of the GW amplitudes during the PNS accretion that follows the onset of the explosions, our models yield at most a slight increase of the GW amplitudes after the SN shock has taken off (Figs.~\ref{Fig:GW_hydro_signal} and~\ref{Fig:GW_hydro_signal_directions}). These differences could be indicative of considerably stronger neutrino heating in the simulations of Refs.~\cite{Vartanyan:2023sxm,Choi:2024irp}.   

Finally, we also note that we do not witness any short-duration, eruptive GW bursts correlated with transient events of enhanced PNS accretion as found in Refs.~\cite{Vartanyan:2023sxm,Choi:2024irp} at several seconds after bounce in long-term models with successful explosions. Instead, accretion-induced GW emission appears in more extended and less sharply defined events. This difference might be connected to the large angular momentum carried by accretion downflows of matter that originates from the convective oxygen shell in the 3D progenitors used in our simulations. Due to their high angular momentum, these accretion downdrafts usually do not hit the PNS head-on, but approach it more on scraping paths with less forceful impacts on the PNS surface (Fig.~\ref{fig:LateDynamics_s12.28}). Nevertheless, although the downflows are mostly wide-angle structures, we cannot exclude that higher angular resolution than applied in our long-term model s12.28 might be needed to properly follow the downflow dynamics in greater detail near the PNS surface.

\subsection{Comparison of neutrino GW signals}
\label{sec:comparison-neutrinos}

In agreement with other previous works on the topic~\cite{Muller:2011yi, Vartanyan:2020nmt, Richardson+2024, Choi:2024irp, Richardson+2025}, we observe that most of the GW power connected to anisotropic neutrino emission in both of the considered 3D SN models is at low frequencies, $f< 10$\,Hz. Moreover, all simulations including ours agree that SN models of low-mass progenitors (9.0--9.6\,$M_\odot$ in the mentioned references) possess considerably lower anisotropy parameters, amplitudes, and radiated energies for neutrino-induced GWs than SN explosions of more massive progenitors. However, we also spot important and interesting differences compared to previous results from 3D SN models.

The absolute values of the anisotropy parameters $\alpha_{+,\times}$ for the total neutrino emission in our models s12.28 and s18.88 typically reach several 0.01 and roughly 0.05 ($\sim$\,0.1 for $\nu_e$ and $\bar\nu_e$ individually) in extreme peaks after an initial post-bounce period of about 200\,ms with nearly isotropic neutrino emission. These values are compatible with those reported for 3D SN simulations with the \textsc{Chimera} code for 15\,$M_\odot$ and 25\,$M_\odot$ progenitors in Ref.~\cite{Richardson+2025}. In contrast, the authors of Ref.~\cite{Vartanyan:2020nmt} found somewhat lower values that are mostly around 0.01 and up to 0.02 only during extreme excursions in the first second after bounce for their entire set of 3D SN models produced with the \textsc{Fornax} code. Nevertheless, the absolute values of the neutrino GW amplitudes $A_{+,\times}$ in the latter paper as well as in the follow-up publications \cite{Vartanyan:2023sxm,Choi:2024irp} are close to those of our models for similar exploding stars in the first second after core bounce, namely roughly 50--150\,cm (but with large progenitor and viewing-angle variations). Surprisingly, Ref.~\cite{Richardson+2025} with the higher $|\alpha_{+,\times}|$ values obtained much lower amplitudes of only $\lesssim$\,5\,cm within 0.5--0.6\,s post bounce. Our amplitudes also agree reasonably well with those for the non-rotating 15\,$M_\odot$ SN models (computed with the \textsc{CoCoNuT-FMT} code) in \cite{Powell:2024nvv}, where values up to nearly 200\,cm are reached within less than 0.5\,s after bounce (but no results for the anisotropy parameters are provided).\footnote{A comment is in order at this point. Although Ref.~\cite{Muller:2011yi} is meanwhile a standard reference for the SN GW analysis, the neutrino GW signals there should not be understood quantitatively and should not be interpreted in comparison to modern calculations. There are two reasons for that. First, in the 3D SN simulations of this paper the high-density core of the PNS was not included on the computational grid, but instead it was replaced by an inward moving boundary to mimic the contraction of the PNS. Second, at this boundary isotropic neutrino luminosities with chosen, time-dependent values were imposed, and the neutrino transport on the grid was described by a simplified (grey) treatment. For these reasons the neutrino anisotropy parameters tend to be underestimated and the neutrino GW amplitudes are smaller than in state-of-the-art, more sophisticated simulations.}

It is difficult to assess the reasons for these differences and partly contradictory results. On the one hand, the lower amplitudes obtained in Ref.~\cite{Richardson+2025} in spite of anisotropy-parameter values that are similar to those in our SN models could mean that the neutrino luminosities in~\cite{Richardson+2025} are significantly lower (which is unlikely) or that the variations with opposite signs of their $\alpha_{+,\times}$ parameters cancel each other out more perfectly. On the other hand, the lower values of $\alpha_{+,\times}$ in~\cite{Vartanyan:2020nmt} seem to be compensated by higher luminosities to yield GW amplitudes in overall agreement with our findings. 

One might speculate that the differences could be connected to the fact that our \textsc{Vertex} code and the \textsc{Chimera} code use the ray-by-ray-plus (RbR+) approximation for neutrino transport in 3D, which takes into account only radial flux components, whereas the \textsc{Fornax} code employs a multidimensional transport treatment accounting for all components of the neutrino flux vectors. Since the RbR+ method tends to overestimate angular variations of the fluxes~\cite{Sumiyoshi+2015,Glas:2018oyz}, it might lead to larger values of the anisotropy parameters. However, such an argument cannot explain the mentioned differences between our \textsc{Vertex} and the \textsc{Chimera}~\cite{Richardson+2025} results, since both codes use RbR+ transport. Moreover, it is also disfavored as an explanation of our differences compared to Ref.~\cite{Vartanyan:2020nmt} by our own core-collapse simulations with RbR+ vs.\ multi-D transport in Appendix~\ref{app:neutrinosignals}. Our calculations show, as expected, that the neutrino anisotropy parameters $\alpha_{+,\times}$ with RbR+ are indeed slightly larger, in particular in peaks, than those obtained with multi-D transport (see Fig.~\ref{Fig:GW_nu_Anisotropy_test}). However, this effect is considerably smaller than the mentioned differences between our \textsc{Vertex} values for $\alpha_{+,\times}$ and those obtained with \textsc{Fornax}~\cite{Vartanyan:2020nmt}. Moreover, since the excursions of the anisotropy parameters are in both positive and negative directions, the effects of higher peaks are likely to partly cancel each other out. This is verified by our relatively moderate RbR+ vs.\ full multi-D differences in the GW amplitudes of Fig.~\ref{Fig:GW_nu_directions_test}. In addition, the numerical aspects connected to these transport methods cannot explain why the neutrino anisotropy parameters in~\cite{Vartanyan:2020nmt} are significantly smaller than our values for s12.28 and s18.88, but the GW amplitudes are much more similar.

Interestingly, our models do not show the steady, long-term growth of the neutrino GW amplitudes at $t_\mathrm{pb}\gtrsim 1$\,s seen in \cite{Vartanyan:2023sxm,Choi:2024irp}, where quite a number of simulations yield values of 500\,cm and some extreme cases even 1000--2000\,cm, still growing further after many seconds. Instead, our models s12.28 and s18.88 indicate a saturation of the amplitudes after not even one second post bounce. Then a slow, long-term trend of a gradual decline follows at $t_\mathrm{pb}\gtrsim 1$\,s, signaling that a growing contribution to the neutrino energy loss by the PNS is radiated more spherically in a time-averaged sense due to cancellation effects caused by sign changes in the anisotropy parameter.\footnote{This is consistent with the neutrino-kick analysis of Ref.~\cite{Janka:2024xbp}, where relatively large fluctuations of the instantaneous neutrino-emission dipole partly eliminate each other so that the time-averaged value of the emission asymmetry parameter is considerably reduced; see, e.g., Figs.~11 and 12 there.} Since such a behavior seems to be present also in some simulations of Refs.~\cite{Vartanyan:2023sxm,Choi:2024irp}, the long-term growth may depend on the progenitor and on special properties of the explosion; in particular it might require more extreme explosion asymmetries than obtained in our models combined with morphologies of the PNS accretion that remain stable for seconds. Therefore our two simulations are not enough to exclude the possibility of a long-term, continuous growth of the neutrino GW amplitudes. It could also be, however, that 3D perturbations in the pre-collapse progenitor introduce more stochasticity and time variability of the accretion downdrafts to the PNS. Since these inward flows get deflected by their large angular momentum and collide with each other and with outflows, they create a highly turbulent medium around the PNS. The chaotic flows in the vicinity of the PNS might prevent the stable accretion and neutrino emission geometry that is probably needed to obtain the long-lasting increase of the neutrino GW amplitudes over many seconds.  

The seconds-long trend of amplitude growth is a low-frequency effect in the Hz and sub-Hz range. Accordingly, it has a relatively minor influence on the energy radiated in neutrino-induced GWs. Since our neutrino GW amplitudes produced with frequencies above $\sim$\,1\,Hz at $t_\mathrm{pb} \lesssim 1$\,s agree fairly well with those of exploding models in \cite{Vartanyan:2020nmt,Vartanyan:2023sxm,Choi:2024irp}, it is therefore plausible that the neutrino GW energies are also quite compatible, with a slight tendency of somewhat lower values (by several 10\% up to about a factor of 2) on our side. The energy radiated in the neutrino memory by the 9.0\,$M_\odot$ simulations with \textsc{Fornax}, for example, is around $10^{-12}\,M_\odot c^2$, whereas our s9 models emit (2.5-10)$\times 10^{-13}\,M_\odot c^2$, depending on the model run (Fig.~\ref{Fig:GW_nu_Energy_test}). Our s12.28 case with $4.2\times 10^{-11}\,M_\odot c^2$ (Table~\ref{tab:SNmodels}) is close to the $\sim$\,$5\times 10^{-11}\,M_\odot c^2$ of the 15.01\,$M_\odot$ and 18\,$M_\odot$ \textsc{Fornax} models (all using progenitors with relatively similar core compactness) and not too far from the $\sim$\,$8\times 10^{-11}\,M_\odot c^2$ of the 11\,$M_\odot$ \textsc{Fornax} model. Similarly, our model s18.88 with $7.0\times 10^{-11}\,M_\odot c^2$ matches pretty well the $\sim$\,(1--$1.5)\times 10^{-10}\,M_\odot c^2$ of the 18.5, 20, 24, 25\,$M_\odot$ \textsc{Fornax} models as well as the $\sim$\,$7\times 10^{-11}\,M_\odot c^2$ of the 23\,$M_\odot$ \textsc{Fornax} case, all of which used progenitors with compactness values close to our 18.88\,$M_\odot$ progenitor. Also the $\sim$\,$9\times 10^{-11}\,M_\odot c^2$ neutrino GW energy of the 19\,$M_\odot$ \textsc{Fornax} simulation is not far from our s18.88 model. This good compatibility of the GW energy for the neutrino memory is in stark contrast to the big differences that we found for the GW energy of the matter signal (see Sec.~\ref{sec:comparison-matter}). 

Despite these similarities of the GW energies, the ASDs of the neutrino memory signals of our models, when compared to those in Ref.~\cite{Choi:2024irp} (where a Nyquist frequency compatible with our analysis was used), exhibit some prominent differences in their shapes and magnitudes. At intermediate frequencies between roughly 10\,Hz and several 100\,Hz the neutrino ASDs in~\cite{Choi:2024irp} also follow a power-law-like decline, though with quite some variability in the steepness between the explosion simulations of different progenitors. However, there is a clear overall tendency of less power in our models, whose ASDs are best compatible with those of the lowest-mass cases (less than 10\,$M_\odot$) in the model set of~\cite{Choi:2024irp}. Moreover, on both the low-frequency ($f \lesssim 10$\,Hz) and the high-frequency sides ($f \gtrsim 100$\,Hz) the ASDs in~\cite{Choi:2024irp} are substantially higher than ours and display a flattening of their slopes that is not present in our neutrino GW ASDs.

At around 1\,Hz all SN runs in~\cite{Choi:2024irp} for progenitors with more than 10\,$M_\odot$ yield high humps of the neutrino ASDs with values well above $10^{-21}$\,Hz$^{-1/2}$, whereas our ASDs for the total neutrino GW signals are up to one order of magnitude lower and in line with the overall power-law behavior. This difference is not astonishing in view of the long-term growth of the neutrino GW amplitudes for most of the \textsc{Fornax} simulations mentioned before. At the high-frequency end above a few 100\,Hz, the ASDs of the \textsc{Fornax} simulations also become {\em flatter}, in contrast to the steeper {\em downward} slopes of our ASDs. Moreover, the \textsc{Fornax} results also show higher fluctuations than our ASDs. This qualitatively different functional behavior occurs on a very low absolute level and therefore seems to have only a minor impact on the neutrino GW energies, too, similar to the ASD differences at low frequencies. 

The high-frequency differences might be connected to a lower level of short-time variability in our simulation outputs. In this context it could play a role that our \textsc{Vertex} neutrino transport is implicit in time and therefore permits larger time steps, thus suppressing very-high-frequency fluctuations. However, this thought is not really convincing, because the time steps used in \textsc{Vertex} are always smaller than roughly $10^{-5}$\,s (compared to $\sim$\,$10^{-7}$\,s for a time-explicit transport as used in \textsc{Fornax}). Therefore our simulations should still be able to track neutrino signal variations with frequencies up to at least $10^4$\,Hz.

\section{Summary and conclusions}
\label{sec:conclu}

In this work we have analyzed the GW signals of two state-of-the-art SN models, which were started from 3D progenitors and evolved continuously in three dimensions with full energy- and velocity-dependent neutrino transport up to late times after core bounce ($t_{\rm pb} \sim 5.1\,\s$ for a 12.28\,$M_\odot$ model and $t_{\rm pb} \sim 1.7\,\s$ for an 18.88\,$M_\odot$ case) by employing the \textsc{Prometheus-Vertex} neutrino-hydrodynamics code. Our analysis in both time and frequency domains comprised the GW signals associated with nonspherical mass motions as well as anisotropic neutrino emission.

The 3D simulations performed for the final evolution phase of the two progenitor stars had developed oxygen-neon shell mergers during the last hour before iron-core collapse in model s12.28 and within the last 7\,min in the s18.88 model. These shell mergers led to vigorous, large-scale mass flows in the convective oxygen-burning shell, which were conducive to buoyancy-aided, neutrino-driven explosions in our subsequent 3D core-collapse simulations with \textsc{Prometheus-Vertex}. For this reason we were interested in 
the question whether the 3D pre-collapse nature of the progenitors produces any observable features in the GW signals. In connection to investigating this question, we compared our results in quite some detail with similar models in the literature. In this context it is important to realize that, despite the stochasticity of many aspects of the signals, basic properties of the GW amplitudes, energies, and spectral distributions both for matter and neutrinos are extremely sensitive to fundamental characteristics of the hydrodynamic mass motions and of the neutrino emission. A GW analysis is therefore a very fine tool to probe and diagnose differences in the numerical simulations.

Our GW signals display the well-known components caused by prompt postshock convection, hydrodynamic instabilities in the neutrino-heated gain layer behind the stalled shock, asymmetric PNS accretion and oscillation modes, and the long-lasting amplitude excursions (matter and neutrino ``memory'' effects) due to nonspherical mass ejection and neutrino emission. We could not identify any new GW features or specific properties that can be unambiguously linked to the 3D structures in the velocity field and density distribution caused by convective oxygen-neon-shell burning in the progenitors. Nevertheless, we spotted some interesting differences compared to results in the literature for SN simulations with progenitors and treatments of the physics similar to ours (see Sec.~\ref{sec:comparison}). Some of these differences might also be caused by the large-scale 3D progenitor perturbations, but there is no definite evidence for that, and also other explanations including numerical effects are possible.

In the SN simulations of Ref.~\cite{Vartanyan:2023sxm}, for example, shock expansion initiates the explosion in a phase of relatively low GW emission, before the high-frequency GW amplitudes due to turbulent mass motions around the PNS rise to their largest values. In contrast, in our SN models (similar to the non-rotating models in~\cite{Powell:2024nvv}) the shock starts to expand during the loudest phase of matter-induced GW production, which coincides with the collapse of the convectively perturbed oxygen layer through the SN shock. After the onset of the explosion, we witness a long-lasting ``haze'' of GW activity in the frequency range of some 100\,Hz up to $\sim$\,1000\,Hz. It persists beyond the phase of massive accretion onto the PNS, continuing until our s12.28 simulation was stopped at $t_\mathrm{pb} > 5.1$\,s and accompanying the low-frequency ($f \lesssim 100$\,Hz) emission caused by the matter memory associated with the asymmetric ejecta expansion. This haze can hardly be seen in the exploding models of Ref.~\cite{Vartanyan:2023sxm}, but its visibility depends on the plotted quantity and is particularly poor when $dE_\mathrm{GW}^\mathrm{M}/df$ is shown~\cite{Vartanyan:2023sxm}, because this energy spectrum weights the high-frequency contributions more strongly compared to the low-frequency range. This circumstance hampers a detailed comparison of our matter GW spectrogram with those of Ref.~\cite{Vartanyan:2023sxm} in the frequency domain below $\sim$\,1000\,Hz. The intensity of the haze in this region grows during phases of high mass-inflow rates to the PNS, when accretion downdrafts form a turbulent medium around the PNS. These downdrafts carry high angular momentum, collide with each other and with upward rising plumes and spin-up the near-surface layers of the PNS. We speculate that these effects might be particularly strong in our models because of the 3D nature of the pre-collapse progenitors and might be responsible for the hazy, broad-band GW noise.  

Also in the GWs associated with anisotropic neutrino emission we witness signatures that might be connected to the 3D convection flows in the progenitors' oxygen-neon shells. During the first second after bounce (${t_\mathrm{pb}\lesssim 1}$\,s), the neutrino anisotropy parameters $\alpha_{+,\times}$ of a 20\,$M_\odot$ simulation, which we initiated with a 1D progenitor, are lower ---and correspondingly the neutrino GW amplitudes, too--- than those of our 12.28\,$M_\odot$ and 18.88\,$M_\odot$ simulations (using 3D progenitors). These computations employed different codes (\textsc{Alcar} with RbR+ as well as full multi-D transport in the former case and \textsc{Prometheus-Vertex} in both latter ones), but otherwise quite similar physics. This disfavors other explanations for the $\alpha_{+,\times}$ difference than the dimensionality of the progenitor, in particular, because the 20\,$M_\odot$ model does not exhibit the conspicuous, low-frequency, large-amplitude excursions from the zero level seen in the time evolution of $\alpha_{+,\times}$ in s12.28 and s18.88 (where they reach up to $\sim$\,5\% for the total neutrino signal). Such an effect can be expected as a consequence of large-scale convective perturbations in the pre-shock accretion flow, although $\alpha_{+,\times}$ excursions can also be spotted in Refs.~\cite{Vartanyan:2020nmt,Richardson+2025}, where 1D progenitors were employed.  

During the late-time post-bounce evolution (${t_\mathrm{pb}\gtrsim 1}$\,s), the neutrino GW amplitudes in Refs.~\cite{Vartanyan:2020nmt,Vartanyan:2023sxm,Choi:2024irp} tend to be larger than our values despite the overall lower magnitudes of the neutrino anisotropy parameters in their models. This could point to a higher temporal stability of the PNS-accretion and neutrino-emission geometry, which would be compatible with the use of 1D progenitors in~\cite{Vartanyan:2020nmt,Vartanyan:2023sxm,Choi:2024irp} instead of 3D progenitors in our s12.28 and s18.88 simulations. In this context we stress that our models neither yield the extremely high neutrino GW amplitudes of $A_{+,\times} > 500$\,cm, nor the quasi-monotonic, long-term trends of amplitude growth that develop at times later than 1\,s after bounce in some of the models of Refs.~\cite{Vartanyan:2020nmt,Vartanyan:2023sxm,Choi:2024irp}. Instead, our neutrino GW amplitudes display nearly steady absolute values of 100--200\,cm at $t_\mathrm{pb}\gtrsim 0.5$\,s and even a shallow, long-term, slow trend of decrease. This indicates that the energy released in the late-time neutrino emission of the PNS is radiated more isotropically in a time-averaged sense, because short-time fluctuations of the anisotropy parameter partly compensate each other. But since there may be strong variations depending on the stellar progenitors, our two investigated cases do not permit firm and definitive statements on the question whether this behavior is a generic consequence of 3D progenitors, or whether continuous amplitude growth is possible at $t_\mathrm{pb}\gtrsim 1$\,s under special circumstances. 

A larger set of core-collapse simulations based on 3D progenitors is needed to come to more robust conclusions on such effects and, in general, on the impact of pre-collapse asymmetries on the GW signals of successful and failed SNe.

\begin{figure}[t!]
    \centering
    \includegraphics[width=1\columnwidth]{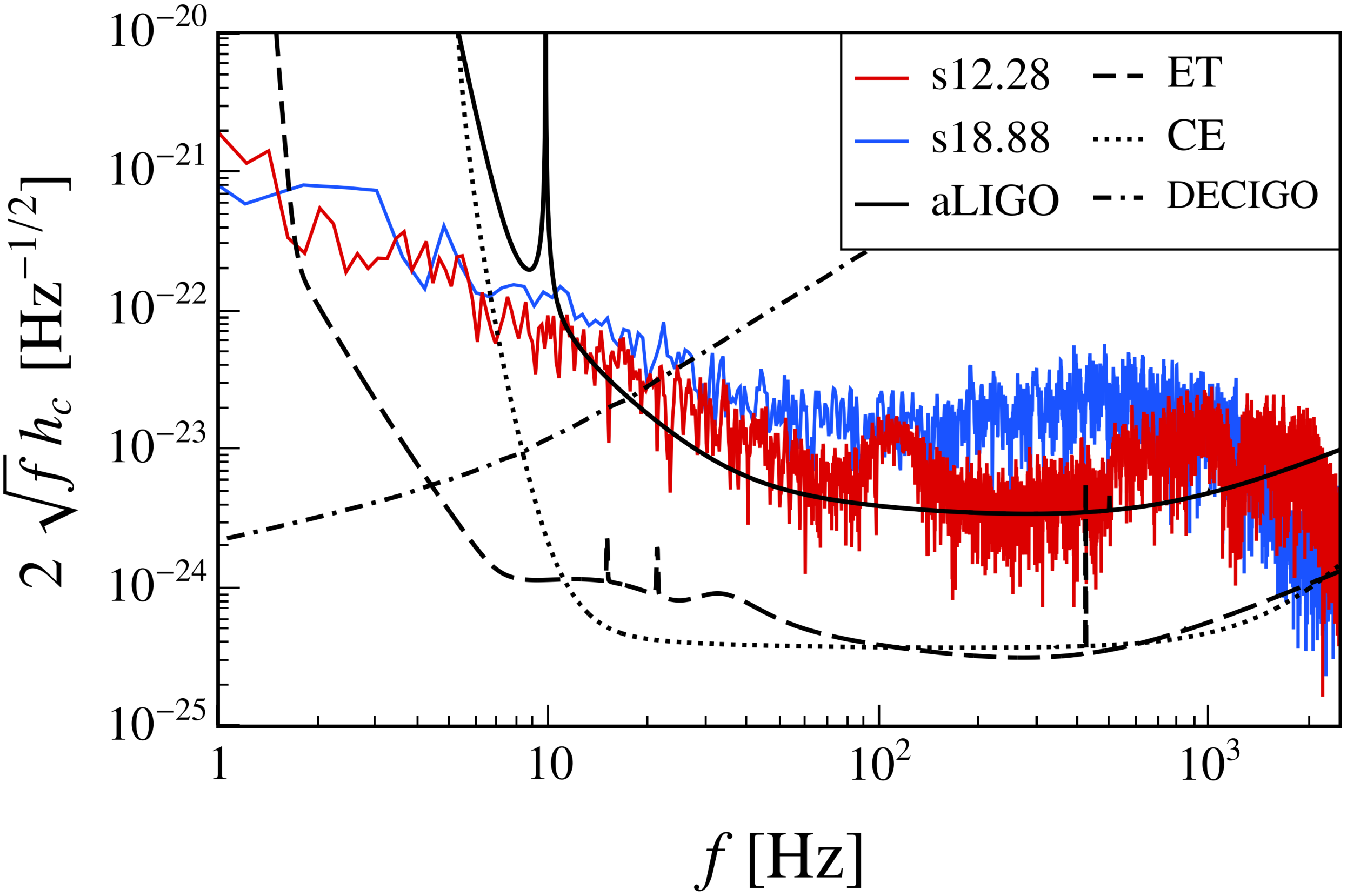}
    \caption{ASDs for the combined neutrino and matter GW signals of both analyzed SN models (colors analogous to Fig.~\ref{Fig:Energy_both} as given in the inset) for an observer located along the $(+x)$-axis in the equatorial plane of the source at a distance of $D=10$\,kpc from a Galactic SN. Black lines depict the sensitivity curves of current (aLIGO: advanced LIGO) and proposed GW experiments (ET: Einstein Telescope; CE: Cosmic Explorer; DECIGO) operating in the frequency range of interest.}
    \label{Fig:Detection}
\end{figure}

As discussed in Sec.~\ref{sec:comparison-neutrinos}, the ASDs for the neutrino-induced GW signals exhibit substantial differences compared to those in Ref.~\cite{Choi:2024irp}. In particular, our ASDs possess considerably less power in the low-frequency range ($f$ below some 10\,Hz) and in the high-frequency wings ($f\gtrsim  100$\,Hz) of the spectra. The steep decline of our neutrino ASDs above a few 100\,Hz is compatible with the Nyquist frequency of $\sim$\,500\,Hz for our neutrino GW analysis, and also the authors of Ref.~\cite{Choi:2024irp} admit that their neutrino GW spectra are not reliable at $f > 500$\,Hz due to their Nyquist sampling of the neutrino data. These differences affect only the low-frequency parts of the combined matter and neutrino GW ASDs, because at frequencies above $\sim$100\,Hz the matter-associated GW emission dominates. The matter GW signals in the simulations of~\cite{Choi:2024irp} contain considerably more energy than those of our models, in particular above 1000\,Hz, where our matter ASDs drop more quickly. This is the case at $\gtrsim$\,1000\,Hz for s18.88 and at $\gtrsim$\,2500\,Hz for s12.28, because the accuracy of our results above these values is limited by too-low output-sampling rates. These different Nyquist frequencies lead to the artifact of a crossing of the GW spectra of our two models in Fig.~\ref{Fig:Detection} at about 1000\,Hz due to a downshifted high-frequency peak and enhanced power around several 100\,Hz in s18.88 caused by aliasing effects. 

Generally, the radiated GW energies, which are dominated by the matter contributions, tend to be significantly lower ---typically by a factor between $\gtrsim$\,10 and several 10--- in our SN models than in those of the \textsc{Fornax} simulations in Ref.~\cite{Choi:2024irp} for progenitors of similar compactness (though our lower Nyquist frequencies might play a role in this comparison). We witness best agreement of our results with the GW energies released by the 9.25\,$M_\odot$ and 9.5\,$M_\odot$ models of~\cite{Choi:2024irp}, which possess very small core-compactness values. This suggests much more vigorous nonspherical matter motions (involving also more mass) and more violent explosions in the \textsc{Fornax} models, probably because of stronger neutrino heating. Therefore, also our ASDs for the total GW emission (matter plus neutrino contributions) in Fig.~\ref{Fig:Detection} are best compatible with the ASDs of the SN models in Ref.~\cite{Choi:2024irp} that have a relatively low production of GW energy, especially the explosions of the lower-mass progenitors between $\sim$\,9.25\,$M_\odot$ and $\sim$\,12.25\,$M_\odot$ and of 16\,$M_\odot$ (if one ignores the higher power of the $>$\,10\,$M_\odot$ models of ~\cite{Choi:2024irp} at frequencies below some 10\,Hz). 

The shape as well as the magnitude of the ASD of s12.28 also agree reasonably well with the results for the non-rotating 15\,$M_\odot$ simulations in Ref.~\cite{Powell:2024nvv}. The latter models, however, radiate over 10 times more GW energy, possibly partly because of the extremely strong signals from prompt postshock convection in those simulations. Correspondingly, the ASD of s12.28 shows somewhat less power at all frequencies $f \gtrsim 50$\,Hz, although the post-bounce evolution was followed in~\cite{Powell:2024nvv} only for about 0.5\,s, which was not sufficient to track the buildup of more power in the low-frequency GW signals due to anisotropic neutrino emission and at high frequencies by accretion-induced GW activity in and around the PNS.

Considering our 9.0, 12.28, and 18.88\,$M_\odot$ simulations, all of which yield successful explosions, the total radiated GW energy increases roughly like $\xi_{1.75}^{2/7}$ with the progenitor compactness $\xi_{1.75}$. The power-law exponent $2/7 \approx 0.286$ is significantly lower than the value of 0.726 given in Ref.~\cite{Vartanyan:2023sxm}. 

Despite the mentioned quantitative differences, our ASDs are generally consistent with those published previously \cite{Vartanyan:2020nmt,Vartanyan:2023sxm,Choi:2024irp,Powell:2024nvv}, including the general finding that higher-compactness progenitors typically lead to more powerful GW emission. For this reason our models permit similar conclusions about the detectability of the GW signals as discussed before. In order to compare our combined ASDs with the sensitivities of current and planned interferometric GW detectors operating in the frequency range of interest, we display in Fig.~\ref{Fig:Detection} the sensitivity curves of the current advanced LIGO~(aLIGO)~\cite{LIGOScientific:2014pky} as well as of future detectors, including the Einstein Telescope~(ET)~\cite{Hild:2010id, Punturo:2010zz}, Cosmic Explorer~(CE)~\cite{LIGOScientific:2016wof} and DECIGO~\cite{Kawamura:2008zz}. For a Galactic SN at a distance of $D = 10$\,kpc, both of our models, in particular their spectral maxima at $\sim$\,100\,Hz and $\sim$\,1000\,Hz, are in reach of aLIGO. Future GW facilities will clearly do even better and will increase the distance of well measurable SN GW signals by at least a factor 10. While the CE is able to capture the complete spectral range of the high-frequency signal associated with mass motions for both s12.28 and s18.88, DECIGO will be sensitive only to the low-frequency band ($f\lesssim 10$\,Hz) mainly fed by the GWs produced by the anisotropic escape of neutrinos. Interestingly, the envisioned ET would be capable of detecting a major part of the combined neutrino and matter GW spectra spanning the range of $3\,\Hz\lesssim f\lesssim2000\,\Hz$. We therefore confirm the common expectation that future GW instruments will offer an excellent perspective to detect the GW signal from a future Galactic SN in coincidence with the concomitant neutrino burst. Thus they show great promise for a next seminal step in multimessenger astrophysics.

\section{Acknowledgments}
We warmly thank Mainak Mukhopadhyay for useful discussions on this topic during the ``Neutrino Frontiers'' workshop. In this regard, we also thank the Galileo Galilei Institute for Theoretical Physics for the hospitality and the INFN for partial support during the work on this manuscript. AM and AL were partially supported by the research grant number 2022E2J4RK ``PANTHEON: Perspectives in Astroparticle and Neutrino THEory with Old and New messengers" under the program PRIN 2022 funded by the Italian Ministero dell’Universit\`a e della Ricerca (MUR). 
Moreover, AL was supported by the Italian MUR through the FIS 2 project FIS-2023-01577 (DD n. 23314 10-12-2024, CUP C53C24001460001), and by Istituto Nazionale di Fisica Nucleare (INFN) through the Theoretical Astroparticle Physics (TAsP) project. The work of GL at SLAC was supported by the U.S. Department of Energy under contract number DE-AC02-76SF00515. At the beginning of this project, GL was supported by the European Union’s Horizon 2020 Europe research and innovation programme under the Marie Skłodowska-Curie grant agreement No 860881-HIDDeN. At Garching, support by the German Research Foundation (DFG) through the Collaborative Research Centre ``Neutrinos and Dark Matter in Astro- and Particle Physics (NDM),'' Grant SFB-1258-283604770, and under Germany’s Excellence Strategy through the Cluster of Excellence ORIGINS EXC-2094-390783311 is acknowledged.
This work also received partial support by ICSC\,--\,Centro Nazionale di Ricerca in High Performance Computing.

\appendix

\section{GW analysis for additional SN simulations of 1D progenitors}
\label{app:code}

\begin{figure*}[t!]
    \centering
    \includegraphics[width=1\textwidth]{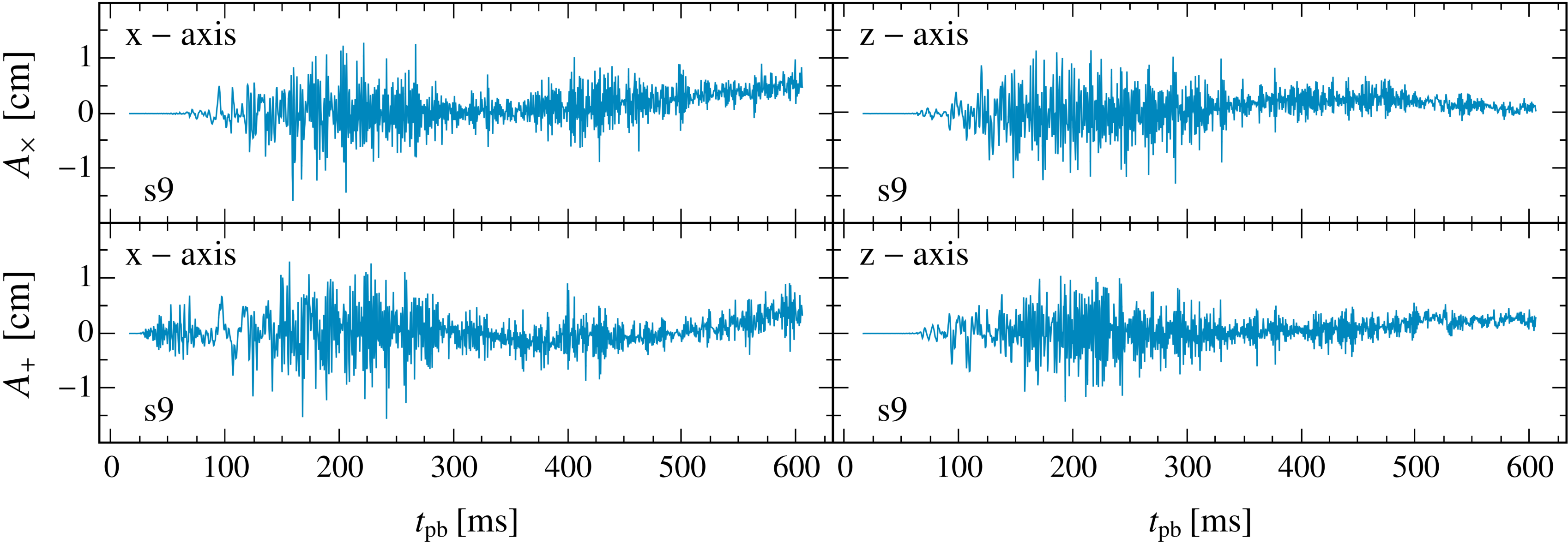}
    \caption{GW amplitudes due to asymmetric mass motions as functions of post-bounce time for model s9-FMD-H computed with the \textsc{Alcar} code~\cite{Glas:2018oyz} and previously analyzed for matter GWs in Ref.~\cite{Andresen:2020jci} (Fig.~2 there). The \emph{upper} and \emph{lower panels} show $A_\times$ and $A_+$, respectively. In the \emph{left panels} the signal is given for an observer along the $(+x)$-axis of the computational grid and in the \emph{right panels} for an observer located along the $(+z)$-axis. Since the s9 model yields a successful explosion, one can witness the development of a long-lasting deviation from the zero level at $t_\mathrm{pb} \gtrsim 400$\,ms due to the matter memory effect caused by the asymmetric SN ejecta.}
    \label{Fig:GW_test_s9}
\end{figure*}

\begin{figure*}[t!]
    \centering
    \includegraphics[width=1\textwidth]{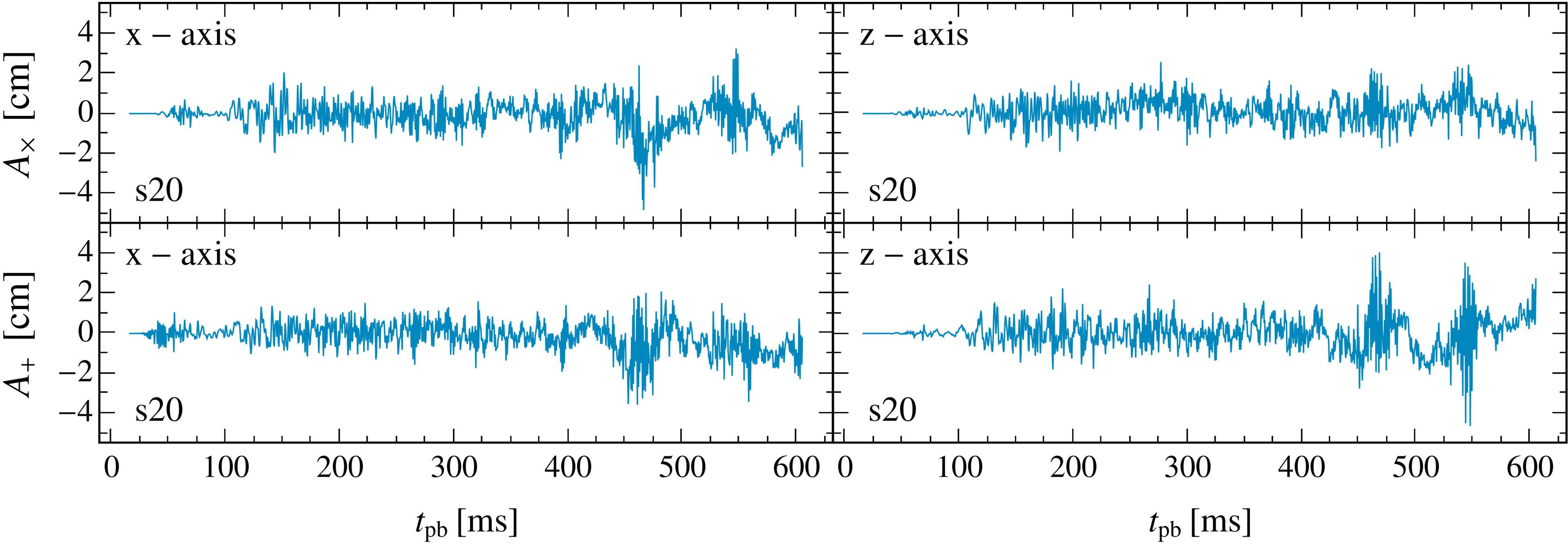}
    \caption{GW amplitudes due to asymmetric mass motions as functions of post-bounce time for model s20-FMD-H computed with the \textsc{Alcar} code~\cite{Glas:2018oyz} and previously analyzed for matter GWs in Ref.~\cite{Andresen:2020jci} (Fig.~4 there). The \emph{upper} and \emph{lower panels} show $A_\times$ and $A_+$, respectively. In the \emph{left panels} the signal is given for an observer along the $(+x)$-axis of the computational grid and in the \emph{right panels} for an observer located along the $(+z)$-axis. The s20 model does not produce an explosion, for which reason the amplitudes fluctuate around zero with high frequency until the end of the simulation.}
    \label{Fig:GW_test_s20}
\end{figure*}

\begin{figure}[t!]
    \centering
    \includegraphics[width=\columnwidth]{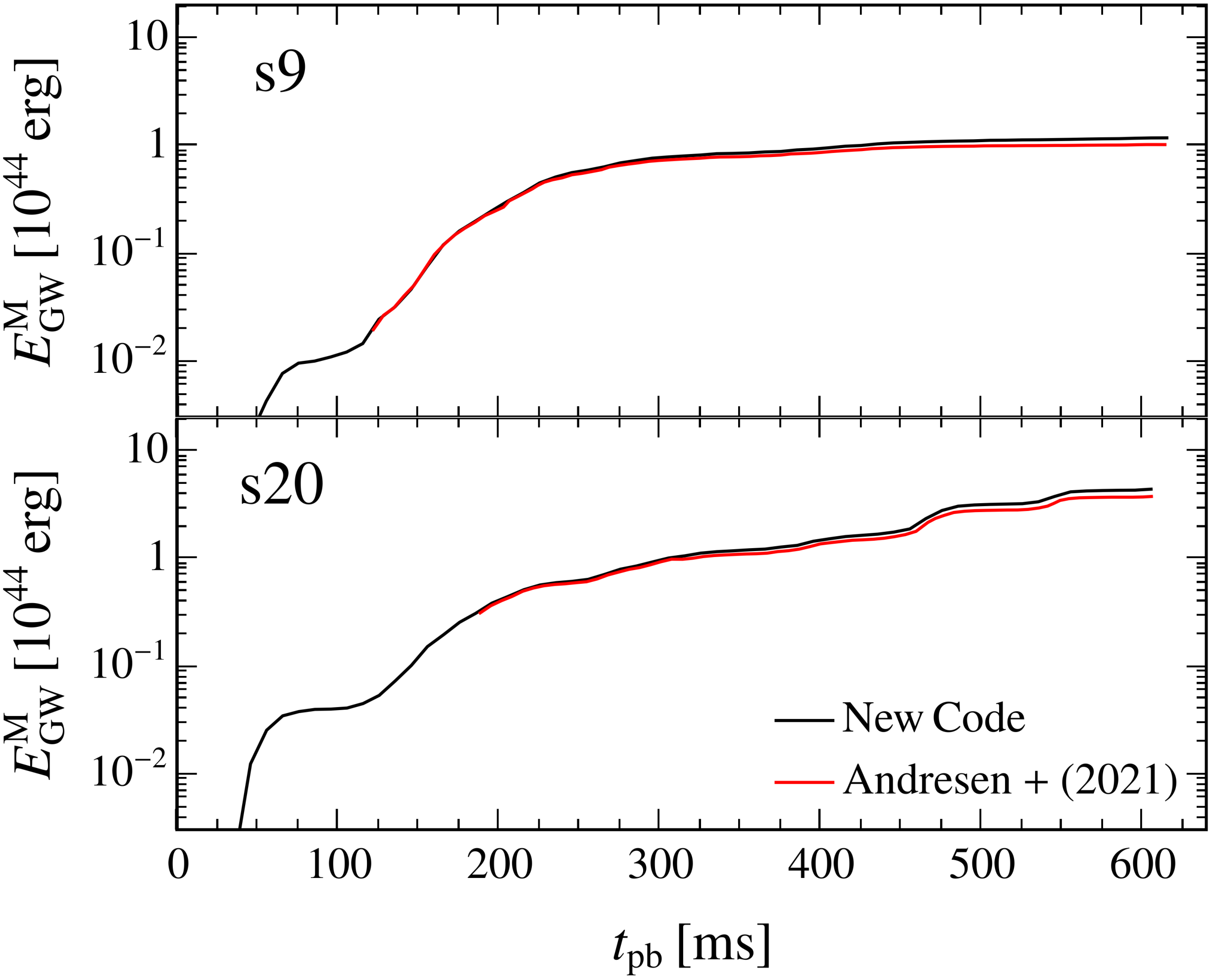}
    \caption{Cumulative radiated GW energy ($10^{44}\,\mathrm{erg} \cong 5.56\times 10^{-11}\,M_\odot c^2$) associated with asymmetric mass motions as function of post-bounce time for models s20-FMD-H (\emph{upper panel}) and s9-FMD-H (\emph{lower panel}). The results of this work (black lines) are compared to the analysis performed in Ref.~\cite{Andresen:2020jci} (red lines). The minor differences are caused by a slightly different numerical evaluation of the GW signals.}
    \label{Fig:GW_test_Energy}
\end{figure}

\begin{figure*}[t!]
    \centering
    \includegraphics[width=1\textwidth]{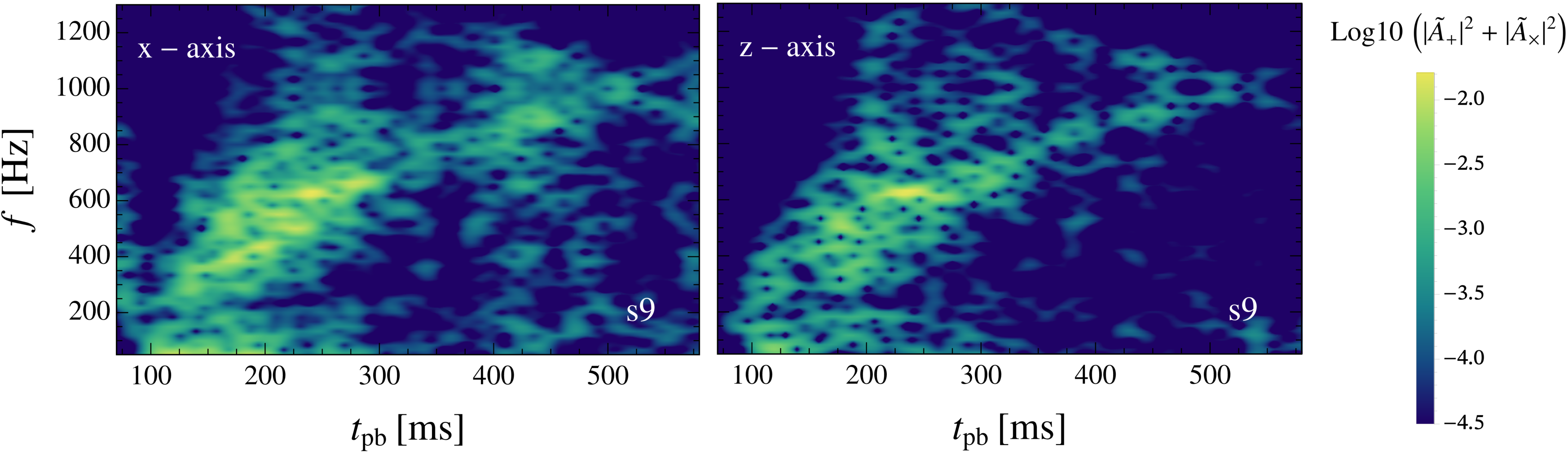}\\
    \includegraphics[width=1\textwidth]{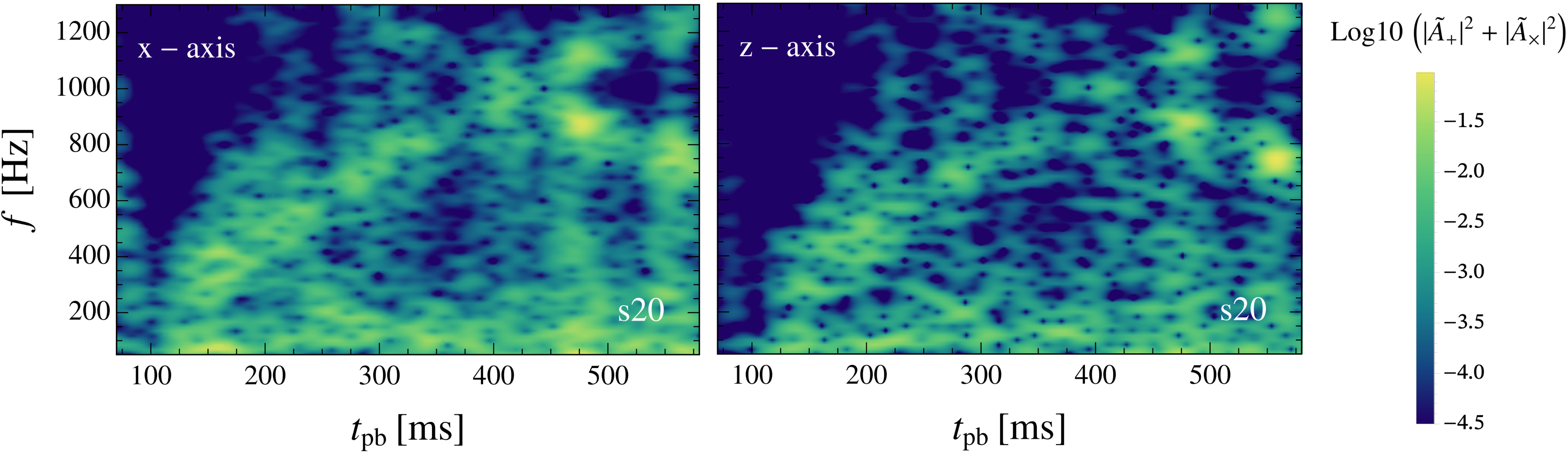}
    \caption{Amplitude spectrograms (in units of cm$^2$s$^2$) as defined in Sec.~\ref{Sec:Hydro_freq} for the GW emission associated with asymmetric mass motions in models s9-FMD-H (\emph{upper panels}) and s20-FMD-H (\emph{lower panels}). The \emph{left} and \emph{right panels} correspond to observers located along the $(+x)$-axis and the $(+z)$-axis of the computational grid, respectively. The upper panels should be compared with Fig.~3 in Ref.~\cite{Andresen:2020jci} and the lower panels with Fig.~5 in this reference. We point out the aliasing artifacts that are caused by data outputs with time intervals of 0.5\,ms and become visible when the GW frequency reaches the Nyquist frequency of 1000\,Hz. }
    \label{Fig:Spectrogram_test}
\end{figure*}

\begin{figure*}[t!]
    \centering
    \includegraphics[width=1\textwidth]{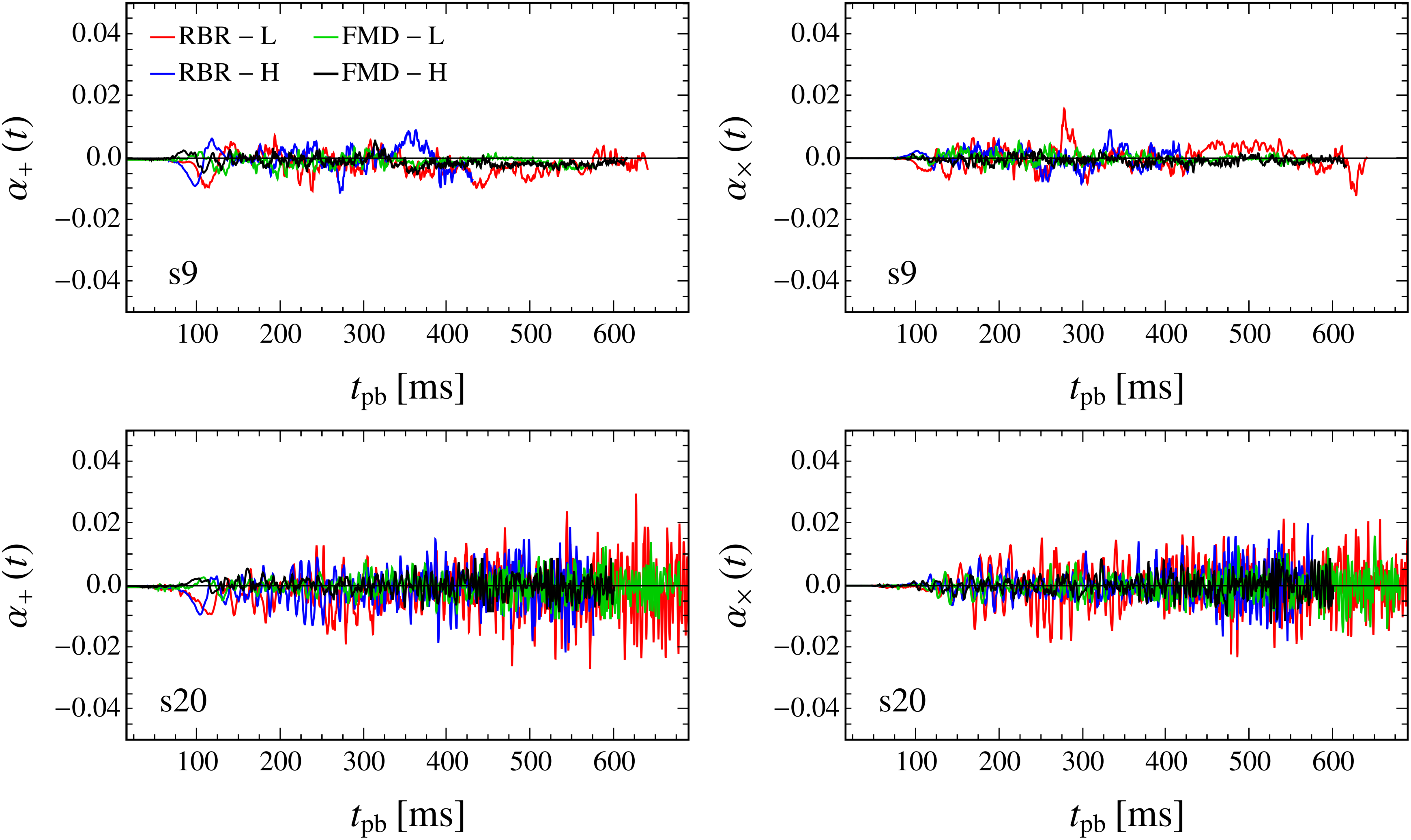}
    \caption{Neutrino anisotropy parameters [Eq.~\eqref{Eq:Anisotropy}] as functions of post-bounce time for the $+$ ({\em left panels}) and $\times$ ({\em right panels}) GW polarization states, assuming an observer along the $(+x)$-axis in the equatorial plane of the source. The \emph{upper panels} and \emph{lower panels} show the results for different SN simulations of a 9\,$M_\odot$ progenitor (models s9) and of a 20\,$M_\odot$ progenitor (models s20), respectively. The colors correspond, as labeled, to simulations with ray-by-ray-plus transport (RBR), full multidimensional transport (FMD) as well as low (L) and high (H) angular grid resolution (see text for details).}
    \label{Fig:GW_nu_Anisotropy_test}
\end{figure*}

\begin{figure*}[t!]
    \centering
    \includegraphics[width=1\textwidth]{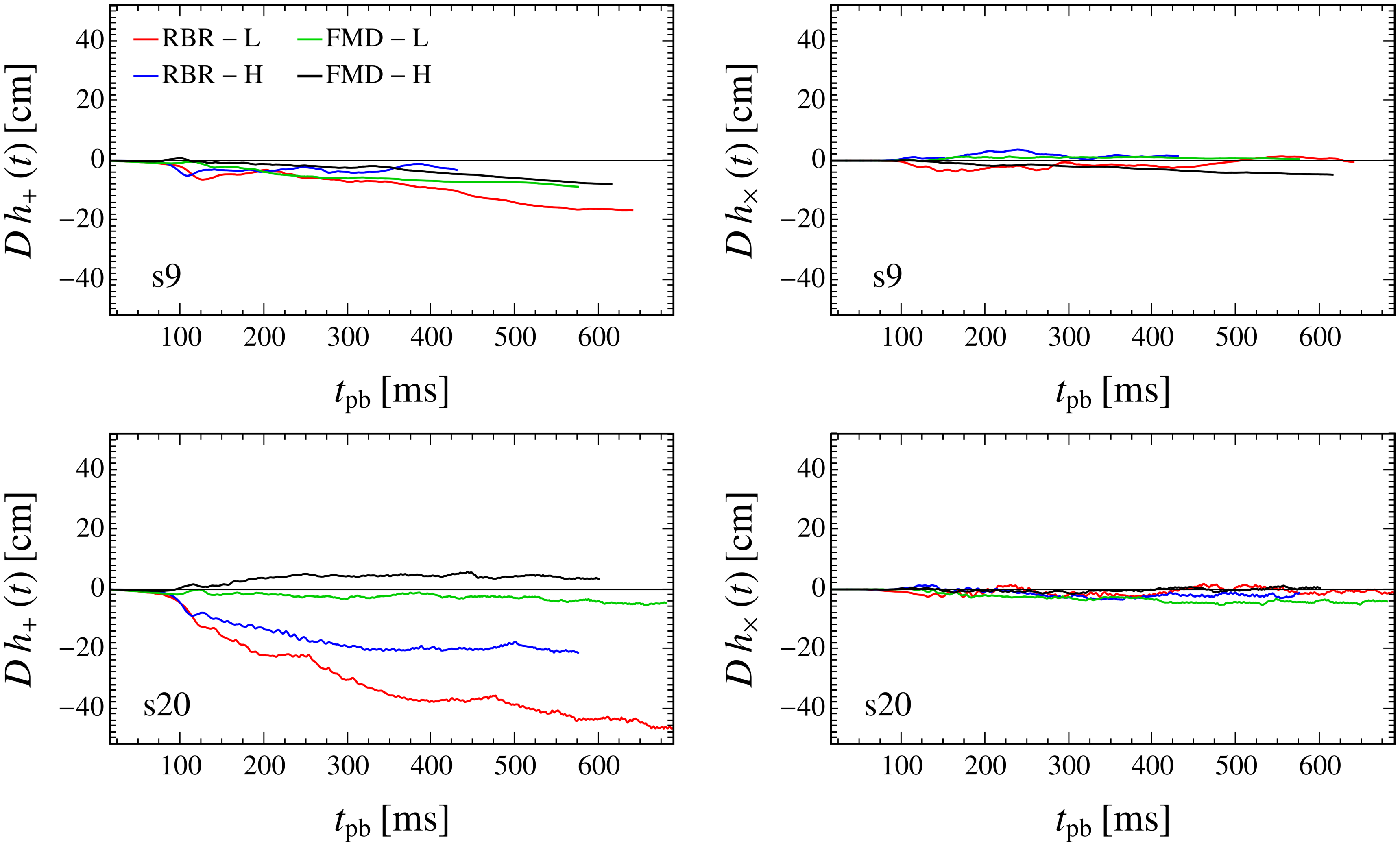}
    \caption{GW amplitudes due to anisotropic neutrino emission as functions of post-bounce time for the $+$ ({\em left panels}) and $\times$ ({\em right panels}) polarization states, assuming an observer along the $(+x)$-axis in the equatorial plane of the source. The \emph{upper panels} and \emph{lower panels} show the results for different SN simulations of a 9\,$M_\odot$ progenitor (models s9) and a 20\,$M_\odot$ progenitor (models s20), respectively. The colors correspond, as labeled, to simulations with ray-by-ray-plus transport (RBR), full multidimensional transport (FMD) as well as low (L) and high (H) angular grid resolution (see text for details). Note that the large differences seen in the lower left panel between RbR+ and full multi-D transport are {\em not} caused by these differences in the numerical treatments, but they are mainly due to stochastic variations for the fixed viewing direction. For the changes of the amplitudes with different viewing angles, see Fig.~\ref{Fig:GW_nu_directions_test}.}
    \label{Fig:GW_nu_signal_test}
\end{figure*}

\begin{figure*}[t!]
    \centering
    \includegraphics[width=1\textwidth]{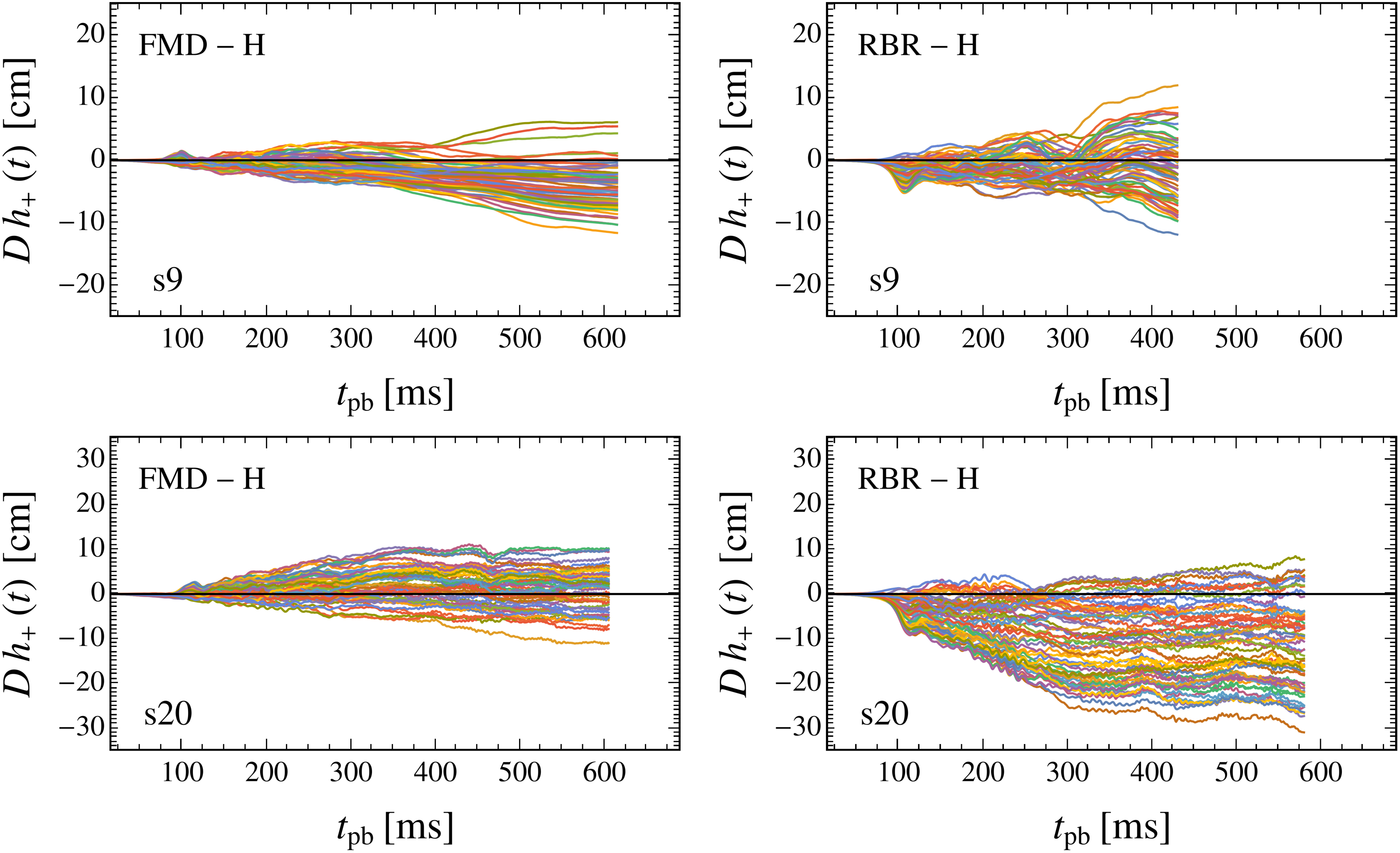}
    \caption{GW amplitudes due to anisotropic neutrino emission as functions of post-bounce time for the $+$ polarization state of our high-resolution s9 (\emph{upper panels}) and s20 (\emph{lower panels}) core-collapse simulations with full multi-D (FMD) neutrino transport (\emph{left panels}) and ray-by-ray-plus (RbR) approximation (\emph{right panels}). The different colors depict different, randomly chosen viewing angles of an observer at a large distance. Note that the scales on the ordinates are different compared to Fig.~\ref{Fig:GW_nu_signal_test} and also between the upper and lower panels. The RbR+ transport approximation appears to be responsible for an earlier development of noticeably larger amplitudes in the simulations for both progenitors. But some of the differences between the results in the left and right panels at later times are likely to be caused by stochastic variations of the model evolution due to the chaotic hydrodynamic instabilities in the postshock region. Therefore they are probably not bigger than case-to-case variations between different simulations with FMD transport.}
    \label{Fig:GW_nu_directions_test}
\end{figure*}

\begin{figure*}[t!]
    \centering
    \includegraphics[width=1\textwidth]{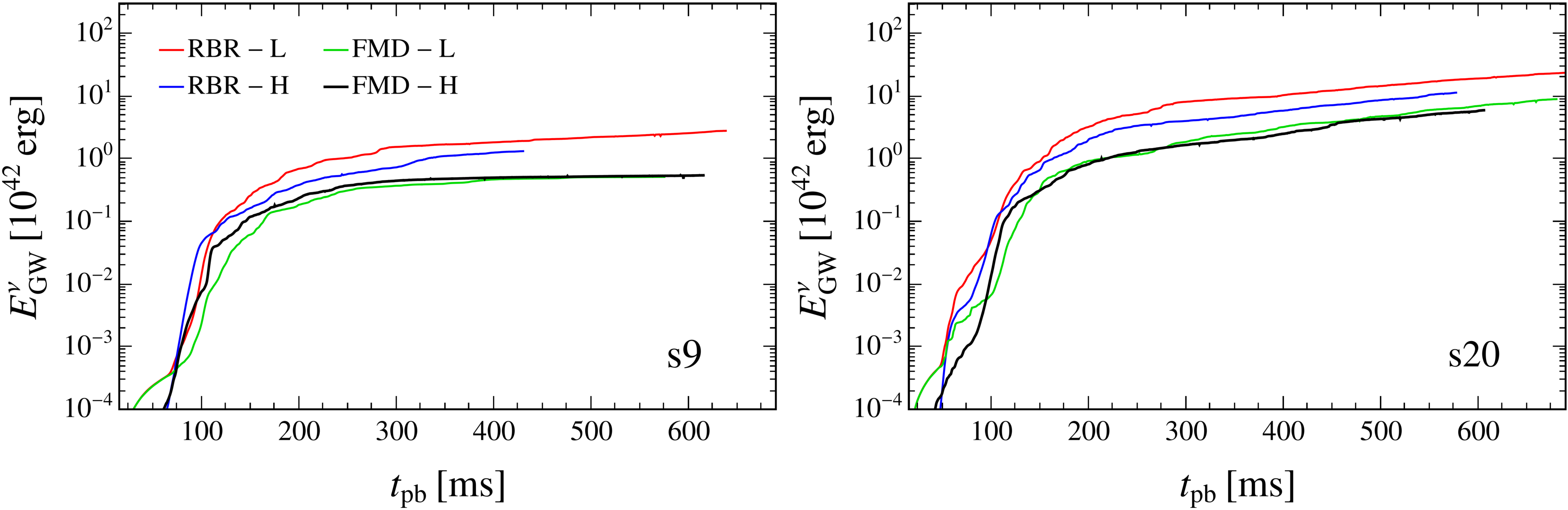}
    \caption{Cumulative energy radiated in neutrino-induced GWs as function of post-bounce time for different SN simulations of a 9\,$M_\odot$ progenitor (models s9; {\em left panel}) and of a 20\,$M_\odot$ progenitor (models s20; {\em right panel}). The colors correspond, as labeled, to simulations with ray-by-ray-plus neutrino transport (RBR), full multidimensional transport (FMD) as well as low (L) and high (H) angular grid resolution (see text for details).}
    \label{Fig:GW_nu_Energy_test}
\end{figure*}

In this Appendix we evaluate the GW signals of additional 3D core-collapse SN simulations of a non-rotating 9\,$M_\odot$ progenitor~\cite{Woosley+2015,Sukhbold+2016} (model s9) and of a non-rotating 20\,$M_\odot$ progenitor~\cite{Woosley+2007} (model s20). In both cases the progenitor data to initiate the core-collapse runs were given as spherically symmetric (1D) outputs from the stellar evolution calculations. These SN simulations were first discussed in Ref.~\cite{Glas:2018oyz} and were performed with the \textsc{Alcar} neutrino-hydrodynamics code, employing a fully multidimensional (FMD) neutrino transport scheme as well as the ray-by-ray-plus (RbR+) approximation for comparison. The former takes all components of the neutrino flux vectors into account, the latter only the radial flux components. Moreover, two different resolutions for the polar coordinate grid of the computations were considered, namely a high-resolution version with 640 radial, 80 lateral, and 160 azimuthal zones (indicated by suffix ``H'' in the model names) and a low-resolution version with 320 radial, 40 lateral, and 80 azimuthal zones (indicated by suffix ``L'' in the model names). All of the model calculations used the SFHo EoS~\cite{Hempel:2011mk,Steiner:2012rk}. In all simulated cases the 9\,$M_\odot$ progenitor exploded successfully, whereas the 20\,$M_\odot$ models failed to yield explosions without any exception. For more details, we refer the reader to Ref.~\cite{Glas:2018oyz}. The names of these models consist of the progenitor mass (s9 or s20), followed by the transport method (FMD or RbR) and by the resolution (L or H), for example s9-FMD-H and s20-FMD-H.

\subsection{Verification of matter GW analysis code}
\label{app:matter}

We used two of these SN simulations to verify the analysis of the matter GW signals for the SN simulations considered in this paper. To do so we tested our newly developed post-processing routine by recomputing the matter GW quantities for models s9-FMD-H and s20-FMD-H, abbreviated as s9 and s20 in this section of the appendix, as representative cases in comparison to the previous evaluation in Ref.~\cite{Andresen:2020jci}. The simulation outputs were stored with a sampling rate of $\delta t=0.5$\,ms, implying a Nyquist frequency of $f_\mathrm{Ny}=1000\,\Hz$.

Figures~\ref{Fig:GW_test_s9} and~\ref{Fig:GW_test_s20} present the GW amplitudes due to asymmetric mass motions obtained by our post-processing of models s9 and s20, respectively. These plots should be compared with the upper panels of Fig.~2 and Fig.~4 in Ref.~\cite{Andresen:2020jci}. The agreement of the results is excellent. Because they were obtained independently with different tools, the outcome is a mutual confirmation of the results and corroborates the proper working of the analysis code employed in the present work. We point out that the contribution from prompt postshock convection is absent in the GW signals of the 9\,$M_\odot$ and 20\,$M_\odot$ SN simulations, because the progenitors were collapsed in 1D and mapped to 3D only at 15\,ms after core bounce, which is too late to follow the convective activity that develops behind the weakening bounce shock.
  
In Fig.~\ref{Fig:GW_test_Energy} the cumulative energy radiated in matter-associated GWs for both the s9 and s20 models is plotted for our analysis versus the results obtained in Ref.~\cite{Andresen:2020jci}. Also here we witness for both models very good agreement between the two independent GW calculations during most of the simulated time evolution. For both SN cases the two lines develop minor discrepancies on the level of 10--15\% only at late times. These deviations are a consequence of small differences in the numerical evaluations, which tend to accumulate when integrating until late times. Independent of that, these results provide further support for the numerical routines applied in our study. 

Finally, in order to check the analysis in the frequency domain carried out on our side, we strove for reproducing the spectrograms of the upper panels in Fig.~3 and Fig.~5 of Ref.~\cite{Andresen:2020jci} for models s9 and s20, respectively. Figure~\ref{Fig:Spectrogram_test} displays our corresponding results. One can recognize the same features and signal components as in the plots of~\cite{Andresen:2020jci}. Both simulations exhibit a low-frequency band ($f \lesssim 200$\,Hz) of activity connected to hydrodynamics instabilities (convective overturn and SASI) in the postshock volume, which is well visible in s9 until the explosion sets in at $\sim$\,300\,ms after bounce~\cite{Glas:2018oyz} but continues in the non-exploding s20 case until the end of the simulated evolution. Moreover, both models possess the high-frequency emission that is associated with the fundamental f/g-mode activity in the PNS and which shows the continuous rise of its dominant frequency. During the period of highest kinetic energies in the neutrino-heating region behind the stalled shock, $100\,\mathrm{ms}\lesssim t_\mathrm{pb}\lesssim 300\,\mathrm{ms}$, model s9 displays broad-band emission between $\sim$\,300\,Hz and $\sim$\,800\,Hz with decreasing intensity towards even higher frequencies. This resembles the extended haze of GW production also seen in the 3D SN simulations of Ref.~\cite{Vartanyan:2023sxm} for the same 9\,$M_\odot$ progenitor. The spectrograms of both of our simulations, however, feature clear aliasing artifacts when the high-frequency strip of GW activity approaches the Nyquist frequency of 1000\,Hz. This prevents our spectrograms to depict the gap-like feature from the avoided mode crossing of PNS pulsations that is well visible at about 1000\,Hz in the plots of Ref.~\cite{Vartanyan:2023sxm}. 

In summary, we can state that the agreement between the results obtained with our GW analysis code and those published previously in Ref.~\cite{Andresen:2020jci} is extremely reassuring and can be interpreted as a convincing verification of the numerical procedures applied in both studies.

\subsection{Neutrino GW signals for additional SN models}
\label{app:neutrinosignals}

In order to complete the GW analysis of the s9 and s20 models considered in this appendix for code testing, we also evaluated the GW signals associated with anisotropic neutrino emission using the treatment outlined in Sec.~\ref{subsec:nu_timedomain}. The neutrino GW signals were not studied before in Ref.~\cite{Andresen:2020jci}. Since we include the entire set of available variants of these models~\cite{Glas:2018oyz,Andresen:2020jci}, i.e., the low-resolution cases as well as those with RbR+ transport in addition to the simulations with high resolution and FMD transport, our analysis also permits us to investigate the effects of the RbR+ vs.\ FMD transport treatment on the properties of the neutrino-induced GW waves. In Ref.~\cite{Andresen:2020jci} such a comparison was carried out only for the GW matter signals. 

Figure~\ref{Fig:GW_nu_Anisotropy_test} presents the time evolution of the neutrino anisotropy parameters for both GW polarization states as observed from the $(+x)$-axis of the simulation frame and for all of the simulation setups available for the s9 and s20 models. Comparing these results to those of models s12.28 and s18.88 in Fig.~\ref{Fig:GW_nu_alpha}, we witness that $|\alpha_{+,\times}|$ in s9 and s20 possess somewhat smaller maxima, ranging within $\sim$\,2\% for s20 in the whole time window under investigation. The anisotropy parameters in the s20 simulations feature the same high-frequency variations on time scales of ${\cal O}(10\,\mathrm{ms})$ as $\alpha_{+,\times}$ in models s12.28 and s18.88, where these high-frequency fluctuations also show peak magnitudes of at most 2--3\%. However, in the latter models they are superimposed on lower-frequency excursions, which happen on time scales of ${\cal O}(100\,\mathrm{ms})$ and drive the maximum neutrino anisotropies farther away from the zero level. The low-frequency modulations are absent in the s20 models. In contrast, in s12.28 and s18.88, they do not only occur during the first second, but with even longer periods (several 100\,ms) also after the onset of the explosion. These low-frequency excursions are mainly responsible for the larger magnitudes of the neutrino anisotropy parameters with peak values up to 5--10\% in extreme maxima. They are caused by time-dependent strengths and directional variations of the neutrino-emitting and neutrino-absorbing accretion flows onto the PNS. In Sec.~\ref{sec:conclu} of the main text we hypothesized that their behavior could be connected to the 3D nature of the pre-collapse progenitors of s12.28 and s18.88, where large-scale convective up- and downdrafts perturb the collapsing oxygen-neon burning layer, in contrast to the spherical structure of the 1D progenitors used as initial conditions in our core-collapse simulations of the s9 and s20 models.

From Fig.~\ref{Fig:GW_nu_Anisotropy_test} it is evident that the peak magnitudes of the neutrino anisotropy parameters in s9 reach marginally half the values they possess in s20 and in the other two simulations discussed in the main part of our paper. This can be understood as a consequence of less massive PNSs and also less massive PNS accretion, which lead to weaker neutrino emission and absorption by accretion flows in the SN models connected to the lowest-mass SN progenitors~\cite{Glas:2018oyz,Andresen:2020jci}.

In more massive progenitors (or progenitors with a higher core compactness), the neutrino-heated postshock layer contains more mass and neutrino heating is stronger. Therefore the non-radial flows in this layer are more vigorous, i.e., their kinetic energy is larger and grows with time. This difference in the intensity of the hydrodynamic mass motions accounts for the hierarchy of GW signals that is reflected by the larger matter GW amplitudes of the s20 model in Fig.~\ref{Fig:GW_test_s20} compared to the s9 case in Fig.~\ref{Fig:GW_test_s9}. The same trend is also visible in the neutrino anisotropy parameters of Fig.~\ref{Fig:GW_nu_Anisotropy_test} and in the neutrino GW amplitudes ---at least $Dh_+$--- shown in Fig.~\ref{Fig:GW_nu_signal_test} for an observer along the $(+x)$-axis of the simulation coordinate frame.

However, this progenitor dependence of the neutrino GW amplitudes mostly disappears at late times (i.e., at $t_\mathrm{pb} \gtrsim 500$\,ms) when the variations for different viewing directions are taken into account (Fig.~\ref{Fig:GW_nu_directions_test}). Despite higher maximum neutrino GW amplitudes in s20 during the first several 100\,ms after core bounce, the amplitudes in both the s9 and s20 models (shown for the high-resolution FMD and RbR+ cases in Fig.~\ref{Fig:GW_nu_directions_test}) are quite similar (around some 10\,cm) at $t_\mathrm{pb} \sim 600$\,ms. This might be caused by the fact that the s9 models explode at $t_\mathrm{pb}\sim 330$\,ms (see Fig.~13 in Ref.~\cite{Glas:2018oyz}), whereas the s20 models do not. In the former case a global asymmetry of the explosion and PNS accretion can develop subsequently, leading to a corresponding, longer-lasting neutrino emission asymmetry. In contrast, in model s20 violent SASI activity takes place during the entire evolution at $t_\mathrm{pb}\gtrsim 200$\,ms (see Fig.~6 in Ref.~\cite{Glas:2018oyz}) and prevents the establishment of a stable geometry for the anisotropic neutrino emission. 

Interestingly, the neutrino GW amplitudes of all s20 simulations exhibit the presence of persistent high-frequency activity, which is absent in the s9 results (Figs.~\ref{Fig:GW_nu_signal_test} and~\ref{Fig:GW_nu_directions_test}). Since SASI mass motions and the corresponding turbulent flows in the postshock volume, which affect PNS accretion and neutrino emission, play a role only in model s20~\cite{Glas:2018oyz}, we consider these hydrodynamic instabilities as the only reasonable explanation for the high-frequency fluctuations of the neutrino GW amplitudes in the s20 cases. Besides the earlier growth to larger neutrino GW amplitudes, these high-frequency contributions explain why the s20 models reach roughly one order of magnitude higher GW energies connected to the anisotropic neutrino emission than the s9 simulations (compare left and right panels of Fig.~\ref{Fig:GW_nu_Energy_test}).
 
Finally, Figs.~\ref{Fig:GW_nu_Anisotropy_test}--\ref{Fig:GW_nu_Energy_test} also permit us to assess the differences of the neutrino GW signals connected to the use of the RbR+ approximation instead of the FMD transport scheme, including resolution effects, in our 3D SN models s9 and s20. The neutrino anisotropy parameters in Fig.~\ref{Fig:GW_nu_Anisotropy_test} suggest that the RbR+ treatment implies somewhat larger peak magnitudes, in particular for the low-resolution cases. This seems to be confirmed by the GW amplitudes in Fig.~\ref{Fig:GW_nu_signal_test}, where the $Dh_+$ for the RbR+ simulations of s20 with both considered resolutions stick out by particularly large absolute values, much larger than the results for FMD transport, the $\times$ polarization amplitudes, and the s9 models. This conclusion, however, is misled by the situation for a single observer position along the $(+x)$-axis of the source frame, which was chosen for consistency with all other plots in this paper. The picture changes when one considers the outcomes for a large set of randomly selected viewing directions in Fig.~\ref{Fig:GW_nu_directions_test}, where only the more relevant high-resolution cases with RbR+ and FMD transport are shown. Now the differences between the amplitudes appear to be less significant. The models with RbR+ method compared to those with FMD exhibit a faster initial growth of the GW amplitudes, a tendency of moderately (several 10\% up to a factor~2) higher absolute values afterwards at intermediate post-bounce times, but no further growth of these differences towards the end of the simulations. 

These differences in the GW amplitudes are reflected by the cumulative energies radiated in neutrino-induced GWs, which are larger by roughly a factor~2 at late times for the simulations with RbR+. But the discrepancy between RbR+ and FMD results clearly shrinks with higher resolution used in the RbR+ models. This can be seen in  Fig.~\ref{Fig:GW_nu_Energy_test}, which also reveals that for SN models of a given progenitor, the GW energies seem to quickly converge to effectively the same value in the low- and high-resolution runs with FMD transport. During early post-bounce times, $t_\mathrm{pb}\lesssim 50$--70\,ms, one can also notice an increased release of neutrino GW energy on a very low level in all low-resolution models compared to the corresponding high-resolution cases. This is a numerical artifact of the post-processing analysis when the integration over all neutrino-emission directions is performed for a reduced number of angular grid points. 
 
Finally, Fig.~\ref{Fig:GW_nu_Energy_test} compared to Fig.~\ref{Fig:GW_test_Energy} shows that the total energy radiated in GWs connected to anisotropic neutrino emission is roughly two orders of magnitude lower than the GW energy associated with asymmetric mass motions in the 9\,$M_\odot$ simulations (see also Fig.~8 in~\cite{Andresen:2020jci}). For the 20\,$M_\odot$ models the difference is a factor of several ten when our calculations were stopped. This is consistent with the factor of $\sim$15 found for both of our s12.28 and s18.88 models (Table~\ref{tab:SNmodels}) in view of the fact that those models were followed for longer evolution times and the neutrino GW energies still continue to grow in Fig.~\ref{Fig:GW_nu_Energy_test} at the end of our simulations.


\bibliographystyle{bibi.bst}
\bibliography{references.bib}


\end{document}